\documentclass[aps,prx,showpacs,notitlepage,twocolumn,superscriptaddress,nofootinbib,preprintnumbers,floatfix,10pt, longbibliography]{revtex4-2}

\usepackage{amsmath,amssymb,amsfonts}
\usepackage{graphicx}
\usepackage{bm}

\usepackage{booktabs}
\usepackage{array}
\usepackage{xcolor}
\usepackage{colortbl}
\usepackage{amsmath}
\usepackage{braket}
\usepackage{amssymb}
\usepackage{amsfonts}
\usepackage{mathtools}

\DeclarePairedDelimiter\floor{\lfloor}{\rfloor}
\usepackage{appendix}
\usepackage{url}

\usepackage{color}
\usepackage{hyperref}
\hypersetup{
  colorlinks=true,
  linktoc=all,
  linkcolor=blue,
}
\usepackage{graphicx}
\usepackage{orcidlink}
\usepackage{soul}
\begin{document}
\title{Non-Abelian Quantum Signal Processing: A Composite Pulse for Fast Analytic Control of Hybrid Oscillator-Qubit Processors}

\author{Shraddha Singh\orcidlink{0000-0002-4921-1410}}\thanks{Corresponding author: shraddha.singh@yale.edu}
\affiliation{Department of Applied Physics and Physics, Yale University, New Haven, Connecticut, USA}
\affiliation{Yale Quantum Institute, Yale University, New Haven, Connecticut, USA}

\author{Baptiste Royer\orcidlink{0000-0001-9695-8552}}
\affiliation{Département de Physique and Institut Quantique, Université de Sherbrooke, Sherbrooke, Québec, Canada}

\author{Steven M. Girvin\orcidlink{0000-0002-6470-5494}}
\affiliation{Department of Applied Physics and Physics, Yale University, New Haven, Connecticut, USA}
\affiliation{Yale Quantum Institute, Yale University, New Haven, Connecticut, USA}
\date{\today}
\begin{abstract}
Quantum Signal Processing (QSP) transforms a unitary parameterized by a classical variable $\theta$ into one governed by a polynomial function $f(\theta)$. Though quantum mechanics is linear, such highly nonlinear transformations arise naturally from the curvature of the qubit Bloch sphere. This capability underpins many quantum algorithms and enables robust control by decreasing sensitivity to errors in a control parameter, or quantum sensing by increasing sensitivity to variations one seeks to measure. Here, we extend QSP from classical, commuting control parameters to noncommuting quantum operators. We call this new multivariate class \emph{non-Abelian QSP}. Rather than treating the noncommutativity of quantum control variables as a complication, non-Abelian QSP exploits their richer commutator algebra as a resource for efficient control. A natural setting is hybrid oscillator-qubit systems realized in superconducting and trapped-ion processors, where the control variables can be the noncommuting position and momentum of a quantum harmonic oscillator. 

We introduce the Gaussian-controlled rotation (GCR), a canonical instance of non-Abelian QSP that cancels the Gaussian quantum-fluctuation error of an oscillator-controlled qubit rotation, matching the robustness of classical composite pulses with a much shorter circuit and an order-of-magnitude lower error. GCR enables fully analytical circuits for preparing squeezed, cat, GKP, and Fock states with performance comparable to state-of-the-art numerically optimized protocols. Building on this primitive, we develop an analytical framework for universal control of GKP bosonic codes, in which logical readout doubles as piecewise gate teleportation, yielding the first high-fidelity, error-mitigated single- and two-qubit logical gates for finite-energy GKP that surpass prior schemes even in the noiseless regime. The same framework supplies mid-circuit error detection and extends in closed form to arbitrary lattices, qudits, and multimode codes relevant to the superconducting-circuit experiments now demonstrating beyond-break-even GKP memory. We further apply these ideas to close a key gap in oscillator-assisted quantum phase estimation. Together, these results show that the non-Abelian structure of quantum control operators can provide a practical resource for robust and efficient quantum control, establishing non-Abelian QSP as both a new theoretical framework and an experimentally accessible control paradigm.
\end{abstract}
\maketitle
\makeatletter
\def\l@subsubsection#1#2{}
\makeatother
\setcounter{tocdepth}{1}
\section{Introduction}

Quantum Signal Processing (QSP)~\cite{low2017optimal,motlagh2023generalized} is a fundamental algorithmic primitive that (roughly speaking) transforms a qubit unitary parameterized by a classical variable $\theta$ into one governed by a polynomial function of that variable, $f(\theta)$.  For example, for the case of a single qubit rotation, we might have
\begin{align}
    U_i[\theta]=e^{i\theta \hat P}\xrightarrow{QSP} U_\mathrm{t}[f(\theta)]\approx e^{i f(\theta) \hat P},\label{eq:QSP1}
\end{align}
where $\hat P$ is a Pauli matrix.  As we will explain, the QSP transformation is achieved via repeated application of the fixed unitary $U_i[\theta]$ alternately with other qubit rotations.
Though quantum mechanics is linear, such highly nonlinear transformations arise naturally from the curvature of the qubit Bloch sphere. QSP is descended from early work in the nuclear magnetic resonance community developing composite pulse sequences that are robust against different types of errors in the pulse controls.   Its broad utility in robust quantum control arises from choosing the polynomial $f(\theta)$ in a way that decreases sensitivity to errors in the control parameter $\theta$. Conversely, in quantum sensing one chooses the polynomial to \emph{increase} sensitivity to the small variations in a parameter $\theta$ that we desire to measure. Importantly and more generally, the QSP primitive has been shown to underpin essentially all known quantum algorithms \cite{martyn2021grand}, thus moving its impact far beyond its roots in magnetic resonance.

In this work, we open a new multivariate regime of QSP---\emph{non-Abelian QSP}---in which the control parameters $\{\hat\theta_1, \hat\theta_2, \ldots\}$ are non-commuting quantum operators. We show that this non-commutativity, rather than being merely a complication, can itself be exploited as a resource for quantum control. Rather than developing a general theory of this regime, we provide its first explicit constructive instance: a single composite pulse, the Gaussian-Controlled-Rotation, from which every result in this paper follows.
Experimental instantiations of this richer algebraic structure are currently being explored in hybrid oscillator-qubit (CV-DV) systems~\cite{brenner2024factoring,ISA,braunstein2005quantum} realized in superconducting~\cite{eickbusch2022fast,haroche2013nobel}, trapped-ion~\cite{bruzewicz2019trapped,fluhmann2019encoding} and neutral atom~\cite{kumar2023quantum,garcia2019single} quantum processors, where the non-commuting variables are quantum harmonic oscillator positions and momenta.  In superconducting systems, the oscillators are the electromagnetic modes of resonators, and for trapped ions, the oscillators are the collective mechanical motional modes of the ions. In neutral atom arrays, the (nearly) harmonic oscillators correspond to the mechanical motion of each atom in its individual optical tweezer trapping potential.

Our work builds on two separate threads of prior research. The abstract necessary conditions for non-commuting multivariate QSP were given in~\cite{nemeth2023variants} but without any sufficiency result, constructive algorithm, or explicit example. Independently, there is a separate history in the quantum control literature of developing gate sequences for hybrid CV-DV systems 
for the preparation of squeezed states, cat states, Fock states, and GKP states in the oscillator \cite{hastrup2021measurement,hastrup2021unconditional,hastrup2022universal,eickbusch2022fast,fluhmann2019encoding,de2022error,rojkov2023two,campagne2020quantum,sivak2023real,matsos_universal_2025,PhysRevLett.133.050602}, or entangled states of the qubits coupled to the oscillator \cite{M-S-gate1,M-S-gate2}, but these gate sequences were not developed from the perspective of NA-QSP. Some of the simplest of these gate sequences, such as the Mølmer-Sørensen gate \cite{M-S-gate1,M-S-gate2} in ion traps, are analytic and `human readable,' but others are only semi-analytic~\cite{hastrup2021measurement,hastrup2021unconditional,hastrup2022universal} or the result of purely numerical optimization from random initial gate sequences~\cite{eickbusch2022fast}.  These are typically difficult to interpret and hard to generalize. Motivated by this, we show in this work that formalizing the key concepts and operations in the language of QSP and NA-QSP permits a deeper understanding and more systematic analytical design of gate sequences that match or outperform state-of-the-art numerical schemes.   In particular, we demonstrate the benefits of interpreting these sequences as providing robustness against quantum fluctuations in the non-commuting control parameters, just as ordinary Abelian QSP can be used to provide control that is robust against classical fluctuations in the control parameters.

In Abelian QSP, the control parameter $\theta$ in Eq.~(\ref{eq:QSP1}) is typically determined by the amplitude of an (essentially classical) electromagnetic pulse applied by the controller to the qubit, or in the case of quantum sensing, the amplitude of a signal to be sensed.  In the hybrid CV-DV systems that we will be considering, the classical control parameter $\theta$ is replaced by a quantum operator $\hat g$
\begin{align}
U_i[\hat g(\hat x, \hat p)]=e^{i\hat g(\hat x, \hat p) \sigma_z}\xrightarrow{\textrm{NA-QSP}}U_\mathrm{t}[\hat f(\hat x, \hat p)]\approx e^{i\hat f(\hat x, \hat p) \sigma_z},\label{eq:U[g]}
\end{align}
where $\hat g$ is an operator that is a linear function of the oscillator position ($\hat x$) and momentum ($\hat p$), that is transformed under non-Abelian QSP (NA-QSP) into a polynomial function $\hat f(\hat x,\hat p)$.

Because of the symplectic structure of the quantum oscillator phase space, it is convenient to parameterize a linear polynomial $\hat g_{\vec u}$ by a fixed two-component vector $\vec u$ via
\begin{align}
    \hat g_{\vec u}&=(u_x,u_p)\left(\begin{array}{cc}
        0 &-1  \\
        +1 & 0
    \end{array}\right)\left(
    \begin{array}{c}
         \hat x  \\
         \hat p
    \end{array}\right),
\end{align}
which makes clear that $\hat g$ is a generator of displacements of the oscillator in phase space in the direction $\vec u$.

If we are dealing with only a single linear function $\hat g_{\vec u}$,
then this is effectively a univariate Abelian case~\cite{sinanan2023single,xsgf-txnq}, which is a trivial extension of the traditional QSP to hybrid CV-DV systems. If, however, we are also dealing with (say) a second such linear function $\hat g_{\vec v}$, 
then we have a bivariate non-Abelian case, provided that $\vec u$ and $\vec v$ are not co-linear so that
\begin{align}
   [\hat g_{\vec u},\hat g_{\vec v}]=\frac{i}{2}\left( u_xv_p-u_pv_x\right)\ne 0,
\end{align}
where, as explained in App.~\ref{app:phase-space}, we are using dimensionless position and momentum variables that obey
\begin{align}
    [\hat x,\hat p]=\frac{i}{2}.\label{eq:xpcommutator}
\end{align}
This illustrates the non-commutative nature of the Weyl-Heisenberg group of translations in phase space.

The non-commutativity of $\hat x$ and $\hat p$, combined with the non-commutativity of DV qubit rotations, yields an even richer algebraic structure that powers NA-QSP.
Interestingly, there are two complementary interpretations of 
the hybrid CV-DV operations\footnote{See Ref.~\cite{ISA} for a concurrent tutorial on hybrid CV-DV systems by overlapping authors; it introduces the phase-space instruction set used here, and cites the present work for non-Abelian QSP.} $U[\hat g(\hat x, \hat p)]$ in Eq.~(\ref{eq:U[g]}).  To illustrate these interpretations, consider the simple case $\hat g(\hat x,\hat p) = u_p\hat x$, where $u_p=k$ is a constant
\begin{align}
    U[k\hat x]&=e^{i(k \sigma_z)\hat x}\label{eq:oscmomboostz}\\
    &=e^{i(k \hat x)\sigma_z}=e^{-i\frac{\hat\theta_{\hat x}}{2}\sigma_z}.\label{eq:Zrotationbyx}
\end{align}
 Eq.~(\ref{eq:oscmomboostz}) invites interpretation as a momentum boost of the oscillator by an amount $\pm k/2$ (see Eq.~(\ref{eq:xpcommutator}) and App.~A to understand the factor of $1/2$) that depends on the state of the qubit.  Conversely, the completely equivalent Eq.~(\ref{eq:Zrotationbyx}) invites interpretation as a rotation of the qubit about the $z$ axis by an angle $\hat\theta_{\hat x}=-2k\hat x$ that is linear in the oscillator position.  Gates of this form are natively available \cite{ISA} in both ion-traps and in superconducting circuits.

\begin{figure*}[t]
  \centering
  \includegraphics[width=\textwidth]{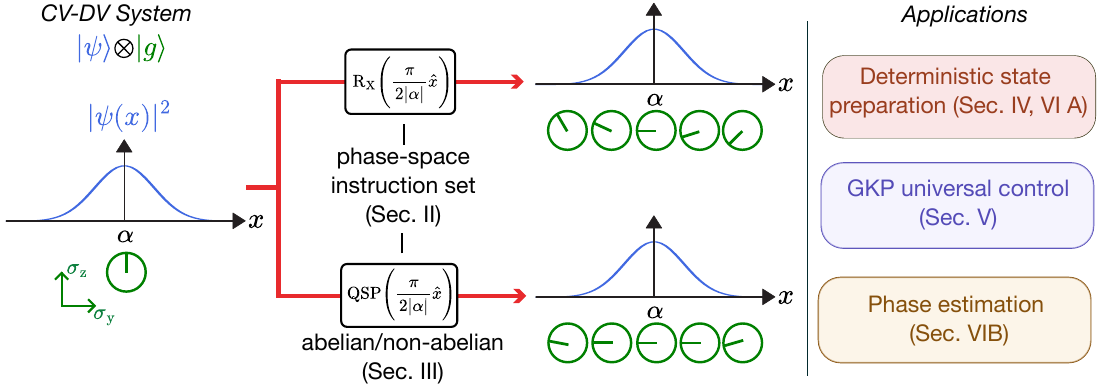}
  \caption{\textbf{Framework of composite pulses in phase space and its 
applications.} A hybrid CV-DV system is controlled via the phase-space 
instruction set, which serves as the substrate for Abelian and 
non-Abelian composite pulse sequences, with applications in oscillator-qubit (CV-DV) control theory. Blue curves show the probability distribution of the 
oscillator position $|\braket{x|\psi}|^2$. The goal is to rotate the 
ancilla qubit spin from $\ket{g}$ to $\ket{\mp i}$ using only 
oscillator-controlled rotations, based on the sign of 
$\braket{\hat{x}}=\pm\alpha$, independent of position uncertainty 
$\Delta$. The qubit state is depicted on the $yz$-plane of the Bloch 
sphere: the initial state $\ket{g}$ points along $+z$ and the target 
state $\ket{-i}$ points along $-y$. \textbf{Left panel:} Depicts the 
hybrid CV-DV system in a product state, controlled via oscillator-controlled qubit 
rotations $\mathrm{R}_\mathrm{X}(2k\hat{x})=e^{-i(k\hat x)\sigma_\mathrm{z}}$ as shown in Eq.~(\ref{eq:Zrotationbyx}). The spread of ticks in the clock around 
the target state on the $yz$-plane of the Bloch sphere illustrates the over- and 
under-rotations arising from Gaussian position uncertainty $\Delta$. The bottom plot introduces our 
key idea --- applying QSP-based Abelian (BB1) or non-Abelian (GCR) composite pulses in oscillator phase 
space to suppress these rotation errors.  \textbf{Right 
panel:} Key applications of this framework include deterministic preparation of oscillator states, universal control of GKP codewords and algorithmic primitives towards universal state preparation and quantum phase estimation.}
  \label{fig:Gaussian-qubit}
\end{figure*}

It is straightforward to show that combining $U[k\hat x]$ with 
$U[-\Delta\hat p]$ 
yields a rich algebraic structure that enables universal control for hybrid CV-DV systems \cite{ISA} due to the non-commutativity of the rotation angles $\hat\theta_{\hat x}$ and $\hat\theta_{\hat p}$.  In particular, NA-QSP sequences can be used to generate higher order polynomials such as
$U[\lambda(\hat x^2+\hat p^2)]$, which can be interpreted as a qubit-dependent rotation of the oscillator in phase space, or as a rotation of the qubit about the $z$ axis by an angle proportional to the energy of the oscillator.  A quartic polynomial can be used to synthetically generate effective anharmonicity in the oscillator to perform quantum Hamiltonian simulations of interacting bosons, spins or fermions coupled to bosons, motion of nuclei in molecules on non-parabolic potential energy surfaces, lattice gauge theories, etc.

The fact that the oscillator position and momentum are non-commuting, of course, leads to Heisenberg uncertainty in these quantities.  We are thus faced with the conundrum that while we may wish to rotate a qubit by an angle proportional to the displacement of an oscillator, there will inevitably be quantum fluctuations in the oscillator position to contend with.  Interestingly, the deleterious effects of these quantum fluctuations can be mitigated using Abelian QSP gate sequences based on purely classical robust control techniques developed by the NMR community \cite{wimperis1994broadband}.

It turns out, however, that the cause of the problem can also be its cure. The very same non-commutativity that produces these quantum uncertainties can be harnessed to construct a simpler and more robust NA-QSP sequence---the Gaussian-Controlled-Rotation (GCR)---that suppresses them. GCR serves as our prototype for explaining NA-QSP and for illustrating its power on experimentally practical finite-energy approximations to the GKP bosonic code, where it significantly improves the mapping of error syndromes and logical states onto the DV qubit for measurement, and enables an improved form of gate teleportation. As we show, this single primitive already yields fully analytical protocols across oscillator state preparation, universal control of GKP error-corrected qubits, and oscillator-assisted phase estimation---domains where comparable results previously required numerical optimization.
\subsection{Outline and Reader's Guide}
We anticipate that this work will be of interest to readers familiar with DV qubits who wish to learn about the extension of QSP ideas to the non-Abelian CV setting, and to readers familiar with CV quantum information who are seeking a more systematic approach to bosonic control problems.
In Sec.~\ref{sec:preliminaries} we will briefly review standard quantum signal processing and introduce NA-QSP, giving an overview in Table \ref{tab:qsp_summary} of the hierarchy of different classes of QSP for hybrid CV-DV systems. We also define the specifics of experimentally realized hybrid gates that can be used for QSP.  Sec.~\ref{sec:comp} shows how traditional NMR composite pulse sequences can be applied to compensate for quantum fluctuations rather than classical fluctuations of the pulse parameters. We then present GCR in detail; this NA-QSP sequence takes particular advantage of the fact that the zero-point fluctuations of oscillators lead to Gaussian distributions of their position and momentum. In Sec.~\ref{sec:state_prep} we present applications of the GCR to the deterministic preparation of squeezed states and non-Gaussian states such as cat states and GKP error correction code words, and compare it against state-of-the-art numerical strategies.  Sec.~\ref{sec:GKP} presents important NA-QSP applications to GKP bosonic error correction codes, giving novel algorithmic primitives for logical readout, error correction, and gate teleportation.  Sec.~\ref{sec:conc} highlights future directions with concrete examples in phase estimation, universal control via preparation of rotation symmetric states like four-legged cat state and Fock state $\ket{1}$, and extensions of hybrid optimization. Sec.~\ref{sec:conclusion} presents a summary, conclusions, and open problems.

\section{Quantum Signal Processing for control of hybrid systems}\label{sec:preliminaries}
In this section, we introduce the key concepts needed to frame the hybrid CV-DV control problem: starting from classical composite pulse sequences, we review quantum signal processing and its extension to hybrid oscillator-qubit systems introducing the two central problem statements in hybrid control, and explain how Gaussian uncertainty in oscillator states creates the control challenge that Sec.~\ref{sec:comp} addresses. For this work, we use $g,e$ to denote the ground and excited states of a qubit, while $\ket{n=0,1,2,..}$ is used to represent the Fock states of an oscillator, with $\ket{0}$ representing the vacuum.

\begin{table*}[htb]
\centering
\begin{tabular}{||l||l||l||l||}
\hline\hline
\textbf{Types of CV-DV QSP} & \textbf{Conditions} & \textbf{Use-case} & \textbf{Refs.} \\ \hline
Univariate QSP & $\hat \theta$ & \begin{tabular}[c]{@{}l@{}}Adaptation of traditional QSP methods \\~\cite{low2016methodology,low2017optimal,martyn2021grand} to quantum control variables \end{tabular} & \cite{sinanan2023single} \\ \hline
\begin{tabular}[c]{@{}l@{}}Multivariate QSP \\ w/ commuting variables\end{tabular} & \begin{tabular}[c]{@{}l@{}}$\hat \theta_1,\hat \theta_2,\hat \theta_3,.. \ \mathrm{s.t.}$ \\ $[\hat \theta_i,\hat \theta_j]=0$\end{tabular} & Control of multiple oscillators & \cite{rossi2022multivariable,nemeth2023variants} \\ \hline
\begin{tabular}[c]{@{}l@{}}Bivariate QSP \\ w/ non-commuting variables\end{tabular} & $[\hat \theta_1,\hat \theta_2]\neq 0$ & \begin{tabular}[c]{@{}l@{}}High-fidelity control of single oscillator\\with low circuit-depth \end{tabular} & This work\\ \hline\hline
\end{tabular}
\caption{\textbf{The landscape of CV-DV QSP.} This table organizes QSP protocols by whether their control variables commute, progressing from classical composite pulses to the non-Abelian regime opened in this work. Overview of different types of QSP techniques, with commuting and non-commuting quantum variables, found in the literature, developed towards universal oscillator control. The bottom two rows belong to the largely unexplored territory of non-Abelian quantum signal processing. In hybrid oscillator-qubit systems, the non-commuting variables are the oscillator position $\hat x$ and momentum $\hat p$. This work presents explicit analytical constructions of non-Abelian QSP sequences with applications in the arena of CV-DV control theory.}
\label{tab:qsp_summary}
\end{table*}
\paragraph{Composite pulses for single-qubit control.} The QSP formalism is the generalization of composite pulses--compositions of multiple single-qubit rotations, developed by the NMR community, to correct for systematic errors in control parameters. As an example, suppose that the target rotation is
\begin{align}
    U_\mathrm{t}&=e^{-i\frac{\theta_\mathrm{t}}{2}\sigma_\phi},\\
     \sigma_\phi&\equiv\cos{\phi} \ \sigma_x+\sin{\phi}\ \sigma_y.
\end{align}
For the experimentally common case where the pulse controller has well-controlled phase, $\phi$, but poorly controlled amplitude, the actual rotation applied will be  
\begin{align}
    U^\textrm{err}_\phi(\theta)&=e^{-i\frac{\theta}{2}\sigma_\phi},\,\textrm{with}\\
      \theta &=\theta_\mathrm{t}(1+\varepsilon),
\end{align}
where $\varepsilon$ represents a small fractional error in the pulse amplitude.

To suppress such errors, composite pulses $U_{\vec{\phi}}(\theta)$, composed of multiple (erroneous) single-qubit rotations with different phases $\vec{\phi}=\{\phi_0,\phi_1,...\}$, are constructed to cancel all low-order error terms up to $\mathcal{O}(\varepsilon^{n-1})$, thereby boosting the rotation fidelity to
\begin{align}
\big|\braket{g|U_\mathrm{t}^\dagger U_{\vec{\phi}}(\theta)|g}\big|^2 =1-\mathcal{O}(\varepsilon^{2n}).
\end{align}
Several well-known composite pulse sequences have been developed for classical (scalar) errors in $\theta$, including BB1~\cite{wimperis1994broadband}, SCROFULOUS~\cite{cummins2003tackling}, and TYCKO~\cite{tycko1985composite}.  In the context of quantum control of qubits, related but more systematic schemes have been developed \cite{ChuangRobust,ChuangRobustErratum,WhaleyRobust,martyn2021grand}.

For our purposes, the BB1 sequence achieves the simplest and most effective error cancellation, and thus, will serve as our benchmark for comparison. The sequence, designed to implement a robust approximation to the rotation $U_\mathrm{t}$ under classical control errors, is given by
\begin{align}
 \mathrm{BB1}(\theta_\mathrm{t})&=e^{-i\frac{\pi}{2\theta_\mathrm{t}}\theta\sigma_\phi}e^{-i\frac{\pi}{\theta_\mathrm{t}}\theta\sigma_{3\phi}}e^{-i\frac{\pi}{2\theta_\mathrm{t}}\theta\sigma_\phi}e^{-i\frac{\theta}{2}\sigma_\mathrm{x}},
 \label{eq:BB1_classical}
\end{align}
where $\phi_1 = \cos^{-1}\left(-\frac{\theta_\mathrm{t}}{4\pi}\right)$. 
 The net effect of the composite sequence is to produce a unitary of the form shown in Eq.~(\ref{eq:QSP1}) with the polynomial $f$ designed to produce a nearly constant output for all input angles $\theta$ near the target angle $\theta_\textrm{t}$
\begin{align}
    f(\theta) &\approx \theta_\textrm{t} + {\cal O}(\varepsilon^{n}),
\end{align}
for some degree $n> 1$.
For the BB1 sequence, $n=3$, thus yielding a fidelity of $|\braket{g|U^\dagger_\mathrm{t}U_{\vec{\phi}}(\theta)|g}|^2=1-\frac{5}{16}(\theta/2)^6$. 
\paragraph{Review of QSP.} Quantum signal processing emerged as a systematic generalization of composite pulse sequences. In particular, these sequences sandwiched a signal, commonly a single-qubit $X$ rotation, between single-qubit $Z$ rotations~\cite{low2016methodology}. For a signal processing vector $\vec{\phi}=\{\phi_0,\phi_1,...\}$, we can write a single-qubit QSP sequence as
\begin{align}
    U_{\vec{\phi}}(\theta)&=e^{-i\frac{\phi_d}{2}\sigma_\mathrm{z}}\prod^{d-1}_{k=0} e^{-i\frac{\theta}{2}\sigma_\mathrm{x}}e^{-i\frac{\phi_k}{2}\sigma_\mathrm{z}} \\
    &=\prod^{d-1}_{k=0} e^{-i\frac{\theta}{2}\sigma_{\Phi_k}}, \quad \Phi_k=\sum_{j=0}^k \phi_j.\label{eq:single-QSP}
\end{align}
We will use the convention that the product of operators is ordered from right to left so that the $k=0$ term is applied first and the $k=d$ term is applied last. The final expression assumes that $\Phi_d=0$. Thus, a QSP sequence is a series of single-qubit rotations about axes on the equatorial plane of the Bloch sphere targeted to achieve a unitary polynomial $U_{\vec{\phi}}(\theta)$. For example, the BB1(90) composite pulse sequence in Eq.~(\ref{eq:BB1_classical}) can be represented in the QSP form as $\vec{\phi}=(0,\lambda,0,2\lambda,0,0,0,-2\lambda,0,-\lambda)$ where $\lambda=\cos^{-1}(-1/8)$. The QSP sequence produces a matrix that can be expressed using polynomial functions of $\cos\frac{\theta}{2}$~\cite{martyn2021grand}, \begin{align}
  U_{\vec{\phi}}(\theta) &= \left( \begin{array}{cc} U_{gg}&U_{ge}\\U_{eg}&U_{ee}  \end{array} \right).\label{eq:Umatrix-QSP}
\end{align}
 For BB1, the first component of this matrix is $U_{gg}(\theta)=|\braket{g|U_{\vec{\phi}}(\theta)|g}|$.

The QSP framework has been subsequently extended from single-variable to multi-variable settings. A general QSP sequence covering all extensions where the signal processing operators are single-qubit rotations is given by
\begin{align}
    U_\phi(\hat\theta_1,\hat \theta_2,\hat \theta_3,..\hat\theta_n)=\prod^{d-1}_{k=0} e^{i\frac{\hat\theta_{s_k}}{2}\otimes\sigma_{\Phi_{k}}},\label{eq:multi-QSP}
\end{align}
where $s_k$ denotes the signal variables for each of the $k$ signal processing steps. Note that this extension may include multi-qubit systems, hybrid oscillator-qubit systems or multi-oscillator systems. A characterization of the polynomial classes achievable by such sequences is discussed in Table~\ref{tab:qsp_summary}. Recent works in this direction developed QSP with multiple commuting control variables~\cite{nemeth2023variants,rossi2022multivariable} where $[\hat\theta_i,\hat\theta_j]=0$ for all pairs of control signal variables.

While the commuting case is now well understood, in this work we push beyond it and begin to address the broader challenge of synthesizing operator-valued polynomials that depend on non-commuting observables $[\hat\theta_i,\hat\theta_j]\neq 0$ for at least one pair of indices $\{i,j\}$. Although a general constructive framework for non-Abelian QSP remains open --- both the characterization of achievable polynomial classes with non-commuting variables and a constructive algorithm mapping target polynomials to phase sequences are outstanding problems --- the hierarchy outlined in Table~\ref{tab:qsp_summary} offers a roadmap for future developments, and identifies promising intermediate classes of protocols that are tractable with current techniques.

We focus on hybrid oscillator-qubit systems, such that the control variables are quantum operator-valued linear functions of the position ($\hat x$) and momentum ($\hat p$) of the oscillator: $\hat\theta_i=\hat f(\hat x, \hat p)$. Importantly, the assumption on linearity of $f$ is sufficient to achieve universal control~\cite{ISA,eickbusch2022fast}. It is useful to note that the QSP sequence $U_{\vec\phi}(\hat\theta_1,\hat \theta_2,\hat \theta_3,..\theta_n)$ defined above can be written in the form of a $2\times 2$ operator acting on the DV qubit,
\begin{align}
  U_{\vec{\phi}}(\hat\theta_1,\hat \theta_2,\hat \theta_3,..\theta_n) &= \left( \begin{array}{cc} \hat U_{gg}& \hat U_{ge}\\\hat U_{eg}&\hat U_{ee}  \end{array} \right),\label{eq:Umatrix}
\end{align}
where each of the $\hat U_{ij}$ blocks is a CV operator acting only on the oscillator. For example, $\hat U_{gg}=\braket{g|U_{\vec{\phi}}|g}$ is the effective operation applied to the oscillator when the qubit outcome is $|g\rangle$, and similarly, $\hat U_{ge}=\langle g| U_{\vec{\phi}}(\hat\theta_1,\hat \theta_2,\hat \theta_3,...,\hat \theta_n)|e\rangle$, etc.  It is important to note that these four block-encoded operators acting on the oscillator need not be unitary--a feature that can be used to advantage in various quantum algorithms.

\paragraph{Problem statements for control of hybrid systems.} For each problem below, we take as the goal that the qubit should be completely unentangled from the oscillator after $U_{\vec{\phi}}$ is applied to the starting state $|g,0\rangle$ (qubit in $|g\rangle$ and oscillator in the vacuum state $|0\rangle$). That is, we want $U_{\vec{\phi}}$ to be block diagonal and $\hat U_{gg}$ to perform a specified target (unitary) operation $\hat U_\mathrm{t}$ on the oscillator. If these conditions are not perfectly satisfied, then we have several important measures of fidelity. 

First, it is useful to know the success probability for the post-selection on the qubit to end up in $|g\rangle$
\begin{align}
  P_g &= 1-P_e = \langle 0|\hat U^\dagger_{gg}\hat U_{gg}|0\rangle = 1-\langle 0|\hat U^\dagger_{eg}\hat U_{eg}|0\rangle.\label{eq:prob}
\end{align}
The second quantity is the hybrid fidelity $F_\mathrm{H}$, which, for the problem of state preparation, is given by
\begin{align}
  F_\mathrm{H}=|\braket{0|\hat U_{\mathrm{t}}^\dagger \hat U_{gg}|0}|^2.\label{eq:fid}
\end{align}
This gives the probability that the ancilla starts and ends in $|g\rangle$ and the oscillator starts in $|0\rangle$ and ends in the target state we want to prepare, $\ket{\psi}$. 
We can also define a third quantity, the post-selected fidelity, given that the qubit did end up in $\ket{g}$, which from Bayes rule is
\begin{equation}
    F_\mathrm{ps}=\frac{F_\mathrm{H}}{P_\mathrm{g}}.
\end{equation}
There are additional fault-tolerance metrics one can consider in the case when the ancilla qubit can raise a flag indicating a leakage error has occurred~\cite{ReinholdErrorCorrectedGates,ma2020path,ma2022algebraic} but this is beyond the scope of the present work.

In defining fidelities for operators, it is necessary to truncate the countably infinite Hilbert space dimension to a finite dimension $d$. We will assume that $d$ is large enough that errors due to this truncation are negligible~\cite{ISA}. If we allow $\mathcal{O}(\eta)$ upper bound on the qubit-oscillator entanglement error, then the problem statements for oscillator-qubit control are framed as follows:

  \textbf{(Problem 1) Universal State Preparation} to realize an arbitrary oscillator state $\ket{\psi}$ starting from vacuum. We require a hybrid unitary $U_{\vec{\phi}}(\hat v_1,\hat v_2,..)$ such that:
  \begin{itemize}
      \item $1-|\braket{\psi|\hat U_{gg}|0}|^2=1-F_\mathrm{H}=\mathcal{O}(\eta)$, and
    \item $||\hat U_{eg}\ket{0}||=\sqrt{|\braket{0|\hat U_{eg}^\dagger \hat U_{eg}|0}|}=\sqrt{P_e}=\mathcal{O}(\eta)$,
  \end{itemize}
  where $||.||$ is the state norm, and $\hat U_{gg}$ ($\hat U_{eg}$) is the effective oscillator operation conditioned on the outcome of measuring the qubit in $|g\rangle$ ($|e\rangle$).

  \textbf{(Problem 2) Universal Control} to synthesize a polynomial Hamiltonian $\hat H (\hat x,\hat p)$ that realizes an arbitrary oscillator unitary ($\hat U_\mathrm{t}=e^{-i\hat Ht}$): Defining $||\hat A^\dagger \cdot \hat B||=\frac{1}{d^2}|\mathrm{Tr}(\hat A^\dagger \hat B)|^2$ as the operator fidelity between operators $\hat A,\hat B$, we need $U_{\vec{\phi}}(\hat v_1,\hat v_2,..)$ to obey:
  \begin{itemize}
  \item $1-||\hat U_{gg}^\dagger \cdot \hat U_\mathrm{t}||=1-F_\mathrm{H} =\mathcal{O}(\eta)$, and
  \item $||\hat U_{eg}||=\mathcal{O}(\eta)$.
  \end{itemize}

With these problem statements in place, we now introduce the instruction set and discuss the physical origin of the control errors that QSP must suppress.

\paragraph{Phase-space instruction set.} Quantum signal processing for hybrid architectures, where the control operators $\hat \theta=f(\hat x,\hat p)$ are linear in the position and momentum of the oscillator, employs the phase space instruction set~\cite{ISA} (see App.~\ref{app:phase-space} for details). The signal is a displacement of the oscillator conditioned on the qubit state, or alternatively, an oscillator-controlled qubit rotation,
\begin{align}
\mathrm{CD}(\beta,\sigma_\phi)&=e^{i2\hat v \otimes \sigma_\phi},\quad\hat v = \mathrm{Im}(\beta)\hat x - \mathrm{Re}(\beta)\hat p.\label{eq:hatvbetadef}
\end{align}
We use the shorthand notation $\hat v\sigma_\mathrm{x}\equiv \hat v\otimes \sigma_\mathrm{x}$ for this operator. These gates have been experimentally realized via dispersive and sideband interactions in superconducting and trapped-ion systems~\cite{eickbusch2022fast,fluhmann2019encoding,de2022error}. Conditional displacements have recently been used to very efficiently compile quantum singular value transformation (QSVT) oracles for hybrid CV-DV hardware~\cite{LukeBellQSVT}, illustrating the broader algorithmic reach of this instruction set. The oscillator position $\hat x$ and momentum $\hat p$ do not commute --- a fact that distinguishes the hybrid CV-DV setting from all purely classical or commuting-variable control problems, and which we exploit directly in Sec.~\ref{sec:GCR}.

\paragraph{Quantum fluctuations.} Quantum states in an oscillator cannot have perfectly defined position and momentum due to the uncertainty principle. A coherent state $\ket{\alpha}$ is a minimal example, represented in the position basis by a Gaussian wavefunction:
\begin{align}
\braket{x|\alpha_\Delta} = \alpha_\Delta(x) = \Big(\frac{2}{\pi\Delta^2}\Big)^{1/4} e^{-\frac{(x - \alpha)^2}{\Delta^2}}, \label{eq:Gaussian}
\end{align}
with mean position $\braket{\hat x}=\alpha \in \mathbb{R}$ and Gaussian uncertainty $\delta x = \Delta/2$. This uncertainty is just the natural uncertainty associated with the zero-point fluctuations of the oscillator ground state or vacuum (for which $\Delta=1$), but may be smaller or larger in squeezed states (see Sec.~\ref{sec:squeezing}). For unsqueezed coherent states, we write $\ket{\alpha} \equiv \ket{\alpha_1}$. For our purposes, the key phase-space information in squeezed coherent states lies primarily in the mean position and momentum encoded in $\alpha$, and secondarily in the uncertainty $\Delta$.

This Gaussian uncertainty is the central practical obstacle in solving Problems~1 and~2: since the oscillator-controlled rotation angle is proportional to the oscillator position, and the position has a Gaussian spread of width $\Delta$, the qubit accumulates a distribution of rotation angles rather than a single definite one. In Sec.~\ref{sec:comp}, we show how classical composite pulse sequences from NMR can be adapted 
to suppress these errors, and introduce GCR as a prototypical non-Abelian composite pulse that 
exploits the non-commutativity of $\hat{x}$ and $\hat{p}$ to outperform BB1, the best composite 
pulse sequence from the traditional NMR literature, while using strictly lower circuit depth.

\section{Composite pulses in phase space}\label{sec:comp}

Oscillator states important for bosonic error correction, like cat states and GKP states, are superpositions of wave functions with the same spread in their Gaussian uncertainty. This associated Gaussian uncertainty causes over- and
under-rotations in the oscillator-controlled qubit gate CD: because the rotation angle is proportional 
to the oscillator position, and the position has a Gaussian spread of width $\Delta$, the qubit 
accumulates a distribution of rotation angles rather than a single definite one. The goal of this 
section is to design QSP sequences that suppress these errors, as illustrated in Fig.~\ref{fig:Gaussian-qubit}.

In the spirit of Eq.~(\ref{eq:U[g]}), we define a target transformation
\begin{align}
    U_\mathrm{t} \approx e^{ik\braket{\hat{x}}\sigma_\mathrm{x}}, \label{eq:targetU}
\end{align}
such that, when $U_\mathrm{t}$ is applied to the hybrid oscillator-qubit state, measurement of the 
qubit in the $\sigma_\mathrm{y}$ basis yields one bit of information about the mean position 
$\braket{\hat{x}}$. The simplest case is distinguishing $\braket{\hat{x}} = \pm|\alpha|$. For example, choosing 
$k = -\theta_\mathrm{t}/2|\alpha|$ with $\theta_\mathrm{t} = \pi/2$ yields
\begin{align}
    U_\mathrm{t} \ket{g} \otimes \ket{\pm\alpha_\Delta} \approx \ket{\mp i} \otimes 
    \ket{\pm\alpha_\Delta}, \label{eq:checkU}
\end{align}
mapping the sign of $\braket{\hat{x}}$ onto two orthogonal qubit states. Note that $U_\mathrm{t}$ 
can only distinguish states with minimal overlap, that is, when $|\alpha|/\Delta$ is large and the 
states $\ket{\pm\alpha_\Delta}$ become nearly orthogonal.

More generally, for states consisting of a superposition of Gaussian peaks located at 
$x = m|\alpha|$, $m \in \mathbb{Z}$, the rotation angle accumulated by the qubit on the $m$th peak in the superposition is $m\theta_\mathrm{t}$. 
Since the qubit can only encode one bit of information per measurement, a single application of 
$U_\mathrm{t}$ can extract only one binary property of $m$ --- for instance, the parity 
$(-1)^m$, obtained by choosing $|\theta_\mathrm{t}| = \pi/2$. This is precisely the measurement 
primitive used in later sections for the control of cat states and GKP states. For clarity, we first focus on real 
$\alpha$ values only.

A naive approach is the single-qubit position-controlled rotation $e^{-i\frac{\theta_\mathrm{t}}{2|\alpha|}\hat x\sigma_\mathrm{x}}$. This pulse achieves the target $U_\mathrm{t}$ only at the peak of the Gaussian, $|x|=\alpha$, but undergoes over- and under-rotations away from the peak due to the uncertainty $\Delta$:
\begin{align}
    e^{-i\frac{\theta_\mathrm{t}}{2|\alpha|}\hat x\sigma_\mathrm{x}}&=\exp\Bigg[{-i\frac{\theta_\mathrm{t}\alpha}{2|\alpha|}\sigma_\mathrm{x}\Bigg]\exp\Bigg[-i\frac{\theta_\mathrm{t}}{2|\alpha|}(\hat x-\alpha)\sigma_\mathrm{x}}\Bigg],\\
    &=U_\textrm{t}\exp\Bigg[-i\frac{\theta_\mathrm{t}}{2|\alpha|}(\hat x-\alpha)\sigma_\mathrm{x}\Bigg].
    \label{eq:cDerror}
\end{align}
The first factor is the target unitary and the second factor represents the erroneous part of this rotation, shown in the blue panel (left figure). Due to the Gaussian envelope $\braket{x|\alpha_\Delta}$, the probability of finding the oscillator far from the peak is small, which limits the effect of these errors for large $|\alpha|/\Delta$. 

To faithfully learn about the oscillator while preserving its state, these errors must be corrected 
in a QND manner. The error operator here takes the same form as in the classical NMR rotations, discussed in 
Sec.~\ref{sec:preliminaries}, as the error is proportional to the target amplitude $\theta_\mathrm{t}$. 
We can therefore employ traditional composite pulse sequences from NMR to cancel these over- and 
under-rotation errors --- classical composite pulses thus serve as a natural precursor to our hybrid 
oscillator-qubit QSP approach. All previous 
approaches in oscillator control~\cite{eickbusch2022fast,hastrup2021improved,hastrup2021measurement,hastrup2021unconditional} 
used numerically optimized CD-based circuits to achieve this error cancellation using 
ancillary qubits. In contrast, the QSP constructions are fully analytical and intuitive.

\paragraph{Quantum control variables.}
One can replace the classical control variable $\theta$ in traditional QSP composite pulses with the quantum operator $\hat{\theta} = (\theta_\mathrm{t}/|\alpha|)\hat{x}$, as long as all rotation arguments in the pulse sequence commute. This substitution enables the direct application of univariate QSP techniques with commuting variables~\cite{low2016methodology,low2017quantum} to quantum systems. A natural starting point is the BB1 sequence, which can be promoted to a hybrid quantum setting by applying the same pulse logic to oscillator-controlled rotations. In its original context, BB1 suppresses the sensitivity of $\braket{g|U_{\vec{\phi}}|g}$ to fluctuations in the classical variable $\theta$ that arise from pulse amplitude errors. Crucially, when the argument is promoted to a quantum variable, $\theta\rightarrow\hat f(\hat x)$, the same sequence can be repurposed to read out the mean position of the oscillator --- now suppressing the sensitivity to fluctuations induced by the Heisenberg uncertainty principle rather than classical noise, while retaining access to encoded phase-space information. Specifically, to correct the rotation $e^{i\frac{\theta_\mathrm{t}}{2|\alpha|}\hat x\sigma_\mathrm{x}}$, using the classical BB1 sequence from Eq.~(\ref{eq:BB1_classical}), we define
\begin{align}
\mathrm{BB1}(\theta_\mathrm{t}) ={}&e^{-i\frac{\pi}{2|\alpha|}\hat x\sigma_{\phi_1}}e^{-i\frac{\pi}{|\alpha|}\hat x\sigma_{3\phi_1}}e^{-i\frac{\pi}{2|\alpha|}\hat x\sigma_{\phi_1}}e^{-i\frac{\theta_\mathrm{t}}{2|\alpha|}\hat x\sigma_\mathrm{x}},\label{eq:BB1}
\end{align}
where $\phi_1 = \cos^{-1}\left(-\frac{\theta_\mathrm{t}}{4\pi}\right)$, directly translating into a sequence of position-controlled rotations. This hybrid composite pulse sequence enhances the fidelity of the target oscillator-controlled qubit rotation by mitigating the quantum fluctuation error due to Gaussian uncertainty, thereby more precisely communicating the sign of the mean position $\braket{\hat x}=\pm \alpha$ of the oscillator to the auxiliary qubit. This sequence has also found its use case in quantum interferometry~\cite{sinanan2023single}.

Traditional QSP sequences like BB1 can only be adapted to Abelian, univariate settings for hybrid control --- leaving potential gains from non-commuting control variables entirely untapped (see blue panel, right figure of 
Fig.~\ref{fig:Gaussian-qubit}). This raises a natural question: can non-Abelian QSP outperform its Abelian counterpart? We answer this affirmatively in Sec.~\ref{sec:GCR}.

\subsection{Gaussian-Controlled-Rotation (GCR): The Non-Abelian Composite Pulse}
\label{sec:GCR}

With the goal of creating a good approximation to the target unitary given in Eq.~(\ref{eq:checkU}), we now introduce the Gaussian-Controlled-Rotation $\mathrm{GCR}(\theta_\mathrm{t})$, a non-Abelian QSP composite pulse sequence, defined as
\begin{align}
\mathrm{GCR}(\theta_\mathrm{t})
  &= e^{-i\frac{\theta_\mathrm{t}}{2|\alpha|}\hat{x}\sigma_{\mathrm{x}}}
     e^{-i\frac{\theta_\mathrm{t}\Delta^2}{2|\alpha|}\hat{p}\sigma_{\mathrm{y}}}
     \label{eq:GCR}.
\end{align}
This is a bivariate QSP sequence with $d=2$ components in the multi-QSP sequence of Eq.~(\ref{eq:multi-QSP}), using the hybrid gates defined in Eq.~(\ref{eq:hatvbetadef}). Compared to the Abelian BB1 sequence in Eq.~(\ref{eq:BB1}), which uses four position-controlled rotations, GCR uses only two gates --- one conditioned on $\hat{x}$ and one on $\hat{p}$. Further, it differs from conventional multivariate QSP~\cite{rossi2022multivariable} in that its two angular variables are non-commuting. As we will show, this non-commutativity is precisely what enables error cancellation with fewer gates when the error fluctuations are known to be Gaussian. Necessary conditions for QSP with non-commuting variables are given in Ref.~\cite{nemeth2023variants}, but no sufficiency result or explicit construction is provided. Numerically optimized pulses using alternating sequences of control parameters $\hat x\sigma_\mathrm{x}$ and $\hat p\sigma_\mathrm{y}$ have previously been used to achieve control of CV-DV architectures \cite{hastrup2021improved,hastrup2021measurement}, however, without any analytical explanation in the context of composite pulses or bivariate NA-QSP.   We provide GCR as a concrete and analytically optimized instance of a NA-QSP sequence.

\begin{figure}[t]
  \centering
  \includegraphics[width=\linewidth]{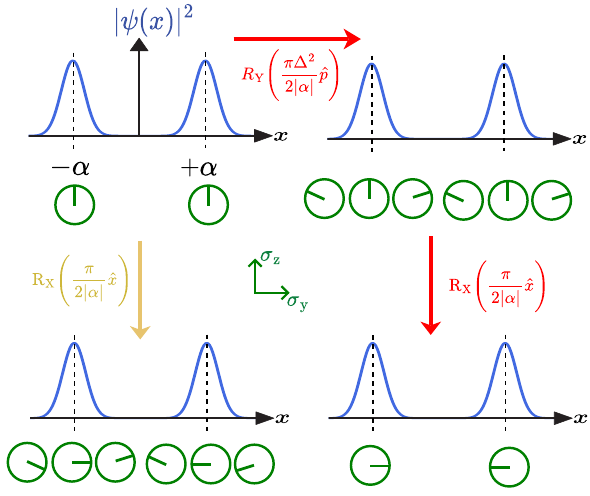}
  \caption{\textbf{Gaussian-controlled-rotation (GCR).} The naive approach (shown by yellow arrow) of applying $R_\mathrm{X}(k\hat x)=e^{-i\frac{k}{2}\hat x\sigma_\mathrm{x}}$ yields over-and under-rotations about the target rotation of $k\braket{\hat x}=\pm k\alpha$. With GCR (shown by red arrows), a momentum-controlled rotation is applied as pre-correction which rotates the qubit about the mean position in the exact way that cancels the over- and under-rotations to yield the final rotation of the qubit predominantly due to the mean position value of each peak ($\pm \alpha$).}
  \label{fig:GCR}
\end{figure}

The working of GCR is illustrated in Fig.~\ref{fig:GCR}. We begin our analysis by noting that the first term in Eq.~(\ref{eq:GCR}) is the simple position-controlled qubit rotation discussed previously.  As we will demonstrate further below, the second term is a `pre-rotation' correction factor designed to approximately cancel the errors from the first term for the specific case of known Gaussian uncertainty. Using Eq.~(\ref{eq:cDerror}), we can rewrite Eq.~(\ref{eq:GCR}) as
\begin{align}
\textrm{GCR}(\theta_\textrm{t})&=U_\textrm{t}\exp\Bigg[-i\frac{\theta_\mathrm{t}}{2|\alpha|}(\hat x-\alpha)\sigma_\mathrm{x}\Bigg] e^{-i\frac{\theta_\mathrm{t}\Delta^2}{2|\alpha|}\hat{p}\sigma_{\mathrm{y}}}.\label{eq:cDerror2}
\end{align}

When acting on the input state $\ket{g,\alpha_\Delta}$, the pre-rotation cancels the error term to first order so that GCR yields an effective polynomial that rotates the qubit by an angle determined by the mean position of the Gaussian wave function:
\begin{align}
  \mathrm{GCR}(\theta_\mathrm{t})\ket{g,\alpha_\Delta}&\approx
  U_\mathrm{t}\ket{g,\alpha_\Delta},
  \label{eq:GCR_action}
\end{align}
that is, the target unitary given by Eq.~(\ref{eq:checkU}): a qubit rotation about an arbitrary equatorial axis by a fixed angle $\pm\theta_\mathrm{t}$ whose sign is determined by the mean position of the oscillator. The key step is that the conditional displacement pre-correction effectively performs a rotation about $\sigma_\mathrm{x}$ that cancels the rotation error (see Eq.~(\ref{eq:cDerror})) arising from Gaussian uncertainty in position to first order in $\frac{\Delta}{\alpha}$. Unlike BB1, which depends only on the mean position $\alpha$, GCR also depends on the uncertainty $\Delta$. The error suppression of this circuit is strictly better than BB1, with circuit duration reduced by at least $4.5\times$ and circuit depth reduced by $2$. Table~\ref{tab:algorithm_comparison} summarizes the full comparison.

We now show how the conditional displacement $e^{i\frac{\lambda}{2}\hat{p}\sigma_{\mathrm{y}}}$ acts as a \emph{pre-correction} to the main position-controlled rotation: it performs a rotation whose amplitude depends on the derivative of the oscillator state.
\begin{align}
    \braket{x|e^{i\frac{\lambda}{2}\hat p\sigma_y}|\alpha_\Delta}&\approx \left[I + \frac{\lambda}{4}\frac{d}{dx}\sigma_{\mathrm{y}}\right]
     e^{-\frac{(x-\alpha)^2}{\Delta^2}}
     \label{eq:pos_basis}\\
     &= \left[I - \frac{\lambda}{2}\frac{x-\alpha}{\Delta^2}\sigma_{\mathrm{y}}\right]
     e^{-\frac{(x-\alpha)^2}{\Delta^2}},
\end{align}
and requires the qubit to be initialised in $\ket{g}$, since $\sigma_\mathrm{y}\ket{g}=i\sigma_\mathrm{x}\ket{g}$, which implies
\begin{align}
e^{i\frac{\lambda}{2}\hat p\sigma_y}\ket{g,\alpha_\Delta}&\approx \left[I - i\frac{\lambda}{2}\frac{\hat x-\alpha}{\Delta^2}\sigma_{\mathrm{x}}\right]\ket{g,\alpha_\Delta}\nonumber\\
     &\approx e^{-i\frac{\lambda}{2}\frac{\hat x-\alpha}{\Delta^2}\sigma_\mathrm{x}}\ket{g,\alpha_\Delta}.\label{eq:GCR-rot1}
\end{align}
These calculations assume $|\lambda|\ll 1$ so that truncating the Taylor expansion to first order in Eqs.~(\ref{eq:pos_basis},~\ref{eq:GCR-rot1}) is justified. Thus, we have proved that the conditional displacement $e^{i\frac{\lambda}{2}\hat p\sigma_\mathrm{y}}$ acts as a position-controlled rotation with amplitude proportional to $(x-\alpha)$. Now, setting $\lambda=-\frac{\theta_\mathrm{t}\Delta^2}{|\alpha|}$ and plugging Eq.~(\ref{eq:GCR-rot1}) into Eq.~(\ref{eq:GCR}) yields the target effect of the unitary $U_\mathrm{t}$, described in Eq.~(\ref{eq:GCR_action}), confirming that the pre-correction exactly cancels the first-order position uncertainty error.

\subsection{Error Bounds}

\begin{figure*}[htb]
   \centering
   \includegraphics[width=\textwidth]{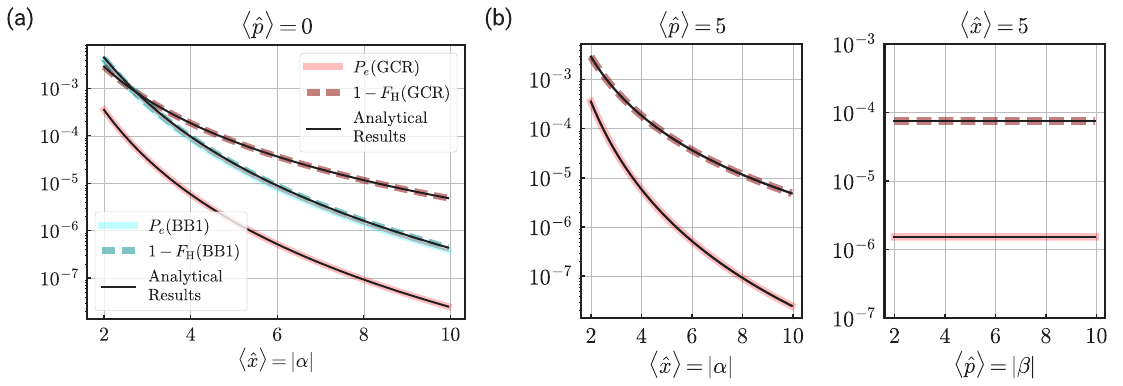}
   \caption{\textbf{Performance of non-Abelian composite pulse sequence $\mathrm{GCR}$ in quantum phase space for $\theta_\mathrm{t}=\pi/2$ and $\Delta=1$.} \textbf{(a)} Comparison against $\mathrm{BB1}(\theta_\mathrm{t})$ for $\chi=\frac{\pi}{4|\alpha|}$. Colored lines: simulated failure probability (solid) and infidelity (dashed) from QuTiP~\cite{Johansson2013}; black lines: analytical expressions Eqs.~(\ref{eq:fail_gcr}--\ref{eq:reset_fid_gcr}) (see App.~\ref{app:comp_err}). Infidelities of $\mathrm{GCR}(\theta_\mathrm{t})$ scale as $\chi^4$; failure probabilities of both schemes scale as $\chi^6$. The $18.5\times$ advantage in $P_e$ is due to the ratio of prefactors at equal
$\chi^6$ scaling. \textbf{(b)} Performance for the coherent basis $\{\ket{\alpha_\Delta+i\beta_\Delta}\}$ with $\alpha\neq0$, $\beta\neq0$. (Left) Varying $|\alpha|$ at fixed $|\beta|=5$ confirms the same scaling. (Right) Varying $|\beta|$ at fixed $|\alpha|=5$ shows independence from $|\beta|$ (Eq.~(\ref{eq:pnot0})), as only a simple rotation is needed to handle states with $\langle\hat{p}\rangle\neq0$.}
   \label{fig:Correctness}
\end{figure*} 

\begin{table}[t]
\centering
\begin{tabular}{>{\centering\arraybackslash}p{2.5cm}
                >{\centering\arraybackslash}p{0.5cm}
                >{\centering\arraybackslash}p{1.0cm}
                >{\centering\arraybackslash}p{1.5cm}
                >{\centering\arraybackslash}p{2.0cm}}
\toprule
\textbf{Pulse} & \textbf{N} & \textbf{T} & $1-F_\mathrm{H}$ & $P_\mathrm{e}$ \\
\midrule
\textbf{BB1} & 4 & $3.6\,\mu$s & $7\times10^{-3}$ & $7\times10^{-3}$ \\
\addlinespace[0.5em]
\rowcolor{lightgray!30}
\textbf{GCR} & 2 & $0.8\,\mu$s & $3\times10^{-3}$ & $4\times10^{-4}$ \\
\bottomrule
\end{tabular}
\caption{Performance comparison of BB1 and GCR for $\theta_\mathrm{t}=\pi/2$ with coherent state $|\alpha|=2$ (first point in Fig.~\ref{fig:Correctness}), at the worst case $\Delta=1$ for GCR. Metrics: number of CD gates (N), circuit duration (T), hybrid infidelity ($1-F_\mathrm{H}$), and failure rate ($P_\mathrm{e}$). CD of amplitude~1 takes $1\,\mu$s~\cite{eickbusch2022fast}. }
\label{tab:algorithm_comparison}
\end{table}

The key small parameter controlling the correctness is,
\begin{align}
  \chi = \frac{\theta_\mathrm{t}\Delta}{2|\alpha|},\label{eq:chi}
\end{align}
a dimensionless parameter that sets the scale of rotation errors due to uncertainty in the signal $\frac{\theta_\mathrm{t}}{2|\alpha|}\hat x$. To bound errors in the pulse, we note that $U_\mathrm{t}^\dagger \mathrm{GCR}(\theta_\mathrm{t})\ket{g}\ket{\alpha_\Delta} = U^\mathrm{Err}\ket{g}\ket{\alpha_\Delta}$, and ideally $U^\mathrm{Err}$ should be simply an identity operation on the oscillator-qubit state.  To quantify this effect, we use the following figures of merit, derivations for which are given in App.~\ref{app:error-analysis}.
\begin{itemize}
    \item \textit{Failure Probability} quantifies qubit rotation errors as $1-|U_{gg}^\mathrm{Err}|^2$ (see Eq.~(\ref{eq:Umatrix})). Odd-order terms ($\chi^3$, $\chi^5$) in the expansion of $U^{\mathrm{Err}}\ket{g}\ket{\alpha_\Delta}$ are the sole contributors to qubit rotation errors (App.~\ref{app:error-analysis}), giving 
\begin{equation}
P_e(\mathrm{GCR})\sim 0.1\chi^6+\mathcal{O}(\chi^{8}),
\quad \chi\ll 1.\label{eq:fail_gcr}
\end{equation}
For comparison (see Apps.~\ref{app:no_corr},~\ref{app:err_BB1}):
\begin{align}
P_e(\mathrm{no\text{-}QSP}) = 0.25\chi^2,
\qquad
P_e(\mathrm{BB1}) = 1.85\chi^6.
\end{align}
GCR and BB1 achieve the same $\chi^6$ scaling (analogous to the classical BB1 scaling in $\varepsilon$; see Eq.~(\ref{eq:BB1_classical})), but BB1's prefactor is $18.5\times$ worse.
\item\textit{Hybrid Infidelity} $|\braket{g,\alpha_\Delta|\mathrm{Err}|g,\alpha_\Delta}|^2$ quantifies oscillator back-action. When $P_e$ is sufficiently small, the hybrid state fidelity governs performance. The residual infidelity of GCR,
\begin{equation}
1-F_\mathrm{H}(\mathrm{GCR})=\chi^4/8+\mathcal{O}(\chi^6),
\label{eq:reset_fid_gcr}
\end{equation}
scales as $\chi^4$ rather than $\chi^6$. This softer scaling reflects unwanted back-action of the momentum-controlled rotation in GCR --- a cost absent in BB1 as it uses only position-controlled rotations and therefore satisfies 
\begin{align}
    1-F_\mathrm{H}(\mathrm{BB1})\sim P_e(\mathrm{BB1})=1.85\chi^6.
\end{align}
 GCR trades a softer $\chi^4$ scaling for lower absolute infidelity in the operating regime for $\chi$ such that $|\alpha|\ge 2$ (see Fig.~\ref{fig:Correctness}). For example, we explicitly compute the values for several metrics at $\alpha=2$ in  Table~\ref{tab:algorithm_comparison}.
 \item \textit{Circuit depth.}
Gate duration scales with displacement amplitude ($T_{\mathrm{CD}(\alpha,\sigma_\phi)}\propto\alpha$), so total displacement amplitude is the appropriate circuit-depth metric. This gives $T_\mathrm{GCR}\propto\frac{\pi}{4|\alpha|}(1+\Delta^2)$ and $T_\mathrm{BB1}\propto\frac{9\pi}{4|\alpha|}$, hence
\begin{align}
  \frac{T_\mathrm{BB1}}{T_\mathrm{GCR}}\ge 4.5,
\end{align}
with equality at $\Delta=1$, the worst case for GCR.\footnote{For $\Delta<1$ the correction pulse is smaller. For $\Delta>1$ position fluctuations are anti-squeezed; applying GCR with $\hat{x}\to\hat{p}$, $\hat{p}\to-\hat{x}$ replaces $\Delta$ by $1/\Delta$, again giving a smaller correction. Both cases confirm $\Delta=1$ as the worst case.}
Since DV ancilla decay --- the dominant error source in hybrid CV-DV architectures --- scales with circuit duration, this reduction directly improves high-fidelity performance.
\end{itemize}

Numerical simulations confirm all analytical results (Fig.~\ref{fig:Correctness}(a)): curves match for both GCR and BB1, and for unsqueezed coherent states ($\Delta=1$, $\theta_\mathrm{t}=\pi/2$) an advantage requires $|\alpha|>2$. Absent qubit or cavity decay, GCR is effectively deterministic --- ancilla measurements serve only to herald extrinsic failures such as qubit or cavity decay errors. 

\subsection{Generalizations of GCR}\label{ssec:pnot0}

The ability of this sequence to remove errors arising from Gaussian position uncertainty inspired 
the name \emph{Gaussian-controlled-rotation}, or GCR. The sequence generalizes straightforwardly: 
the qubit operators may be rotated by any angle $\phi$ about the $z$-axis 
($\sigma_{\mathrm{x}}\to\sigma_\phi$, $\sigma_{\mathrm{y}}\to\sigma_{\phi+\pi/2}$), and the 
phase-space quadratures may be replaced by any pair of perpendicular directions $\hat{x}\to\hat{v}$, 
$\hat{p}\to\hat{v}_\perp$ (see Eq.~(\ref{eq:hatvbetadef})). There are two further cases to consider.

For momentum-squeezed states with $\Delta>1$ and $\braket{\hat{x}}=0$, $\braket{\hat{p}}\neq 0$, 
the position fluctuations are anti-squeezed and the correction should act along $\hat{p}$ rather 
than $\hat{x}$. This is achieved by simply choosing $\hat{v}=\hat{p}$, replacing $\Delta$ by $1/\Delta$, confirming that $\Delta=1$ remains 
the worst case for GCR.

For states with both $\braket{\hat{p}}\neq 0$ and $\braket{\hat{x}}\neq 0$, the wavefunction 
takes the form $\braket{x|\alpha_\Delta+i\beta_\Delta}=e^{i\beta\hat{x}}e^{-(x-\alpha)^2/\Delta^2}$ 
(up to normalization and a phase). The momentum offset $\beta$ introduces an additional 
position-dependent phase that the standard GCR sequence does not account for. This is corrected 
by a simple initial qubit rotation before applying the standard GCR sequence:
\begin{align}
    \mathrm{GCR}(\theta)\ket{g,\alpha_\Delta+i\beta_\Delta}
    = e^{i\beta\hat{x}}\,\mathrm{GCR}(\theta)\,
    e^{-i\frac{\theta\Delta^2}{2|\alpha|}\beta\sigma_\mathrm{y}}
    \ket{g,\alpha_\Delta}.
    \label{eq:pnot0}
\end{align}
Numerical confirmation is shown in Fig.~\ref{fig:Correctness}(b), confirming that GCR extends 
naturally to coherent states along arbitrary phase-space vectors.

A possible extension to non-Abelian composite pulses is the concatenation of Abelian and non-Abelian pulses. GCR suppresses errors from Gaussian position uncertainty but degrades when the oscillator experiences a small displacement error $\langle\hat{x}\rangle\neq\pm\alpha$; BB1, by contrast, is resilient to such displacement errors but carries no information about $\Delta$. Concatenating the two --- replacing each BB1 rotation with the corresponding GCR pulse --- would appear to combine these complementary strengths, but the deterministic (unitary) composition performs poorly: the physically realizable pre-correction cancels the position-uncertainty error only to leading order in the rotation angle, so BB1's large-angle $(\pi,\ 2\pi, \ \pi)$ pulses break this condition and wash out the response. The exact GCR correction can instead be realized in this composition probabilistically by heralding an ancilla, recovering a near-ideal response on a $\lesssim\!17\%$ post-selected fraction. The issue with this pulse and its heralded  analogue has been described in App.~\ref{app:bb1gcr}. We do not pursue this composition, as the concatenated sequence is long and unfavorable for noise resilience.

Based on these results, we use GCR to design analytical preparation schemes for states with favorable representation in phase space, like Squeezed, Two-Legged Cat, and GKP states in Sec.~\ref{sec:state_prep}, while the preparation of rotation-symmetric states, like four-legged cat and Fock states, is given in Sec.~\ref{ssec:universal}. We also highlight its utility in bosonic error correction with a use case in stabilization as well as logical gates for universal control of GKP states in Sec.~\ref{sec:GKP}.
\section{Deterministic Bosonic State Preparation}\label{sec:state_prep}
In this section, we leverage the Gaussian Controlled Rotation (GCR)
composite pulse sequence developed in Sec.~\ref{sec:GCR} to derive
deterministic preparation schemes for structured, non-Gaussian oscillator
states — particularly superpositions of non-overlapping Gaussian
wavefunctions.  We introduce a modular set of hybrid quantum control
routines — or ``gadgets'' — constructed using GCR, and use them to design
fully analytical schemes for deterministic preparation of two-legged cat
states (Sec.~\ref{ssec:Cat_States}), squeezed vacuum states
(Sec.~\ref{sec:squeezing}), and GKP codewords
(Sec.~\ref{ssec:GKP-States}).  These protocols are analytically derived,
match or exceed the performance of numerically optimized schemes, and are
directly implementable in current superconducting and trapped-ion
platforms.

The modular structure of these gadgets extends naturally beyond the
experiment-ready results presented here.  In the applications and outlook
section (Sec.~\ref{sec:conc}), we use these same tools as a starting
point for analytical preparation of rotationally symmetric states —
including $N$-legged cat and Fock states — as concrete steps toward the
broader open problem of universal oscillator state preparation formulated
in Problem~1 of Sec.~\ref{sec:preliminaries}.  Those examples illustrate
both the reach of the phase-space ISA and the fundamental constraints that
arise when it is pushed beyond its natural domain.

Throughout the rest of the section, we adopt the tensor product convention $\ket{\mathrm{osc}}\otimes\ket{\mathrm{qubit}}$ for the joint system, and refer to the total amplitude of conditional displacements (CDs) in a pulse sequence as the circuit duration (reported in $\mu$s). The precise mapping from CD amplitude to runtime in $\mu$s is detailed in App.~\ref{app:squeezing}. We focus on noiseless state preparation in this section, and performance against realistic noise parameters is discussed in the next section.

Before introducing our state preparation protocols, we define two gadgets which will be used throughout all the methods below. Consider states of the form
\begin{equation}\label{eq:stateSumOfGaussians}
    \ket{\psi_\Delta} = \sum_{m \in \mathbb  Z} c_m \ket{(m \alpha)_\Delta},
\end{equation}
where $\{c_m\}$ are complex coefficients. States with this description correspond to evenly-spaced superpositions of Gaussian peaks, such as cat or GKP states. The gadgets we introduce allow us to entangle ($\mathcal E$) or unentangle ($\mathcal U$) a qubit with the individual peaks (indexed by $m$), without being sensitive to the position or momentum fluctuations of each peak, and will allow preparation of states of the form of Eq.~(\ref{eq:stateSumOfGaussians}).

\paragraph{$\mathcal{E}$:} A Gaussian-peak-dependent entangling gadget can be defined as 
\begin{equation}
  \mathcal{E}(\theta,\alpha,\Delta)\,:\,\mathrm{GCR}(\theta)  \textrm{ or } \mathrm{BB1}(\theta).
\end{equation}
Applying the entangling sequence to a state of the form of Eq.~(\ref{eq:stateSumOfGaussians}), the entangling gadget has an action
\begin{equation}
    \mathcal E(\theta,\alpha,\Delta)\ket{\psi_\Delta}\ket{q} \approx \sum_{m \in \mathbb  Z} c_m \ket{m \alpha_\Delta}\otimes e^{-i \frac{m\alpha}{2|\alpha|}\theta_\mathrm{t}\sigma_\mathrm{x}}\ket{q},\label{eq:ent}
\end{equation}
where $\ket{q}$ is any qubit state. Note that this entangling gadget depends on $\alpha$ and $\Delta$, which requires prior knowledge of the state on which this sequence is applied.

\begin{figure}[htb]
  \centering
  \includegraphics[width=\linewidth]{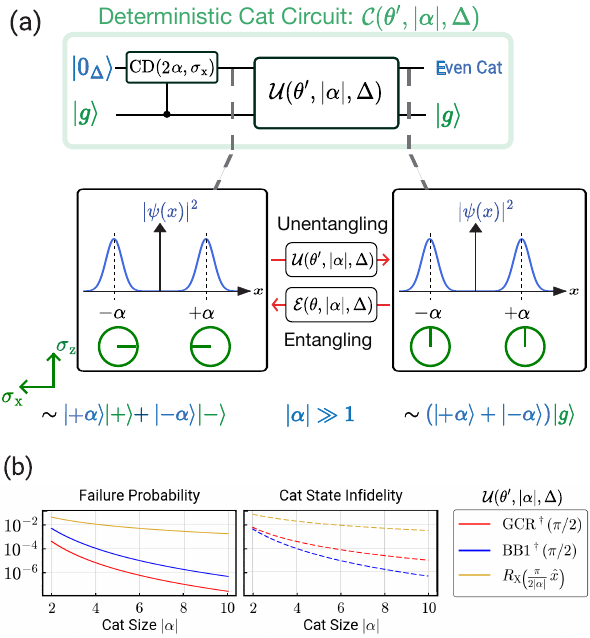}
  \caption{\textbf{Deterministic preparation of two-legged cat states.} \textbf{(a)} Deterministic cat state preparation requires an un-entangling sequence given by $\mathcal{U}$. \textbf{(b)} We show numerical results with options of no correction ($\mathcal{U}=\mathrm{R}_\textrm{y}(\theta'\hat x/|\alpha|)$ in yellow), univariate or traditional QSP correction ($\mathcal{U}=\mathrm{BB1}$ in cyan), bivariate non-Abelian QSP correction ($\mathcal{U}=\mathrm{GCR}$ in red). (Left) Failure probability of ancilla ending in state $\ket{g}$. (Right) Infidelity of output oscillator state with the desired cat state upon success.}
  \label{fig:Cat_states}
\end{figure}

\paragraph{$\mathcal{U}$:} Starting in an entangled hybrid oscillator-qubit state of the form above, an unentangling circuit fixes the qubit state at the end of the circuit (see Fig.~\ref{fig:Cat_states}(a)). We define the unentangling gadget as any circuit of the form,
\begin{equation}
  \mathcal{U}(\theta',\alpha,\Delta):\mathrm{GCR}^\dagger(\theta')  \textrm{ or } \mathrm{BB1}^\dagger(\theta').
\end{equation}
This gadget is used to un-entangle an entangled oscillator-qubit pair, such that
\begin{equation}
    \ket{\psi_\Delta'}\ket{q'} \approx \mathcal U(\theta',\alpha,\Delta)\sum_{m \in \mathbb  Z} c_m \ket{m \alpha_\Delta}\otimes e^{-i \frac{m\alpha}{2|\alpha|}\theta_\mathrm{t}\sigma_\mathrm{x}}\ket{q},\label{eq:unent}
\end{equation}
with $\ket{\psi_\Delta'}$ of the form of Eq. (\ref{eq:stateSumOfGaussians}) and $\ket{q'}$ a fixed qubit state.
Evidently, the choice $\theta = -\theta'$ achieves this goal (i.e. $\mathcal U = \mathcal{E}^{-1}$). However, any choice of $\theta'$ such that
\begin{align}\label{eq:piConditionUE}
  \theta+\theta'=k\pi\quad k\in\mathbb{Z},
\end{align}
also achieves the desired objective, potentially resulting in $\ket{q} \neq \ket{q'}$. In particular, choosing $\theta' \neq -\theta$ leads to a situation where $\mathcal U \mathcal E$ applies a trivial action only on states of the form of Eq.~(\ref{eq:stateSumOfGaussians}), which as we show below in Sec. \ref{sec:GKP} leads to interesting error-correction properties. 

From these gadgets, we now introduce our state preparation protocols for various oscillator states that follow the representation of Eq.~(\ref{eq:stateSumOfGaussians}).

\subsection{Two-legged cat states}\label{ssec:Cat_States}
The superposition of two coherent states located at diametrically opposite locations in the phase space of an oscillator is known as a two-legged cat state.
We define
\begin{align}
  \ket{C_{+\alpha}}&\propto(\ket{\alpha}+\ket{-\alpha})\quad \text{Even Cats}\\
  \ket{C_{-\alpha}}&\propto(\ket{\alpha}-\ket{-\alpha})\quad\text{Odd cats},
\end{align}
and if the basis states $\ket{\pm\alpha_\Delta}$ with $\Delta\neq 1$ are used in this definition, then $\ket{C_{\pm\alpha}}$ are squeezed cat states. These are of the form of Eq.~(\ref{eq:stateSumOfGaussians}), with $c_1 = \pm c_{-1}$ and $c_{k}=0$ for $k \notin \{\pm1\}$.

\par\textbf{Protocol:} The protocol starts with the qubit in the ground state and the oscillator in a (squeezed) vacuum state. An even cat state is then obtained through the sequence
\begin{equation}
    \ket{C_{+\alpha}} \ket{g} \approx \mathcal U(-\pi/2,\alpha,\Delta) \mathrm{CD}(2\alpha,\sigma_\mathrm{x}) \ket{0_\Delta}\ket{g}.
\end{equation}
For simplicity, we assume below $\Delta=1$, i.e. an initial vacuum state, and choose $\mathcal U = \mathrm{GCR}$. After the first conditional displacement, the state is given by
\begin{align}
  \mathrm{CD}(2\alpha,\sigma_\mathrm{x})\ket{0}\ket{g}&\propto\ket{\alpha}\ket{+}+\ket{-\alpha}\ket{-}.
\end{align} 
This leaves the oscillator-qubit in an entangled state. At this point, even or odd cat states can be prepared probabilistically by measuring the qubit in the $\sigma_\mathrm{z}$ basis~\cite{sun2014tracking,wineland2013nobel,haroche2013nobel}. However, for a deterministic process, one can instead apply a $\pm \pi/2$ qubit rotation, where the sign of the rotation depends on the associated oscillator state. This is precisely the effect of the $\mathcal U$ sequence introduced above, which applies on the qubit a rotation $R(\pi/2 \times \mathrm{sign}(\alpha))$, see Eq.~(\ref{eq:GCR}). After the full sequence, the resulting state is given by
\begin{equation}\label{eq:cat-state}
\begin{aligned}
  &\mathrm{GCR}^\dagger(\pi/2) \mathrm{CD}(2\alpha,\sigma_\mathrm{x}) \ket{0_\Delta}\ket{g}\\
   &\quad\propto e^{-i\frac{\pi}{4|\alpha|}\hat p \sigma_\mathrm{x}}e^{i\frac{\pi}{4|\alpha|}\hat x \sigma_\mathrm{y}}(\ket{\alpha}\ket{+}+\ket{-\alpha}\ket{-})\\
  &\quad\approx (\ket{\alpha}+\ket{-\alpha})\otimes\ket{g}+O(\chi^2)\ket{\psi'}\otimes\ket{g}\\
  &\quad +O(\chi^3)\ket{\psi''}\otimes\ket{e},
\end{aligned}
\end{equation}
where $\chi$ is the error parameter defined in Eq.~(\ref{eq:chi})~\footnote{GCR used here is a generalization where the main vector is conditioned on $\sigma_\mathrm{y}$ and pre-correction is conditioned on $\sigma_\mathrm{x}$.}. Odd cats $\propto(\ket{\alpha}-\ket{-\alpha})$ can similarly be prepared by starting in qubit state $\ket{e}$ (see App.~\ref{app:odd-cat}).

\par\textbf{Discussion:} In Fig.~\ref{fig:Cat_states}(b) we fix $\Delta = 1$, vary $\alpha$ and plot two quantities: (i) $1-P_e$ which is marked by the measurement of ancilla in $\ket{e}$ and (ii) $1-F_\mathrm{H}$, the infidelity of the hybrid output state against the desired even cat state with the qubit in $\ket{g}$. Low $P_e$ for $\mathrm{GCR}$ and $\mathrm{BB1}$ indicates that ancilla measurement is not required. As shown in Sec.~\ref{sec:GCR}, the un-entanglement error $P_e$ decreases with increasing $|\alpha|$ and decreasing $\Delta$. We can see the same trend as given by the corresponding correctness metrics for our framework in Fig.~\ref{fig:Correctness}. To study the effectiveness of QSP, we also show the respective curves when choosing the simpler unentangling step as $\mathrm{CD}(\pi/4|\alpha|,\sigma_\textrm{y})$. The analytical expression $P_e=\mathcal{O}(\chi^2)$ for this curve has been derived in App.~\ref{app:cat_I}. The analytical fidelity for the $\mathrm{GCR}$ and $\mathrm{BB1}$ has been computed in App.~\ref{app:comp_err}. Thus, we have shown orders of magnitude improvement achieved from our framework of using composite pulses in phase space when correcting continuous-variable rotation errors on qubits.

\subsection{Squeezed states}\label{sec:squeezing}
\begin{figure*}[ht]
  \centering
  \includegraphics[width=\linewidth]{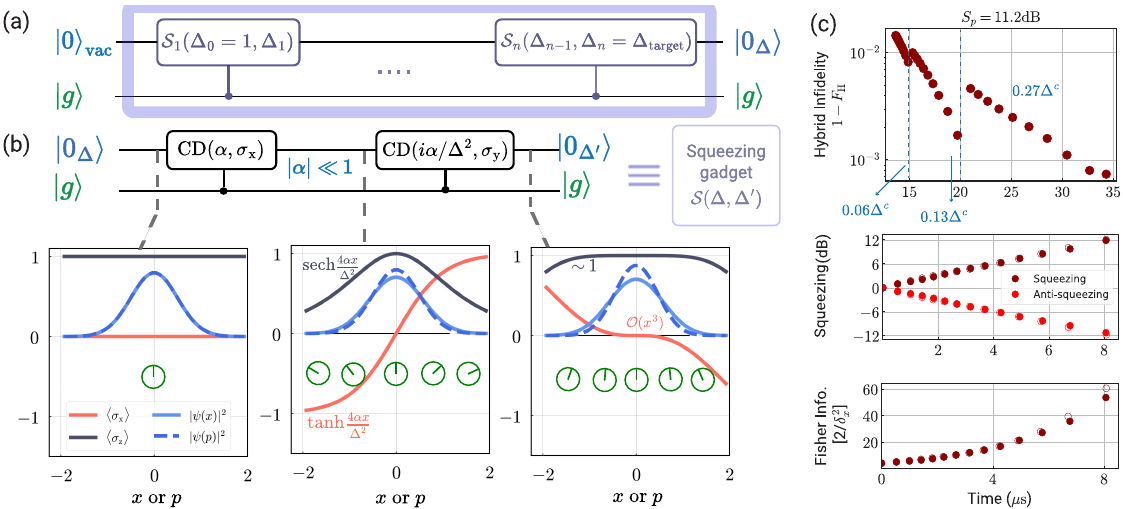}
  \caption{\textbf{Deterministic preparation of squeezed states.} \textbf{(a)} Deterministic squeezing protocol with incremental $\mathrm{GCR}$. \textbf{(b)} Interpretation of $\mathrm{GCR}$ as a squeezing gadget $\mathcal{S}(\Delta,\Delta')$. The plots show how this sequence introduces a small amount of squeezing while unentangling the qubit from the final state for $\Delta=1,|\alpha|=0.25$. \textbf{(c)} All plots share the x-axis legend: Time($\mu$s). (Top) Variation in fidelity and circuit duration with varying squeezing rate $|\alpha|_{k+1}=a\Delta_{k}^c$ with 10 points in the range $c\in[-2,0]$, for three different protocols with $a\in\{0.06,0.13,0.27\}$ aimed at a target squeezing of $11.2 \ \mathrm{dB}$. (Middle) Squeezing (maroon) and anti-squeezing (red) are shown as a function of the circuit duration for the faster protocol with $c=2$. See App.~\ref{app:squeezing} for definitions of $S_x,S_p$ in terms of $\Delta$. (Bottom) Fisher information for the faster protocol. The empty circles in (c) represent a plot of the results for the case when post-selection is activated. Here, $\mu$s denotes the total length of conditional displacements; conversion to actual runtime can be found in App.~\ref{app:squeezing}.
  }
  \label{fig:squeezing}
\end{figure*}

We present a novel approach for generating squeezed states. This result is comparable to state-of-the-art schemes in Refs.~\cite{hastrup2021unconditional,eickbusch2022fast} without the need for numerical optimization tools. In addition, our construction allows optimization of the fidelity with respect to circuit duration. Our protocol has a basic unit composed of $\mathrm{GCR}$, which can be understood as a deterministic small-even-cat preparation circuit or, equivalently, a slightly squeezed state. 

\paragraph{Squeezing gadget:}  Our gadget $\mathcal{S}_k(\Delta_{k-1},\Delta_k)$ takes a state with variance $\Delta_{k-1}^2$ and outputs a state with variance ${\Delta_k}^2$. We consider oscillator state wavefunctions,
\begin{align}
\psi(x)\propto e^{-\frac{(x-\beta)^2}{\Delta^2}},
\end{align}
where the position and momentum uncertainties are given by $\delta x=\Delta/2,\delta p=1/(2\Delta)$, respectively. For simplicity, we focus henceforth on the case $\beta= 0$ and momentum squeezing, $\Delta_{\mathrm{target}} > 1$.

As shown in Fig.~\ref{fig:squeezing}(b), our squeezing gadget at the $k$th step of the protocol consists of two conditional displacements,
\begin{equation}
    \mathcal S_k(\Delta_{k-1},\Delta_{k}) = \mathrm{CD}(i\alpha/\Delta,\sigma_\mathrm{y})\mathrm{CD}(\alpha,\sigma_\mathrm{x}),
\end{equation}
where the choice of $0 < \alpha \ll 1$ sets the resulting fluctuations, $\Delta'$, and the distance of the final state to a true squeezed state. This sequence mirrors the position-based rotation from $\mathrm{GCR}$ in the momentum basis. 

To gain intuition about the process, it is easier to interpret the effect of a CD gate as a rotation of the qubit state conditioned on the oscillator state. Accordingly, after the first conditional displacement $\mathrm{CD}(\alpha,\sigma_\mathrm{x})=e^{i\alpha\hat p\sigma_\mathrm{x}}$ is applied to the joint state $\ket{0_\Delta}\otimes\ket{g}$, the expectation values of qubit operators conditioned on the oscillator position are given by,
\begin{align}
 \braket{\sigma_\textrm{x}}_x&=\tanh{\frac{4\alpha x}{\Delta^2}},\\
 \braket{\sigma_\textrm{y}}_x&=0,\\
 \braket{\sigma_\textrm{z}}_x&=\mathrm{sech}{\frac{4 \alpha x}{\Delta^2} }.
\end{align}
Note that $[\mathrm{CD}(\alpha,\sigma_\mathrm{x}),\sigma_\mathrm{x}] = 0$, such that for a qubit initialized in $\ket{g}$ we have
\begin{align}
\braket{\sigma_\mathrm{x}} = \int_{-\infty}^\infty \ dx \ \braket{\sigma_\mathrm{x}}_x=0,
\end{align} 
as should be the case. However, for a specific value of $x$, $\braket{\sigma_\mathrm{x}}_x \propto \braket{\ket{x}\bra{x}\sigma_\mathrm{x}}\neq 0$ is possible. Keeping $\alpha$ small ensures that $\braket{\sigma_\mathrm{x}}_x \propto x$ on the interval where $\psi(x)$ has significant support, which enables better disentanglement of the qubit. To disentangle the qubit and restore $\braket{\sigma_\mathrm{z}} = 1$, we apply a $\sigma_\mathrm{y}$-rotation of the qubit by an angle $\sin^{-1}(\tanh{\frac{4\alpha x}{\Delta^2}}) \approx \frac{4\alpha x}{\Delta^2}$ for $4|\alpha|x/\Delta^2 \ll 1$, i.e., $\mathrm{R}_\mathrm{y}(-4\alpha \hat{x}/\Delta^2)=\mathrm{CD}(i\alpha/\Delta^2, \sigma_\mathrm{y})$. This corrects the linear part of $\braket{\sigma_\textrm{x}}_x$, setting
\begin{align}
\braket{\sigma_\textrm{x}}_x\rightarrow 0,\braket{\sigma_\textrm{z}}_x\sim 1,
\end{align}
in the range where $x\ll \Delta^2/4|\alpha|$. Exact expressions for $\braket{\sigma_\textrm{x}}$ and $\braket{\sigma_\textrm{z}}$ are given in App.~\ref{app:squeezing}. After disentanglement of the qubit, the oscillator state corresponds to a small even (squeezed) cat state, i.e., a superposition of two Gaussian states which approximate a single Gaussian state with increased variance $\Delta_k^2 \approx \Delta_{k-1}^2+\alpha_k^2$. The repeated application of this squeezing gadget can yield arbitrary squeezing of the vacuum state.

\par\textbf{Protocol:} In Fig.~\ref{fig:squeezing}(a), we illustrate repeated application of the
squeezing gadget $\mathcal{S}_k$ yielding a desired target variance
$\Delta_\mathrm{target}^2$.  Careful selection of $\alpha_k$ for successive
steps $k$ is crucial: for optimal squeezing, $\alpha_k$ should be as large
as possible while ensuring (1) the slope of $\langle\sigma_\mathrm{x}\rangle_x$
is linear over $|x|\le\Delta_k$, and (2) the resulting sum of two
overlapping Gaussians can be well approximated by a single Gaussian.  For
the slope to remain linear over the FWHM of the state, we derive
$\alpha_k\ll 0.13\Delta_{k-1}^{1/2}$ (App.~\ref{app:squeezing}).
To systematically explore the tradeoff between convergence speed (circuit
duration) and disentanglement (oscillator-qubit fidelity), we parameterize
$\alpha_k=a\Delta_{k-1}^c$ for various $a$ and $c$, targeting
$S_p=11.2\ \mathrm{dB}$ (Fig.~\ref{fig:squeezing}(c)).  For $c<0$,
increasing $a$ or $c$ improves fidelity but slows convergence.  For $c>0$,
convergence is faster but the state becomes increasingly non-Gaussian,
eventually deviating too far from a true squeezed state.  These cases were studied for a pedagogical understanding of the competition between convergence speed and Gaussianity, captured via infidelity. For the bottom two plots of Fig.~\ref{fig:squeezing}(c), we use a semi-optimized protocol at the boundary case of $c=2$. The baseline $|\alpha|_{k+1}\ll 0.13\Delta^{1/2}$ protocol gives a $\Delta$-independent slope of
$\langle\sigma_\mathrm{x}\rangle_x$ and the fastest convergence, but for
momentum squeezing with $\Delta\ge 1$ the nonlinearity in
$\langle\sigma_\mathrm{x}\rangle_x$ grows, complicating disentanglement.
We handle this by numerically fitting a straight line to
$\langle\sigma_\mathrm{x}\rangle_x$ to track deviations in slope from
$4\alpha/\Delta^2$ as $\Delta$ increases (App.~\ref{app:squeezing}), yielding
the accelerated protocol used for middle and bottom panels of Figs.~\ref{fig:squeezing}.  As a
result, squeezing in dB scales linearly with circuit duration,
outperforming the baseline.

Squeezed states are useful for measuring net displacement or momentum boost,
with sensitivity quantified by the Fisher information
$F=2/\delta x^2$~\cite{paris2004quantum}.  Although this formula is strictly
valid only for Gaussian states, we apply it here because our states are
closer to Gaussian than those of other
works which use this method to quantify the squeezing~\cite{hastrup2021unconditional,eickbusch2022fast}, enabling direct
comparison.  The highest Fisher information $F=53.5$ is achieved for
$S_p=11.2\ \mathrm{dB}$, $S_x=-11.9\ \mathrm{dB}$ at infidelity $0.007$
in $8.06\,\mu\mathrm{s}$.

Post-selection on the qubit in $\ket{g}$ (empty circles) after each
$\mathcal{S}_k$ does not significantly improve results in the absence of
qubit errors, confirming that the choices of $\alpha_k$ achieve good
disentanglement at each step.  In the presence of errors, these mid-circuit
measurements serve as ancilla error detection.
\par\textbf{Discussion:} Our analytically derived circuit performs on par with the semi-analytical and numerically optimal methods~\cite{hastrup2021unconditional,eickbusch2022fast}. We plot results against circuit duration instead of circuit depth, given that the speed and errors in a conditional displacement gate depend significantly on the length of the amplitude~\cite{eickbusch2022fast}. Our protocol achieves squeezing levels, with $S_p=8.5 \, \mathrm{dB}$ of squeezing and $S_x=-8.4 \, \mathrm{dB}$ of anti-squeezing, alongside an infidelity of $\sim \mathcal{O}(10^{-3})$ in $5.8\,\mu\mathrm{s}$ while Ref.~\cite{hastrup2021unconditional} reports $\delta x=8.5\, \textrm{dB}$ and $\delta p=-9.9\, \textrm{dB}$ with an infidelity of $\sim \mathcal{O}(10^{-2})$. The performance of our scheme is also on par with numerically optimized schemes~\cite{eickbusch2022fast} (see Fig.~\ref{fig:numerical_squeezing} in the appendix). More details can be found in App.~\ref{app:squeezing}, where we also discuss the benefits of our scheme against Trotterization. Finally, our analytical approach uses incremental values of conditional displacement amplitudes, keeping the state close to Gaussian at all steps in contrast to Ref.~\cite{hastrup2021unconditional} which considers a large initial conditional displacement.

\subsection{GKP states}\label{ssec:GKP-States}
\begin{figure*}[t]
  \centering
  \includegraphics[width=\linewidth]{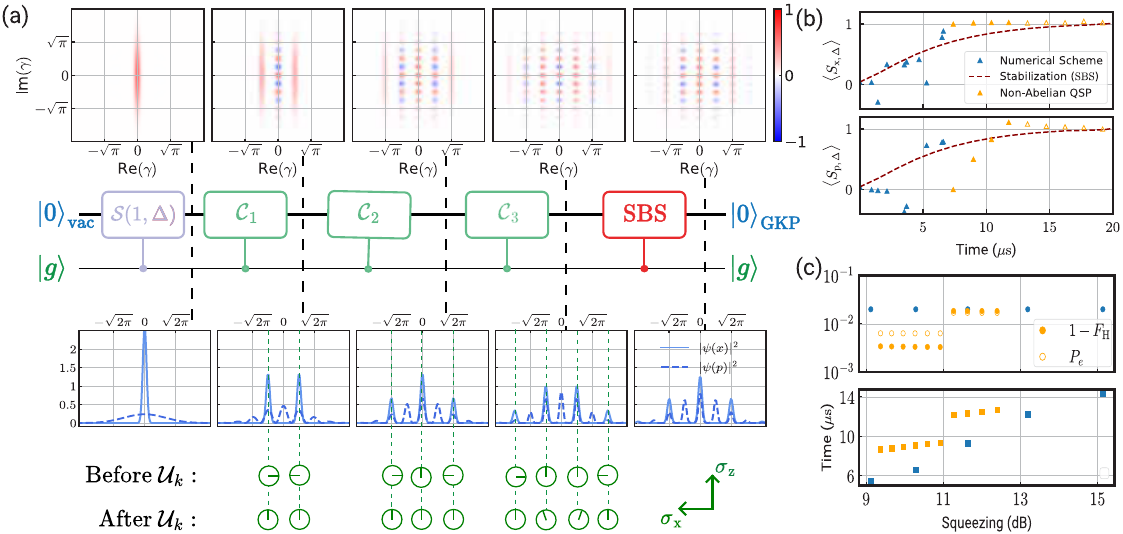}
  \caption{\textbf{Deterministic GKP logical $\ket{+Z}$ state preparation.} \textbf{(a)} Circuit components for GKP preparation with $\Delta=0.34$ (as used in recent experiments~\cite{eickbusch2022fast,sivak2023real}). $\mathcal{S}$ denotes the squeezing circuit from Fig.~\ref{fig:squeezing}, $\mathcal{C}_k = \mathcal{U}_k e^{-i\sqrt{2\pi}\hat{p}\sigma_\mathrm{z}}$ represents the iterative cat preparation sequence given by Eq.~(\ref{GKP-circuit}) in the main text, and $\mathrm{SBS}$ is one round of the small-big-small protocol. See Table~\ref{tab:GKP-prep-circuit-depth} for circuit-depth justification. (Top) Rescaled Wigner distribution functions are shown after each step. Note the change in the state before and after $\mathrm{SBS}$. (Bottom) Wave functions and corresponding spin polarizations are shown before and after each $\mathcal{U}_k$. After $\mathcal{U}_3$, the qubit remains slightly entangled, with $\braket{\sigma_\mathrm{z}} = 0.9937$ and GKP fidelity $\mathcal{F}_\mathrm{GKP} = 0.9989$.
\textbf{(b)} Convergence of various protocols, tracked using the expectation values of finite-energy stabilizers $\braket{S_{\mathrm{x},\Delta}}$ and $\braket{S_{\mathrm{p},\Delta}}$ (see Eqs.~(\ref{eq:GKP-stab-p}-\ref{eq:GKP-stab-x})). The non-Abelian QSP protocol using $\mathrm{GCR}$ (via $\mathcal{C}_k$'s) starts at $8.06\mu\textrm{s}$, accounting for the time to achieve $11.2\ \mathrm{dB}$ of squeezing (see Fig.~\ref{fig:squeezing} and Sec.~\ref{sec:squeezing}). Empty triangles mark the section where SBS is appended. Our protocol matches the performance of the numerical scheme~\cite{eickbusch2022fast} while significantly outperforming stabilization-based methods (using only $\mathrm{SBS}$~\cite{royer2020stabilization}).
\textbf{(c)} Hybrid infidelity, failure probability (top), and circuit duration (bottom) versus squeezing (or $\Delta$) for the numerical and analytical (non-Abelian QSP) circuits. Our protocol achieves fidelity and circuit duration comparable to the numerical scheme, while maintaining a low failure probability. All simulations here neglect physical errors such as damping, heating, and decoherence, which are addressed in Sec.~\ref{sec:GKP}.}
  \label{fig:GKP-prep}
\end{figure*}
GKP codes have applications in quantum sensing and bosonic error correction~\cite{sivak2023real,campagne2020quantum,fluhmann2019encoding,brady2024advances}. In this section, we only discuss the preparation of logical codewords, leaving other details to Sec.~\ref{sec:GKP}, where we discuss universal control of these codewords. While we focus on the square GKP (qubit) codewords to give explicit constructions for preparation, our protocols are generalizable to other lattices such as hexagonal or qudit GKP codes. Such generalizations are not easily accessible to numerically optimized circuits~\cite{eickbusch2022fast}, where for each qudit codeword a new optimization needs to be run. The states prepared using the method below have Gaussian peaks whose amplitudes are binomial coefficients, in contrast to the usual definition of GKP states, which uses a Gaussian envelope:
\begin{align}
  \ket{\mu}_\mathrm{GKP}=\mathcal{N}_\mu \sum_{m=-\frac{N+\mu}{2}}^{\frac{N-\mu}{2}+1} b_{\mu,m}\mathrm{D}\big([m+\mu/2]\sqrt{2\pi}\big)\ket{0_\Delta},\label{eq:GKP-bin-logical}
\end{align}
where $\mu \in \{0,1\}$, $b_{\mu,m}={N\choose m+\mu+\floor{N/2}}$, $N=\floor{0.32/\Delta^2}$
and $\mathrm{D}(\alpha)=e^{\alpha \hat a^\dagger-\alpha^*\hat a}=e^{2i(\mathrm{Im}(\alpha)\hat x-\mathrm{Re}(\alpha)\hat p)}$ denotes an unconditional displacement of the oscillator by $\alpha$ generated by the operator $\hat v(\alpha)=2\mathrm{Im}(\alpha)\hat x-2\mathrm{Re}(\alpha)\hat p$. We use the Newton-Raphson method to find the optimal number of steps $N$ required for a given envelope size $\Delta$ (see App.~\ref{app:GKP-prep}).

\par\textbf{Protocol:} In some sense, the preparation of GKP states is an extension of the preparation of squeezed two-legged cat states, where it is required to create superpositions of multiple squeezed coherent states instead of only two. A sketch of the preparation scheme is given in Fig.~\ref{fig:GKP-prep}(a).

The protocol starts with an initial squeezed vacuum state which can be prepared using the protocol in Sec.~\ref{sec:squeezing}. Then the cat-state preparation circuit $C_k$ is applied repeatedly, with $\mathcal{C}_1$ corresponding to the protocol described in Sec.~\ref{ssec:Cat_States}, while the general definition of gadgets in the circuit corresponds to the gate sequence, 
\begin{equation}  \mathcal{C}_k:\mathcal{C}\Big(\frac{\pi}{4k},\sqrt{\frac{\pi}{2}},\Delta_{k-1}\Big)=\mathcal{U}\Big(\frac{\pi}{4k},\sqrt{\frac{\pi}{2}},\Delta_{k-1}\Big)e^{-i\sqrt{2\pi}\hat p\sigma_\mathrm{z}}\label{GKP-circuit}.
\end{equation}

Each $\mathcal C_k$ leaves the oscillator and the qubit (approximately) unentangled, and adds one Gaussian to the peak in a fashion conceptually similar to the so-called ``cat-breeding" protocols~\cite{PhysRevA.97.022341}. At the $k^\mathrm{th}$ step the resulting state exhibits $k+1$ peaks. The $\alpha$ argument of the unentangling gadget $\mathcal U$ within $\mathcal{C}_k$ is set by the GKP lattice spacing, in this case we use $\sqrt{2\pi}$, corresponding to the Wigner units for the square-GKP qubit code. Finally, after applying a series of $\mathcal C_k$ circuits, we apply a single round of error correction, such as the SBS protocol, optionally, can refine the state.

\paragraph{Fidelity and circuit-depth:} 
Arguably, a preparation routine should prepare a state close to the GKP manifold such that subsequent rounds of error correction can correct the residual (yet correctable) errors. To avoid unimportant variations in performance due to a particular choice of fidelity, in Fig.~\ref{fig:GKP-prep}(b) we also benchmark the prepared states according to the expectation values of the finite energy stabilizers~\cite{royer2020stabilization},
\begin{align}
  S_\mathrm{x,\Delta}&=e^{i2\sqrt{2\pi}(\hat x\cosh{\Delta^2}-\hat p\sinh{\Delta^2})}\label{eq:GKP-stab-x}\\
  S_\mathrm{p,\Delta}&=e^{i2\sqrt{2\pi}(\hat p\cosh{\Delta^2}-\hat x\sinh{\Delta^2})}\label{eq:GKP-stab-p}.
\end{align}

In Fig.~\ref{fig:GKP-prep}(b), we see that our scheme (orange) converges to a GKP state faster than preparing the logical state via stabilization (dashed line), and is on par with the numerical preparation (blue). Fig.~\ref{fig:GKP-prep}(c) yields the corresponding infidelity against the target GKP state and the circuit-depth in time for both protocols, validating our claim that the analytical scheme discussed in this work performs on par with the state-of-the-art numerical schemes. Note that we have accounted for the circuit depth of squeezing in Fig.~\ref{fig:GKP-prep}(c), as the first point for non-Abelian QSP starts at $8.06\,\mu\textrm{s}$.

\paragraph{Success probability:} For the circuits $\mathcal{C}_k$ we need $\mathcal{U}(\theta',\alpha,\Delta)$ with $\theta'=\pi/4$ for $k<3$. For $k=3$, as can be seen in Fig.~\ref{fig:GKP-prep}(a), the angle required to rotate the qubits at peaks on the farther end is $\pi/12$. Thus, the un-entanglement gadget $\mathcal{U}_k$ at this point rotates the qubits by $\theta=\pm\pi/12$ at the two extreme peaks as required. However, it also rotates the qubits entangled with the peaks in the middle, ones that did not require any rotation. While the rotation at the central peaks is not significant, there is a different angle of rotation compared to $\pi/4k$ which could produce better un-entanglement for $k\ge 3$. We compute the optimal angle of rotation using the procedure given in App.~\ref{app:GKP-prep}. With this protocol, we obtain Figs.~\ref{fig:GKP-prep}(a,b) for $\Delta=0.34$ (used in recent experiments~\cite{eickbusch2022fast,sivak2023real}). The state after $k=3$ yields $4$ peaks with a fidelity of $F_\mathrm{H}=0.9989$ to the target GKP state, while the success probability is $P_g=0.99$. Thus, we may also reset the ancilla after each $\mathcal{C}_k$. Such high success probability justifies using the measurements to keep ancilla errors in check (see Sec.~\ref{ssec:GKP-errors}). 

\paragraph{Other GKP lattices:} Finally, to achieve different rectangular and hexagonal lattice-based GKP codes, we will only need to change the lattice spacing $l$ and the rotation angle $\theta$ according to the position of the deformed lattice peaks $k\alpha$, in the circuit components $\mathcal{C}_k$,
\begin{align}
  \mathcal{C}_k=\mathcal{U}\Bigg(\frac{\pi}{4k},\frac{l}{2},\Delta_{k-1}\Bigg)e^{-i\sqrt{2}\alpha\hat p\sigma_\mathrm{z}}.
\end{align}

Arbitrary GKP states (other than Pauli eigenstates) can be prepared using the gate-teleportation circuit discussed in Sec.~\ref{ssec:piecewise-teleportation}.

\par\textbf{Discussion} We show a comparison of our scheme against two different GKP preparation methods~\cite{eickbusch2022fast,royer2020stabilization,de2022error} in Figs.~\ref{fig:GKP-prep}(b,c). We find that our scheme is twice as fast as the stabilization scheme \emph{small-big-small}~\cite{royer2020stabilization,de2022error}, described later in detail in Sec.~\ref{sec:GKP}. Moreover, our circuit depth and fidelity (see Table~\ref{tab:GKP-prep-circuit-depth} and Fig.~(\ref{fig:GKP-prep})) are on par with the optimized E\textrm{CD} circuits in~\cite{eickbusch2022fast}, used in recent experiments. Importantly, our scheme yields a natural method to make the scheme tolerant to circuit errors and achieve higher fidelity in the presence of faults. After each $\mathcal{C}_k$, in the absence of errors, the qubit is in a known pure state untangled from the oscillator with a very high probability ($>0.99$) as indicated by the low failure probability in Fig.~\ref{fig:GKP-prep}(c). At this point, GKP states can be post-selected, given the qubit is found in the desired state. Thus, we can also keep qubit errors in check with this scheme. 

Moreover, in Ref.~\cite{hastrup2021measurement}, the authors propose to prepare GKP states using the same pattern of alternating conditional displacements and momentum boosts. However, due to the lack of a non-Abelian QSP pulse, in this case, the qubit needs to be unentangled using numerical schemes. Our scheme uses smaller displacements at each step, inducing fewer ancilla errors, and potentially allows for mid-circuit error detection.

 \section{Universal Control of Bosonic Error-Corrected Qudits}\label{sec:GKP}
The phase-space instruction set is most suited for the translationally invariant grid codes (GKP codes), and we focus on these codes for high-fidelity universal control of finite-energy error-correction~\cite{royer2020stabilization,brock2024quantum,eickbusch2022fast,sivak2023real}. These codes have shown promise in autonomous error correction, achieving break-even gain for quantum memory, which has only been matched so far by the latest surface code experiment~\cite{acharya2024quantum} using $\sim 100$ DV qubits. 
Our schemes are generalizable for arbitrary lattice spacing, thus yielding universal control for square, hexagonal, and rectangular GKP qubits, GKP qudits, and multi-mode GKP codes. We first summarize our GKP control results before diving into the details. 

We start by extending the state preparation results of Sec. \ref{ssec:GKP-States} to the noisy case in Sec. \ref{ssec:GKP-errors}. Then we tie together the ancilla-assisted finite-energy GKP readout of Ref.~\cite{hastrup2021improved} and stabilization scheme in Refs.~\cite{royer2020stabilization,de2022error} within a single framework in Sec.~\ref{ssec:GKLP-framework}. 
In addition, we give an end-of-the-line readout circuit in Sec.~\ref{ssec:logical-readout} which could yield better readout fidelity for GKP qubits with correctable displacement errors using the Abelian QSP sequence $\mathrm{BB1}$. 

A key result of this paper is an error-corrected gate teleportation scheme (Sec.~\ref{ssec:piecewise-teleportation}) that applies a target logical operation while simultaneously correcting oscillator displacement errors---to our knowledge the first high-fidelity such gates, both single- and two-qubit, native to finite-energy GKP, emerging directly from recognizing readout and stabilization as two halves of the same GCR-based circuit. A piecewise version of this design further mitigates ancilla dephasing, suppressing otherwise uncorrectable ancilla-induced faults.
As a result, our scheme enables high-fidelity, universal single-qubit logical rotations even with biased-noise ancillae like cat qubits~\cite{mirrahimi2014dynamically,puri2019stabilized,grimm2020stabilization,ding2024quantum}. To our knowledge, it is the most effective approach for non-Clifford operations on GKP qubits to date~\cite{hastrup2021unsuitability}.

We restrict the derivation and discussion to the case of square GKP, where $x$ and $p$ are treated symmetrically, for Secs.~\ref{ssec:GKP-errors}-\ref{ssec:piecewise-teleportation}, but note that these circuits can be extended to qubit and qudit lattices.

\subsection{High-fidelity error-detected state preparation}\label{ssec:GKP-errors}
Fault-tolerant preparation of logical GKP states is an important resource for bosonic quantum error correction. Previously in this work, we described a protocol that prepares GKP states in small steps $\mathcal{S}(1,\Delta),\mathcal{C}_k$, at the end of which the ancillary qubit was left in a deterministic state. In this section, we investigate the performance of this scheme in the presence of oscillator and qubit noise, which provides an understanding of how our preparation schemes, discussed in Secs.~\ref{sec:squeezing}-\ref{ssec:GKP-States}, perform on an experimental platform such as a superconducting hardware platform. 

\paragraph{Noisy state preparation.} For the protocol presented in Sec.~\ref{ssec:GKP-States}, we introduce mid-circuit detection on ancilla errors via post-selection upon outcome $\ket{g}$ after every round; resetting the qubit to $\ket{g}$ and starting over if the step failed. During each gate in the circuit, we add photon loss at the rate $\kappa/2\pi=1/1000 \ \mu\textrm{s}^{-1}$, ancilla decay at the rate $\gamma/2\pi=1/200 \ \mu\textrm{s}^{-1}$, and ancilla dephasing at the rate $\gamma_\phi/2\pi=1/200 \ \mu\textrm{s}^{-1}$. We repeat the protocol for $\Delta=0.34$~\cite{sivak2023real} for $10^5$ rounds. For each preparation round, we execute the GKP state preparation circuit $\mathcal{S}(1,0.34)-\mathcal{C}_1-\mathcal{C}_2-\mathcal{C}_3-\mathrm{SBS}$ shown in Fig.~\ref{fig:GKP-prep} with measurements at every $'-'$. We discard any round where we encounter a $-1$ outcome for $Z$ measurement of the ancilla qubit at any of these points and start over with the oscillator in the vacuum state. 

\paragraph{Performance.} The success probability (fraction of rounds that are kept) for this simulation is $0.94$, with an average fidelity of $0.96$, in comparison to $0.9969$ in the noiseless case. The numerical scheme in Ref.~\cite{eickbusch2022fast} reports a simulated (and experimental) fidelity of $0.85$ for a numerically optimized circuit with the same $\Delta$ without post-selection. Post-selection with their circuit is possible, but only at the end of the entire circuit, the length of which is comparable to our complete preparation circuit. Thus, the efficiency with which such a circuit can detect errors with post-selection would be low. Relative to Ref.~\cite{hastrup2021improved}, our scheme performs better due to a similar argument. In comparison to the first step of that protocol, the length of our snippet $\mathcal{C}_1$ is much smaller, reducing the probability of errors and thus improving the success probability and fidelity.

\subsection{GKP protocols in the non-Abelian QSP framework}\label{ssec:GKLP-framework}
\begin{figure}[tbh]
  \centering
  \includegraphics[width=0.5\textwidth]{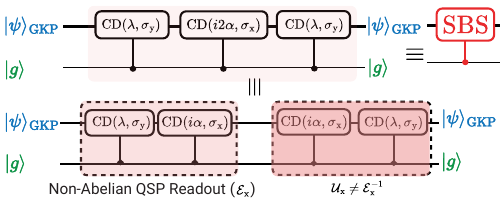}
  \caption{\textbf{Finite-energy GKP readout~\cite{hastrup2021improved} and stabilization~\cite{royer2020stabilization} protocols in non-Abelian QSP framework.} Interpretation of $\mathrm{SBS}$ circuit along the position quadrature as logical identity on the GKP codewords. The circuit is divided into the entangling and unentangling gadgets. The first half of this circuit $\mathcal{E}$ is the GKP readout circuit. Here, $\alpha=\frac{\sqrt{\pi}}{2\sqrt{2}}$ and $\lambda=-\alpha\Delta^2$.}
  \label{fig:GKp-framework}
\end{figure}
\paragraph{Analytical explanation for numerically-optimized readout scheme in Ref.~\cite{hastrup2021improved}.} Ref.~\cite{hastrup2021improved} shows a numerically-optimized circuit for improved GKP readout, which 
is a rotation gadget which intends to rotate the state by $2\pi$ ($\pi$) if the logical codeword is $\ket{0}_{\mathrm{GKP}}$ ($\ket{1}_{\mathrm{GKP}}$) using $e^{i\sqrt{\frac{\pi}{2}}\hat x\otimes\sigma_\mathrm{x}}$ with the qubit starting in $\ket{g}$. Thus, a readout circuit is nothing but an entanglement gadget. For finite-energy GKP, we can employ our entanglement gadget with the small GCR pre-correction $e^{i\sqrt{\frac{\pi}{2}}\Delta^2\hat p\otimes\sigma_\mathrm{y}}$. Thus, we give the logical readout circuits for logical bases X ($\hat v=\hat p$), Y ($\hat v=\hat x+\hat p$), and Z ($\hat v=\hat x$),
\begin{align}
  \mathcal{E}_{\hat v}\left[\frac{\pi}{2},\sqrt{\frac{\pi}{2}},\Delta\right]=e^{i\sqrt{\frac{\pi}{2}}\hat v\sigma_\mathrm{x}}e^{i\sqrt{\frac{\pi}{2}}\hat v_\perp\Delta^2\sigma_\mathrm{y}},\label{eq:GKP-readout}
\end{align}
see Sec.~\ref{sec:preliminaries} for the definition of $\hat v_\perp$. Notice that the displacement required to implement a logical $Y$ operation is longer than the corresponding displacements required for logical $X$ or $Z$ operations by a factor of $\sqrt{2}$. 

The GCR correction given here of the entangling gadget $\mathcal E$ is exactly the numerically-optimized correction in Ref.~\cite{hastrup2021improved}.
\paragraph{Dissipation-engineering based stabilization and correction of displacement errors.} 
Dividing the stabilization circuit $\mathrm{SBS}$ from Ref.~\cite{royer2020stabilization} into two halves (see Fig.~\ref{fig:GKp-framework}), the first half can be identified exactly as the entangling gadget used for readout as described in Eq.~(\ref{eq:GKP-readout}). The second half satisfies the constraints identified for the un-entangling gadget $\mathcal{U}(\pi/2,\sqrt{\pi/2},\Delta)$ in Sec.~\ref{ssec:Cat_States} since,
\begin{align}
  \pi/2\textrm{ (from }\mathcal{U})+\pi/2\textrm{ (from }\mathcal{E})=\pi,
\end{align}
as required by Eq.~(\ref{eq:piConditionUE}). And thus, 
\begin{align}
\mathrm{SBS}_{\hat v}=\mathcal{U}_{\hat v}\left[\frac{\pi}{2},\sqrt{\frac{\pi}{2}},\Delta\right]\mathcal{E}_{\hat v}\left[\frac{\pi}{2},\sqrt{\frac{\pi}{2}},\Delta\right].  
\end{align}
Note that here $\mathcal{U}\neq\mathcal{E}^{-1}$, such that the circuit has a trivial back action only if the oscillator is in the GKP code space, which is key to its success for error correction.

The error correction properties of this circuit were experimentally verified and qualitatively discussed in~\cite{sivak2023real}. Here, we provide a quantitative argument using our framework to compute its effects on the GKP code space affected by a displacement error, say $\epsilon$. 

In that situation the GKP states are positioned at $m\sqrt{2\pi}+\epsilon$, where $m$ is any odd (even) integer for the peaks of $\ket{0_\Delta}_\mathrm{GKP}$ ( $\ket{1_\Delta}_\mathrm{GKP}$). Now, the effect of the entangling part of SBS is to rotate the qubit by $e^{i\sqrt{\frac{\pi}{2}}(m\sqrt{2\pi}+\epsilon)\sigma_\mathrm{x}}$. Thus, for an erroneous state $\ket{\psi}_\epsilon=\alpha\ket{0}_\epsilon+\beta\ket{1}_\epsilon$, where $\ket{0}_\epsilon,\ket{1}_\epsilon$ denote erroneous GKP states $\ket{0_\Delta},\ket{1_\Delta}$ with displacement error $\epsilon$, we have,
\begin{align}
  \mathcal{E}_{\hat x} \ket{\psi}_\epsilon\ket{g}&\approx e^{i\epsilon\sqrt{\frac{\pi}{2}}\sigma_\mathrm{x}}(\alpha\ket{0}_\epsilon\ket{g}+\beta\ket{1}_\epsilon\ket{e}).
\end{align}
The second half of the circuit, $\mathcal{U}_{\hat x}$, has a correction $e^{i2\lambda\hat p\sigma_\mathrm{y}}$ (with $\lambda=\alpha\Delta^2$) on which depends the final qubit state. The un-entanglement routine $\mathcal{U}_{\hat x}$ correctly rotates the qubit state back to $\ket{g}$ for the first terms in each row. However, for the second term, where the qubit will be rotated to $\ket{e}$, the sign of the finite-energy correction is wrong. And hence, here the finite energy correction of $\mathcal{U}_{\hat x}$ will apply a back action of $e^{i4\lambda\hat p\sigma_\mathrm{y}}$. Thus, after $\mathcal{U}_{\hat x}$, the hybrid oscillator-qubit state takes the form,
\begin{align}    
&=(\cos{\epsilon\sqrt{2\pi}}\ket{g}+ie^{i4\lambda\hat p\sigma_\mathrm{y}}\sin{\epsilon\sqrt{2\pi}}\ket{e})(\alpha\ket{0}_\epsilon-\beta\ket{1}_\epsilon).
\end{align}
If $\epsilon=0$, this state is equal to $(\alpha\ket{0}_\epsilon-\beta\ket{1}_\epsilon)\ket{g}$, and a deterministic logical Pauli operation has been applied to $\ket{\psi}$. This Pauli operation can be tracked and hence does not play any role in stabilization. The next step of the protocol is to measure the qubit, and the probability for outcome $\ket{e}$ is given by
\begin{align}
  P_e&=\int_{-\infty}^\infty \ dp \ \sin^2{(\epsilon\sqrt{2\pi})}\cos^2{(4\lambda p)}|\psi(p)|^2.
\end{align}
Here the back-action on the oscillator state depends on the result of the qubit measurement.
In the event that the qubit is projected to state $\ket{g}$, the back-action is given by 
\begin{align}
&(\cos{(\epsilon\sqrt{2\pi})}\mathrm{I}+i\sin{(4\lambda\hat p)}\sin{(\epsilon\sqrt{2\pi})})\ket{\bar\psi}_\epsilon\\
&\approx(\mathrm{I}+i4\lambda\epsilon\sqrt{2\pi}\hat p)\ket{\bar\psi}_\epsilon\nonumber\\&\quad +(\mathcal{O}(\epsilon^2)+\mathcal{O}(\lambda\epsilon^2)\hat p+\mathcal{O}(\lambda^2\epsilon)\hat p)\ket{\bar\psi}_\epsilon\\
&\approx \mathrm{D}(-\pi\epsilon\Delta^2)\ket{\bar\psi}_\epsilon.
\end{align}
Considering that initially $\ket{\bar \psi}_\epsilon = D(\epsilon)\ket{\bar \psi}$, in the event of a $g$ outcome a single round of SBS partially corrects the error by applying a displacement of $-\pi\epsilon\Delta^2\approx-0.36\epsilon$ for $\Delta=0.34$. 

The back action in the event when the qubit is projected to state $\ket{e}$ is given by,
\begin{align}
\cos{(4\lambda\hat p)}\ket{\bar\psi}_\epsilon
\end{align}
This back action is independent of the error parameter $\epsilon$, but its probability increases with $\epsilon$. This operator applies a symmetrized displacement along the position axis. In the momentum basis, this is equivalent to a cosine envelope on the state. To second order in $\hat p$, the cosine envelope can be approximated as a Gaussian,
\begin{align}
  \cos{4\lambda\hat p}\approx e^{-8\lambda^2\hat p^2},
\end{align}
where $8\lambda^2=8\alpha^2\Delta^4=\pi\Delta^4$. We compare this envelope correction with the momentum part of the target Gaussian envelope $e^{-\Delta^2\hat n}=\exp{[-\Delta^2(\hat x^2+\hat p^2)]}$ and note that an $e$ outcome reduces the momentum uncertainty by $15\%$. Currently, we do not have an intuitive explanation for how this back action supports error correction or stabilization. The probability of both the back actions increases with an increase in the error $\epsilon$, as should be the case for any autonomous error correction scheme. Similarly, the stabilization of the logical $X$ basis corrects displacement errors and envelope errors along the momentum axis.

In summary, we have shown that $\mathrm{SBS}$ is an example of probabilistic (or autonomous) error correction. 

\subsection{Logical readout with correctable errors}\label{ssec:logical-readout}

\begin{figure*}
  \centering
  \includegraphics[width=\linewidth]{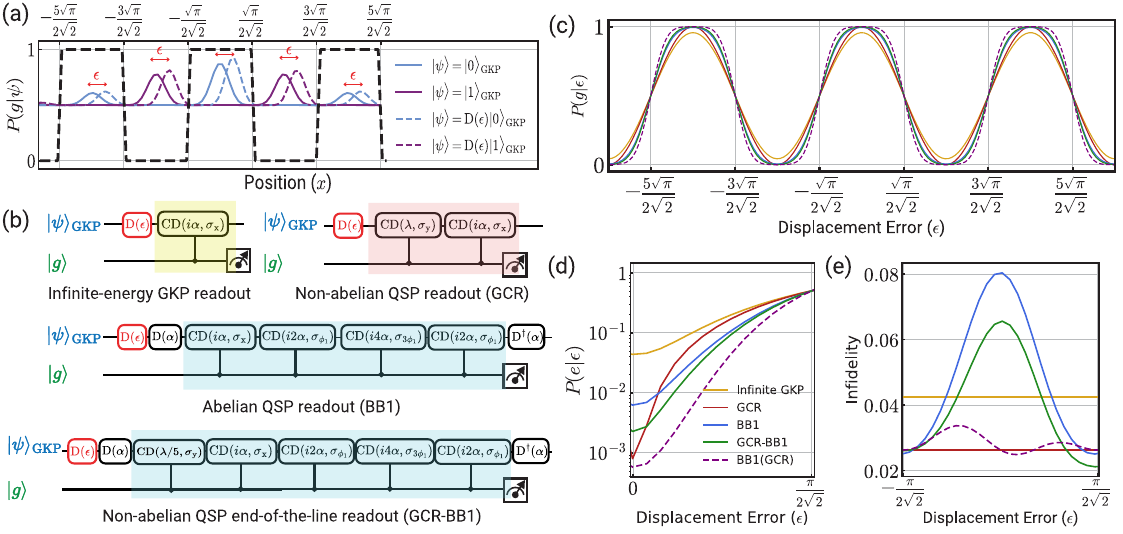}
 \caption{\textbf{Logical readout of GKP states with correctable errors.} The GKP readout procedure maps the logical state onto the ancilla qubit states $g$ and $e$, which are then measured.
\textbf{(a)} Solid curves show finite-energy GKP codewords; dotted curves show the corresponding displaced codewords. An ideal end-of-sequence pulse for reading out a state with a correctable displacement error $\epsilon$ is a square wave (black dashed line). It yields a correct measurement probability $P(g|0,\epsilon)$ (for $\sigma_\textrm{z}$ measurement) across all logical $\ket{0,\epsilon}$ states with $|\epsilon|<\sqrt{\pi/2}$, and similarly for $P(e|1,\epsilon)$ for $\ket{1,\epsilon}$.
\textbf{(b)} GKP readout schemes: the infinite-energy scheme corrects only peak locations; the non-Abelian QSP readout (also found numerically in Ref.~\cite{hastrup2021improved}) corrects Gaussian broadening; and the Abelian BB1(90) QSP sequence corrects peak location ambiguity by flattening the cosine response.
\textbf{(c)} Readout probability of getting the right outcome $P(g|0,\epsilon)$ as a function of displacement error $\epsilon$ for different schemes, without circuit-level errors: infinite-energy (no correction), GCR (Gaussian correction), and BB1 (peak location correction). Color legend is in (d).
\textbf{(d)} Zoom-in of (c) shows $1-P(g|0,\epsilon)$ within the Voronoi cell of $\ket{0}_\mathrm{GKP}$. Here, GCR-BB1 denotes enhancing BB1(90) with a conditional displacement, while BB1(GCR) corresponds to an optimized version of the composed sequence derived in Eq.~(\ref{eq:GCR-BB1}). In BB1(GCR), the blue region circuit is replaced according to this equation. These sequences correct Gaussian uncertainty while producing a square response. Parameters $\alpha$ and $\lambda$ follow Fig.~\ref{fig:GKp-framework}. \textbf{(e)} Readout infidelities: color coding matches (b,d). BB1 and GCR-BB1 show the highest infidelities, while finite-energy readout and BB1(GCR) achieve the lowest, mainly limited by envelope errors that could be further stabilized. All curves correspond to a GKP state width set by $\Delta=0.34$.}
  \label{fig:bb1-readout}
\end{figure*}

At the end of any quantum circuit, logical qubits are measured to determine their state. Inevitably, there can be residual correctable errors in the logical code states being measured. For the GKP encoding, such errors might be correctable random displacements, that is, displacements smaller than $\sqrt{\pi}/2\sqrt{2}$.

The problem at hand is described by Fig.~\ref{fig:bb1-readout}(a) which suggests that we need to extract the one bit of information $\{0_\mathrm{GKP},1_\mathrm{GKP}\}$ where $0_\mathrm{GKP}$ ($1_\mathrm{GKP}$) corresponds to all states in the Hilbert space which are closer to logical $\ket{0}_\mathrm{GKP}$ ($\ket{1}_\mathrm{GKP}$) than to the opposite logical state. In the infinite-energy GKP case read out using a single CD (one-bit phase estimation protocol), the readout fidelity follows a cosine curve (shown in panel (c) of Fig.~\ref{fig:bb1-readout}) as a function of displacement $\epsilon$. The fidelity value is less than unity when applying a single CD as a readout scheme for a finite-energy GKP state (see Fig.~\ref{fig:bb1-readout}). The finite-energy readout, described in Sec.~\ref{ssec:GKLP-framework}, yields a better readout fidelity that follows a similar curve with a peak value (case of no error) closer to unity. However, since small errors are correctable in principle, we would like a readout sequence that yields a square wave response shown by the dotted black lines in Fig.~\ref{fig:bb1-readout}(a), a problem which we recast in the QSP framework.

\paragraph{Abelian QSP readout.} We can achieve something close to the required square wave using our adaptation of BB1(90) in Eq.~(\ref{eq:BB1}) for hybrid oscillator-qubit control. See circuit in Fig.~\ref{fig:bb1-readout}(b). This circuit yields improved readout fidelity for all correctable error states compared to the finite-energy readout circuit, as shown by Fig.~\ref{fig:bb1-readout}(c). A zoomed-in version is shown in Fig.~\ref{fig:bb1-readout}(d) to assess the situation in the no error case of $\epsilon=0$. Note that the Helstrom bound for the given value of $\Delta=0.34$ is $\frac{1}{2}\big(1-\sqrt{1-|\braket{0|1}|^2}\big)\approx 10^{-6}$, far below the BB1 protocol readout error at $\epsilon=0$. Hence, we do not need to account for the non-orthogonality of the GKP logical Pauli states. The Abelian end-of-the-line sequence works desirably well and is relatively robust for $\epsilon\neq 0$ case. However, its performance for the no error case $\epsilon=0$ is worse than the finite-energy readout sequence (labeled GCR in Fig.~\ref{fig:bb1-readout}(c,d)).

\paragraph{Non-Abelian end-of-the-line readout.} A natural, deployable way to combine BB1's peak-location robustness with GCR's finite-energy correction is GCR-BB1, in which a single conditional displacement is prepended to the four bare BB1 rotations, yielding a simple unitary pulse sequence. The amplitude of this GCR-type correction is optimized numerically to take into account the finite-energy corrections of the four rotations in BB1 collectively. GCR-BB1 improves over bare BB1 across the correctable region while adding only a small conditional displacement, and its short length keeps the decoherence it accrues under noise modest. Because it applies one correction collectively rather than one per rotation, however, a residual Gaussian-fluctuation error survives at the no-error point (see Fig.~\ref{fig:bb1-readout}(d)), setting a coherent floor on its readout accuracy. This floor is lowered by the full concatenation of BB1 and GCR, called BB1(GCR), discussed in App.~\ref{app:bb1gcr}, which interleaves a GCR correction with every BB1 rotation and thereby cancels the Gaussian fluctuation error kick by kick, pushing the no-error readout close to the Helstrom bound. The cost is length: BB1(GCR) consists of $8$ conditional displacements, longer than BB1 for some $\Delta$, so it accrues correspondingly more decoherence. The two sequences therefore trade sequence length against coherent readout accuracy rather than one strictly dominating the other. In the presence of noise, both pulses perform worse than the much shorter GCR readout at the no-error point.

\paragraph{Back action.} Note that during each readout sequence, the GKP state is displaced by an amount $|\alpha|$ along the quadrature orthogonal to the one being measured, a deterministic displacement that can be accounted for. Cancelling this displacement, we compute the fidelity of the resulting state with the initial erroneous state. A large infidelity indicates a larger back action from the respective circuit. See Fig.~\ref{fig:bb1-readout}(e) for the back action of all readout schemes discussed in this section. For an end-of-the-line readout sequence, the back action is not important; for schemes where back action is minimal, the readout can be repeated to increase measurement fidelity.
\subsection{Universal qubit rotations with piecewise circuits}\label{ssec:piecewise-teleportation}
\begin{figure*}
  \centering
  \includegraphics[width=\linewidth]{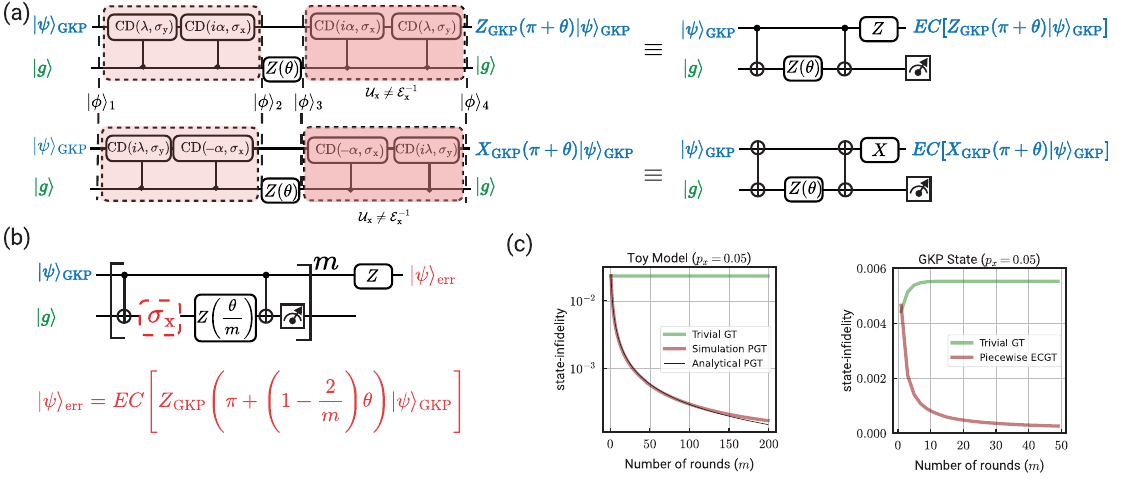}  
  \caption{\textbf{Error-suppressed GKP gate teleportation.} \textbf{(a)} Error-corrected gate teleportation of logical $Z(\theta)$ ($X(\theta)$) gate by an entangling-un-entangling sequence obtained from stabilizer of the logical $\{0,1\}$ ($\{+,-\}$) basis. \textbf{(b)} Toy model of a piecewise circuit to mitigate effects of biased-noise ancilla errors. \textbf{(c)} Comparison of trivial gate teleportation and piecewise gate teleportation for $\theta=\pi/4$ in the presence of ancilla errors. Here, the solid black line in the left panel presents the analytical curve for the state infidelity in the case of piecewise teleportation obtained from Eq.~(\ref{eq:PCGT}). The GKP state used for the simulation results shown in the right panel has an envelope size of $\Delta=0.34$.}
  \label{fig:GKP_Teleportation}
\end{figure*}

Arbitrary operations on finite-energy GKP states are generally not easily available. Recall that Pauli operations are simple phase space translations. However logical rotations are exponentials of Paulis which (by the Pauli-Euler identity) can be written as a coherent superposition of identity and a phase space translation. Unfortunately, it is not possible to apply a classical control pulse that is in a superposition of zero amplitude (to achieve identity) and non-zero amplitude (to achieve the phase space displacement). One method to circumvent this problem is to use ancillary systems to teleport gates into the GKP codespace by use of conditional displacements that entangle the GKP and ancilla qubit, for example through an effective CNOT gate. However, this method limits the logical fidelity of the gate when considering physical errors of the ancilla. Here, we devise a technique to teleport gates that simultaneously 1) correct errors in the oscillator and 2) protect against ancilla decay errors, yielding room for high-fidelity gate operations using a biased-noise ancilla. This protection against errors is different from the general path-independent mechanisms engineered for circuits using SNAP gates~\cite{ma2020path,ma2022algebraic,ReinholdErrorCorrectedGates}. Our construction does not require any hardware engineering feats such as chi matching~\cite{ReinholdErrorCorrectedGates,xu2024fault}, though these might yield further improvements. 
\paragraph{Error-corrected gate teleportation.}
The error-corrected gate teleportation sequence is constructed by realizing that the stabilization circuit $\mathrm{SBS}=\mathcal{U}\mathcal{E}$ is composed of entangling and unentangling gadgets, see Fig.~\ref{fig:GKp-framework}. By introducing a qubit gate in between the two gadgets, we can construct a phase-transfer circuit, as illustrated in Fig.~\ref{fig:GKP_Teleportation}(a). For logical $Z(\theta)$ gate and with an initial GKP state $a\ket{0}_\mathrm{GKP}-b\ket{1}_\mathrm{GKP}$, the circuit obeys the following equations, up to a global phase,
\begin{align}
\ket{\phi}_2&=\mathcal{E}_\mathrm{x}\ket{\phi}_1\nonumber\\
  &=a\ket{0}_\mathrm{GKP}\otimes\ket{g}-b\ket{1}_\mathrm{GKP}\otimes\ket{e}\\
  \ket{\phi}_3&=Z(\theta)\ket{\phi}_2=e^{-i\frac{\theta}{2} \sigma_\mathrm{z}}\ket{\phi}_2\nonumber\\
  &=a\ket{0}_\mathrm{GKP}\otimes\ket{g}-e^{i\theta}b\ket{1}_\mathrm{GKP}\otimes\ket{e}\\
  \ket{\phi}_4&=\mathcal{U}_\mathrm{x}\ket{\phi}_3\nonumber\\
  &=[Z(\pi+\theta)_\mathrm{GKP}(a\ket{0}_\mathrm{GKP}+b\ket{1}_\mathrm{GKP})]\otimes\ket{g}.
\end{align}
To perform logical $X(\theta)$ ($Y(\theta)$) rotations, one only needs to entangle the qubit with the logical $X$ ($Y$) eigenstates of the GKP code. 
For example, the same circuit becomes a logical $X(\theta)$ gate if $\mathcal{E}_\mathrm{x},\mathcal{U}_\mathrm{x}\rightarrow \mathcal{E}_\mathrm{p},\mathcal{U}_\mathrm{p}$. 
The circuit can be easily changed to use ancilla $X(\theta)$ gates, \textit{mutatis mutandis}, if this is an easier gate for the DV ancillary system. 

If we ignore qubit errors, these circuits will yield $\ket{g}$ corresponding to the desired gate operation. The fidelity of these gates in the absence of any physical errors and for $\Delta = 0.34$ is $99.88\%$ while the probability of a successful logical gate operation is $0.9994$. \par 
Note that the error-correction feature of our sequence comes from the fact that it is constructed from a logical identity circuit, $\mathcal{U}_\mathrm{x/p/x+p}\mathcal{E}_\mathrm{x/p/x+p}$ corresponding to the SBS stabilization circuit, that is not a universal identity  (where $\mathcal{U}_\mathrm{x/p/x+p}= \mathcal{E}_\mathrm{x/p/x+p}^{-1}$). That is, the error-corrected gate teleportation circuit, in addition to applying the logical gate on logical GKP codewords, also applies a corrective back action on erroneous GKP states similar to the small-big-small stabilization circuits discussed in Sec.~\ref{ssec:GKLP-framework}.
\paragraph{Protection from biased-ancilla errors.} The teleportation circuit described above is, however, severely affected in the presence of ancillary errors (the same as any teleportation circuit). For our circuit, ancilla errors tend to occur during the (relatively long duration) controlled displacement gates acting on the oscillator. In the case of a biased-noise ancilla, the circuit will only be affected by one type of ancillary error since the other errors are largely suppressed~\cite{grimm2020stabilization,ding2024quantum}. For protection against a single type of ancillary error, we propose the piecewise circuit shown in Fig.~\ref{fig:GKP_Teleportation}(b). For example, without loss of generality, let us imagine a biased-noise ancilla where the dominant error is $\sigma_\mathrm{x}$. The effect of this gate is shown in the top circuit of Fig.~\ref{fig:GKP_Teleportation}(b). The $\sigma_\mathrm{x}$ error on the ancilla propagates to the oscillator by reversing the sign of the logical operation, \emph{i.e.} as a logical rotation angle error of $-2\theta$ on the GKP state. We propose to solve this problem by dividing the circuit into $m$ pieces where each piece applies a rotation by $\mathrm{Z}(\theta/m)$ as shown in the lower panel of Fig.~\ref{fig:GKP_Teleportation}(b). We call this method \emph{piecewise gate teleportation}. In this case, a single $\sigma_\mathrm{x}$ error induces an erroneous rotation of $Z(-2\theta/m)$, with reduced effect on the logical fidelity as $m$ increases. The angle of the GKP logical rotation can effectively be described by a random walk such that the average rotation of the gate is $\theta' =\theta(1-2p_\mathrm{x})$ with a standard deviation of $\sigma_m =\frac{|\theta|}{\sqrt{m}}2\sqrt{p_\mathrm{x}(1-p_\mathrm{x})}$, where $p_\mathrm{x}$ is the probability of $\sigma_\textrm{x}$ errors. Importantly, the standard deviation of the angle decreases as $m$ increases.  
Taking a toy model with effectively pure qubit states, the fidelity of the resulting state is given by
\begin{align}
  \mathcal{F}=\sum_{k=0}^m {m\choose k} (1-p_\mathrm{x})^{m-k}p_\mathrm{x}^k \cos^2{\Big[\theta(p_\mathrm{x}-k/m)\Big]}.\label{eq:PCGT}
\end{align}
The systematic error in the mean rotation angle can be compensated by choosing a larger target angle $\theta^\prime = \theta/(1-2p_\mathrm{x})$.
 The curves in Fig.~\ref{fig:GKP_Teleportation}(c), computed using QuTiP~\cite{Johansson2013}, show that the favorable decrease in infidelity is proportional to $1/m$. In the toy model, we apply errors only just before the $\mathrm{CX}$ gates with probability $p_x=0.05$ to emulate the case of GKP states where the $\mathrm{CD}$ gates are longer and more erroneous compared to the qubit rotation $Z(\theta)$. In the GKP simulation, we apply a $\sigma_\mathrm{x}$ error at a rather large rate of $\gamma\sim 1/22\,\mu\textrm{s}^{-1}$ during all conditional displacements (to emulate the probability $p_x=0.05$ during the large conditional displacements). We use the metric that a conditional displacement by a magnitude of $1$ takes time $\tau=1\,\mu\textrm{s}$ as outlined in App.~\ref{app:squeezing} and use $\tau$ to make all rates dimensionless. We also simulated the performance of the piecewise gate teleportation described above, on the GKP state, and compared it against single-shot trivial gate teleportation followed by $m-1$ stabilization rounds in the right plot of Fig.~\ref{fig:GKP_Teleportation}(c). The first step of the trivial protocol which teleports the gate, however, is not correcting errors in the GKP state. This causes the initial bump in the fidelity curve, see the right plot of Fig.~\ref{fig:GKP_Teleportation}(c). 
 
 Note that, for trivial gate teleportation, one could use the measurement outcome to check for ancilla errors more efficiently\footnote{the corrective back action renders the measurement outcomes less useful in terms of detecting errors.}, but that would make the protocol reliant on measurements which can be the slowest (or, most erroneous) part of the circuit. Just as in the stabilization circuit, the qubit is reset to $\ket{g}$ at the end of every piece in the circuit. This reset could be erroneous and such an error has not been considered explicitly in our simulations. 

\subsection{Entangling GKP qubits: Extension of $\mathrm{GCR}$ to multi-modal operations}
The two-qubit gates suggested in~\cite{gottesman2001encoding} for an infinite-energy GKP code have poor fidelity for the finite-energy code and require a few stabilization rounds to improve the error rate~\cite{rojkov2023two}. 
\paragraph{Single-qubit-ancilla.} For the finite-energy states, Ref.~\cite{rojkov2023two} derived the finite-energy version of the two-qubit entangling gate, which can also be derived using the non-Abelian QSP extension for two modes. We discuss this extension here. For ideal GKP codes with support at positions $m\sqrt{\pi/2}, m\in \mathbb{Z}$, the conditional SUM gate displaces the second mode by the position $\pm x$ of the first mode with the sign of the displacement determined by the state of the ancilla. Equivalently, for each pair of peaks of the two GKP states, in the position (first mode) and momentum basis (second mode), respectively, the conditional SUM gate, $e^{i2\hat x_1\hat p_2\sigma_\mathrm{x}}$, rotates the qubit by $2l\pi$, where $l$ is the product of the two integers defining the positions of the two peaks. This operation is equivalent to $(-1)^l$ phase on the joint oscillator-qubit state, and it applies a $\mathrm{CX}_\mathrm{GKP}$ gate on the two logical GKP codewords with the qubit going back to the original state ($\ket{g}$, in this case). The non-Abelian correction for the entangling half of this operation $e^{i\hat x_1\hat p_2\sigma_\mathrm{x}}$ due to the envelope size $\Delta\neq 0$ with respect to the first (second) GKP qubit is given by $e^{i\Delta^2\hat p_1 \hat p_2 \sigma_\mathrm{y}}(e^{-i\Delta^2\hat x_1\hat x_2 \sigma_\mathrm{y}})$, assuming the ancilla starts in state $\ket{g}$. Thus, the GCR-type pre-correction due to both modes is given by
\begin{align}
  S\equiv e^{-i\frac{\Delta^2}{2}(\hat x_1\hat x_2-\hat p_1\hat p_2) \sigma_\mathrm{y}}.\label{eq:corr_sum}
\end{align}
The corresponding SBS-type circuit, where $B\equiv e^{2i\hat x_1\hat p_2\sigma_\mathrm{x}}$ and $S$ is given by Eq.~(\ref{eq:corr_sum}), rotates the qubit by an angle of $2\pi$ about $\sigma_\mathrm{x}$ and applies the desired logical operation. This operation, similar to the stabilization circuit $\mathrm{SBS}$, is protected against biased-noise ancilla errors. However, the fidelity of the Bell states prepared using this circuit is $\sim 0.90$ for $\Delta=0.34$. Higher fidelity requires either additional correction terms or, as we show next, a two-ancilla construction; but first, we discuss the fast echoed two-mode realization of the single-ancilla gate, then give the two-ancilla scheme. The fast conditional two-mode operations required for this operation can be achieved using Gaussian operations and weak dispersive coupling between the oscillator and qubit. Ref.~\cite{rojkov2023two} suggests that this sequence takes more gates (5 conditional two-mode operations equivalent to a SUM gate). 

However, we claim that this circuit can be achieved in three gate sequences given we can obtain fast $S$ gates using a two-mode extension of the echoed conditional displacement~\cite{campagne2020quantum,eickbusch2022fast,sivak2023real}. This decomposition is given in Ref.~\cite{ISA} for entangling oscillator gates in the weak dispersive regime, and we present it in App.~\ref{app:finite-SUM} in the context of two-qubit GKP operations. Fast echoed conditional displacement was obtained using (weak) dispersive coupling in the displaced frame. Similarly, we can obtain a fast echoed conditional SUM gate using (weak) dispersive coupling in a two-mode squeezed frame. The two-mode squeezing required for this operation can be obtained from single-mode squeezing and beam-splitters using a Bloch-Messiah decomposition~\cite{ISA}. Thus, this circuit involves two single-mode squeezing operations and two beam-splitters to go to the two-mode squeezing frame using TMS$(\alpha,\pi)$ (see definition in Ref.~\cite{ISA}). In this new frame, the circuit uses evolution under the dispersive interaction in this frame for the duration $t_\mathrm{CX}\ge \frac{\Delta^2}{\chi\sinh{2\alpha}}+\frac{2\Delta^2}{\chi}$ where $\chi$ is the strength of the weak dispersive coupling. Thus, by increasing $\alpha$ one can make this process much faster. Hypothetically, $\lim_{\alpha\rightarrow \infty}t_\mathrm{CX}\ge\frac{2\Delta^2}{\chi}$ is allowed, however, in reality, we are limited to finite and much lower values of $\alpha$ due to unwanted state transitions in a nonlinear ancilla dispersively coupled to a resonator (oscillator)~\cite{ding2024quantum,eickbusch2022fast,sivak2023real}. 
\begin{figure}[t]
  \centering
  \includegraphics[width=\linewidth]{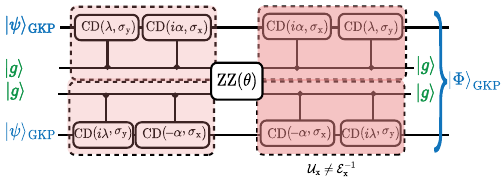}
  \caption[Piecewise GKP entangling operations.]{\textbf{Piecewise GKP entangling operations.} Circuit for logical Pauli operations using two ancillae. Here, $\ket{\Phi}=EC[\mathrm{ZZ}_\mathrm{GKP}(\pi+\theta)\ket{\psi,\psi}_\mathrm{GKP}]$. For $\theta=\pi/4$, we can prepare a Bell state if $\ket{\psi,\psi}_\mathrm{GKP}=\ket{++}_\mathrm{GKP}$. The fidelity of this Bell state preparation is $0.9997$ (as per Sec.~\ref{ssec:piecewise-teleportation}) and the success probability of the qubit outcome $\ket{g,g}$ is $P_g=(0.9993)^2=0.998$.}
  \label{fig:two-qubit-GKP}
\end{figure}

\paragraph{Two qubit ancillae.} In order to avoid conditional Gaussian operations which can be more difficult to realize in practice, another option is to realize a two-qubit version of the gate teleportation circuits discussed in Sec.~\ref{ssec:piecewise-teleportation}, which only uses conditional displacements and (two-) qubit gates.
This circuit, shown in Fig.~\ref{fig:two-qubit-GKP} (see also App.~\ref{app:c-pauli}), yields better fidelity compared to the single qubit ancillae scheme in the absence of errors. 
It is similar to Fig.~\ref{fig:GKP_Teleportation}(a), except the middle gate can be a controlled Pauli operation between the two DV qubits given the initial states and final operation to be implemented. Any controlled Pauli logical operation can be realized in a similar way. However, due to the way errors propagate, most of these gates do not benefit from dividing the rotation into smaller angles.

In the special case of a logical $\mathrm{ZZ}_\mathrm{GKP}(\theta)$ operation, ancilla $X$ errors propagate to the GKP codes as a rotation error of $-2\theta$. This is similar to the one-qubit gate example, and as a result the same piecewise construction can be applied, see App.~\ref{app:c-pauli}, yielding an error-corrected two-qubit entangling operation that is autonomously protected from ancilla errors and photon loss. In the absence of errors, this sequence yields a success probability of $P_g=0.9987$, highest for a perfect entangling logical operation on finite-energy GKP codewords. We believe that the state fidelity of this process will be the same as the single-qubit gate teleportation scheme. For example, computing the fidelity of states prepared by SBS, for a system with two truncated oscillators (each with Hilbert space dimension of $50$) and two qubits, is $0.98$ (upper bounded by truncation issues and not the protocol). We achieve the same fidelity for the Bell pair constructed using our teleportation protocol. This fidelity is much lower than what is expected ($0.998$ from Sec.~\ref{ssec:piecewise-teleportation}), and we believe this is due to the Hilbert space constraints. The logical error probability increases with the use of two-qubit DV gates while it remains the same as the error in conditional displacement increases, similar to the case of single-qubit gate teleportation analyzed in Sec.~\ref{ssec:piecewise-teleportation}.

All our protocols derived in this section have been obtained analytically through the use of NA-QSP, which either explain previously numerically optimized protocols~\cite{hastrup2021improved,hastrup2021unconditional,hastrup2021measurement,hastrup2023analysis} or yield new protocols for universal construction of GKP codewords. One of the problems with GKP codewords is that most works establishing protocols for finite-energy GKP codewords could not reach a fidelity of $>0.90$ for universal control~\cite{rojkov2023two,hastrup2021unsuitability}. The newly found protocols for entangling operations and piecewise gate teleportation yield the best protocols for finite-energy GKP codewords, even in the noiseless setting, compared to all previous works.
\section{Further Directions}\label{sec:conc}
Non-Abelian QSP extends naturally beyond the state preparation and error 
correction procedures introduced above. In this section, we explore three instances of problems where NA-QSP provides advantages, but leave the complete solving of such problems for future work.
We begin with more general state preparation algorithms, i.e., we take steps towards solving Problem~1 in Sec.~\ref{sec:preliminaries}. More precisely, we present analytical tools for the preparation of 
rotationally symmetric states — including $N$-legged cat and Fock states.  We then present a non-Abelian control 
sequence for oscillator-assisted phase estimation and then comment on 
how the analytical structure of NA-QSP can inform and seed hybrid 
analytical--numerical optimization strategies.  Together, 
these results suggest that NA-QSP is not only a practical tool but a 
theoretical framework for identifying and attacking open problems in 
hybrid CV-DV quantum control.
\subsection{Towards universal state preparation}\label{ssec:universal}

The phase-space instruction set is particularly well-suited to states naturally expressed as superpositions of non-overlapping Gaussian wave packets — squeezed, cat, and GKP states — as demonstrated in
the preceding sections.  A natural question is how far this instruction set
extends toward universal state preparation, and what fundamental constraints
emerge when it is pushed beyond its natural domain.  Here we explore two
such cases — rotationally symmetric states and Fock states.  We note that the preparation of a large class of states is likely more efficient with the Fock-space instruction set~\cite{ISA,fong2025engineeringnongaussianbosonicgates}.

\paragraph{Rotationally symmetric codewords.} 
\begin{figure}
  \centering
  \includegraphics[width=\linewidth]{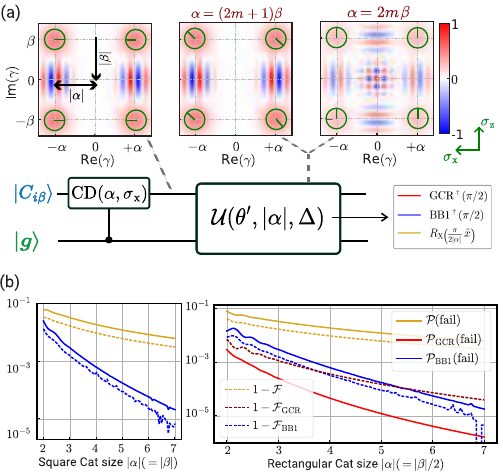}
  \caption{\textbf{Circuit for preparation of four-legged cat using QSP
  sequences.}
  \textbf{(a)} Rescaled Wigner distribution function for the two-legged cat preparation using the circuit in
  Fig.~\ref{fig:Cat_states}(a), starting from a momentum-cat
  state. The protocol does not require an initially unentangled
  two-legged cat; starting from vacuum with two conditional displacements
  along $\hat{x}$ and $\hat{p}$ also suffices. Using a perfect cat here
  simplifies the simulation without underestimating infidelity. Green
  circles indicate qubit spin polarization at each blob; same alignment across
  all blobs is necessary for deterministic preparation. During unentanglement, GCR (right plot) can only
  prepare rectangular cats with even aspect ratio $|\alpha|/|\beta|$ so $m\in\mathbb{Z}$,
  while BB1 has no such constraint (left plot).
  \textbf{(b)} Performance of GCR and BB1 for rectangular (square)
  four-legged cat states; yellow curve shows performance without QSP
  correction.}
  \label{fig:4-legged-cat}
\end{figure}

$N$-legged cat states consist of superpositions of $N$ coherent states at the
vertices of a regular $N$-gon in phase space. To prepare these states, we require a minimum of
$\log_2 N$ conditional displacements, after which the qubit must be
completely disentangled from all $N$ oscillator blobs simultaneously. For example, to prepare a four-legged cat state (see Fig.~\ref{fig:4-legged-cat}), one strategy is to first prepare a momentum two-legged cat state (using the protocol in Sec.~\ref{ssec:Cat_States}), then prepare a superposition of two cat states. However, this reveals a constraint of the GCR pulse which was designed assuming Gaussian states with average momentum $\langle p\rangle = 0$. Lifting this requirement, GCR can work for \emph{rectangular} cat states with an even aspect ratio $\alpha/\beta=2m$, $m\in\mathbb{Z}$ (see App.~\ref{app:4-legged-cat}).
This constraint reflects the fact that the pre-correction in GCR depends on a single
quadrature, whereas for $N$-legged cat states the unentanglement pulses should work for Gaussian states identified by both quadrature coordinates. The Abelian BB1 sequence has no such geometric
constraint and enables the preparation of four-legged cat states with any aspect ratio, at the cost of
additional circuit depth and a local phase $e^{i\alpha\beta}$ on two of
the four coherent states comprising the cat (vanishing when $\alpha \beta=2\pi m$ with $m\in\mathbb{Z}$).  As $|\alpha|$ increases,
both sequences improve in fidelity to their respective target states
(Fig.~\ref{fig:4-legged-cat}(b)). Similar strategies generalize to $N>4$, but the identification of optimal
unentanglement sequences for arbitrary $N$ and the interplay between
lattice geometry and QSP sequence structure remain open questions.

\paragraph{Fock state preparation.}\label{sec:fock} 
\begin{figure}[t]
  \centering
  \includegraphics[width=0.48\textwidth]{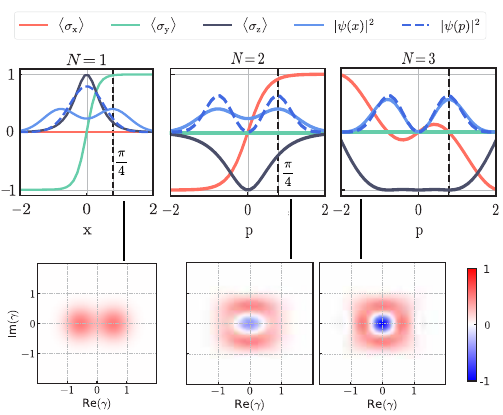}
  \caption{\textbf{Preparation of Fock state $\ket{1}$ using GCR.}
  Wavefunctions $\psi(x)$, $\psi(p)$, and rescaled Wigner distribution functions plotted against position $x$ and
  momentum $p$.
  \textbf{(Left)} Qubit expectation values after
  $\mathrm{CD}(\alpha_1,\sigma_\mathrm{y})$; hybrid fidelity
  $F_\mathrm{H}\sim 0.58$.
  \textbf{(Middle)} After $\mathrm{CD}(\beta_1,\sigma_\mathrm{x})$;
  with $F_\mathrm{H}\sim 0.84$.
  \textbf{(Right)} After $\mathrm{CD}(\alpha_2,\sigma_\mathrm{y})$;
  final fidelity $F_\mathrm{H}\sim 0.99$, matching state-of-the-art numerically optimized
  protocols of Refs.~\cite{ISA,eickbusch2022fast}.}
  \label{fig:fock1}
\end{figure}

Fock states are not superpositions of a finite number of non-overlapping Gaussians, placing
them outside the natural domain of the phase-space ISA. One option is to use controlled displacements to approximate the Law--Eberly protocol~\cite{law1996arbitrary} which
enables preparation of arbitrary superpositions of Fock states given access
to pure JC or AJC evolution. App.~\ref{app:law-eberly} shows that Trotterization of the
Jaynes--Cummings Hamiltonian outperforms numerically optimized circuits in
circuit \emph{duration} but not the circuit \emph{depth}. Another option is to consider a general series of  conditional displacements and momentum boosts. For a series of $N=3$ CD, this corresponds to
\begin{align}
  \mathrm{CD}(\alpha_2,\sigma_\mathrm{y})\,
  \mathrm{CD}(i\beta_1,\sigma_\mathrm{x})\,
  \mathrm{CD}(\alpha_1,\sigma_\mathrm{y}),
  \label{eq:CD_circuit}
\end{align}
which has a connection to quantum random walks~\cite{singh_quantum_2025}. The first CD parameter is chosen to make the smallest odd cat state with maximum success probability and fidelity. As shown in Fig.~\ref{fig:fock1}, $\beta_1$ and $\alpha_2$ are chosen by alternately imposing minima of $\langle\sigma_\mathrm{y}\rangle_x$
and $\langle\sigma_\mathrm{x}\rangle_x$ at the wavefunction maxima in
position and momentum of the oscillator at steps $1$ and $2$, yielding the states at steps $2$ and $3$, respectively. This protocol yields
$\alpha_1=\pi/4$, $\beta_1=1/2$, and
$\alpha_2=(1/\pi)\tan^{-1}\!\sinh(\pi/2)$. See App.~\ref{app:fock_circuit} for details. The algorithm can be generalized to $\ket{n>1}$
and $N>3$. Our construction yields analytical explanation for the remarkable fidelities achieved by numerically optimized protocols for Fock state preparation using only phase-space instruction sets in previous experimental implementations~\cite{eickbusch2022fast}.

\subsection{Quantum phase estimation using ancillary oscillators}
\label{sssec:quantum-phase-estimation}
The sections above considered using the qubit as an ancilla to either control the oscillator or extract information from the oscillator. It is also interesting to consider the reverse situation: can we extract information about a qubit unitary using the oscillator? This use of hybrid QSP comes under Problem~$2$ of hybrid control, described in Sec.~\ref{sec:preliminaries}, focused on achieving arbitrary Hamiltonian evolution of the oscillator.

For example, consider the phase estimation problem in which one has oracle (i.e., black-box) access to the single-qubit rotation $U=e^{i\theta\sigma_\phi}$, and the task is to learn $\theta$. 
Liu et al.~\cite{liu2016power} discuss an approach where, if the information could be
encoded in the displacement of an oscillator (in a squeezed coherent state), a single-shot homodyne measurement can be used to extract multiple bits of information on $\theta$ in a single shot. Their analysis assumes access to (but does not provide) a hybrid control gate $C_U=e^{i\hat{x}\otimes H\tau/x_0}$ as a primitive (which for our problem corresponds to $H\tau/x_0 =\theta\sigma_\mathrm{y}$). Here, we show how to \emph{synthesize} such a primitive from realistic gates -- unitary access to $U=e^{i\theta\sigma_\mathrm{y}}$ together with the phase-space instruction set -- and analyze the resulting precision.

Here, we derive a hybrid QSP sequence which displaces the oscillator by an amount which depends on $\theta$ via access to the phase-space instruction set (see Sec.~\ref{sec:preliminaries}) and $U$. Throughout this section and App.~\ref{app:phase-est} we adopt $[\hat{x},\hat{p}]=i$
 (rather than $i/2$ used elsewhere)\footnote{For this section we work in the convention with $\hat{x}=(a+a^\dagger)/\sqrt{2}$; this matches the convention used in our numerical simulations and yields a clean form for the phase-estimation result. Other sections of the paper use $[\hat{x},\hat{p}]=i/2$.}. The protocol is specified as follows: we aim to learn the eigenvalue $\theta$ of a unitary $U=e^{i\theta\sigma_\mathrm{y}}$ by constructing a hybrid unitary which, when acting on a $\sigma_\mathrm{y}$ eigenstate, displaces the oscillator momentum by an amount proportional to $\sin(2\theta)$.

\begin{figure}[t]
  \centering
  \includegraphics[width=\linewidth]{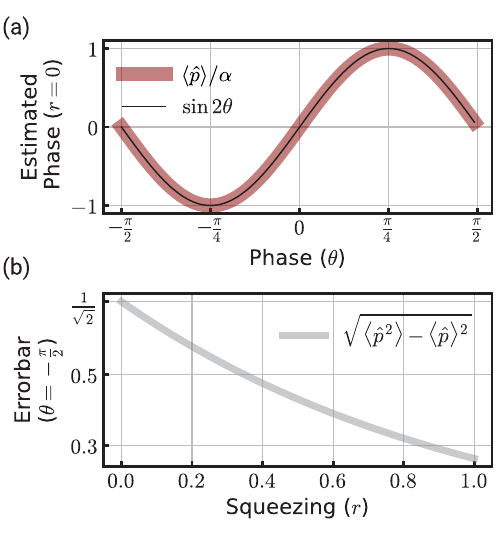}
  \caption{\textbf{Quantum phase estimation in hybrid oscillator-qubit
  architecture.}
  \textbf{(a)} Phase estimation using the control sequence in
  Sec.~\ref{sssec:quantum-phase-estimation}, confirmed numerically: at
  small $\alpha$ the rescaled mean momentum $\braket{\hat{p}}/\alpha$
  approaches $\sin(2\theta)$, giving a direct readout of $\theta$. The
  figure uses $\alpha=10^{-3}$, where the exact result
  Eq.~(\ref{eq:p_exact}) deviates from this leading-order form by less
  than $10^{-6}$.
  \textbf{(b)} The standard deviation is upper bounded by $1/\sqrt{2}$
  at zero squeezing ($r=0$), consistent with Eq.~(\ref{eq:p2_exact}),
  and improves with momentum squeezing $r>0$.}
  \label{fig:phase_est}
\end{figure}

The construction
\begin{align}
e^{i\alpha\hat{x}\sigma_\mathrm{x}}\,
  \sqrt{X}^{\dagger} U \sqrt{X}\,
  e^{-i\alpha\hat{x}\sigma_\mathrm{x}}\,
  \sqrt{X}\, U\, \sqrt{X}^{\dagger}\nonumber\\
  \;\approx\; e^{i\alpha\hat{x}
  (2\sin^2\theta\,\sigma_\mathrm{x}
  +\sin(2\theta)\,\sigma_\mathrm{y})},
  \quad \alpha\hat{x}\ll 1,
  \label{eq:CxU}
\end{align}
where $\sqrt{X}\,U\,\sqrt{X}^{\dagger}$ is used to synthesize the
adjoint of $\sqrt{X}^{\dagger} U \sqrt{X}$ without access to
$U^\dagger$, when applied to $\ket{0,+i}$. At leading order in $\alpha\hat{x}$, the construction reduces to a 
qubit rotation with two components: a $\sigma_\mathrm{y}$ term 
$\alpha\hat{x}\sin(2\theta)$ and a $\sigma_\mathrm{x}$ term 
$2\alpha\hat{x}\sin^2\theta$. The $\sigma_\mathrm{y}$ component imprints a 
momentum-displacing phase $e^{\pm i\alpha\hat{x}\sin(2\theta)}$ on 
the oscillator, from which $\theta$ can be read out via homodyne 
measurement. The $\sigma_\mathrm{x}$ component drives transitions 
between $\ket{+y}$ and $\ket{-y}$, transferring a fraction 
$|\alpha_-|^2$ of shots to a wrong-sign momentum kick, reducing the 
mean and inflating the variance by $4\alpha^2\sin^4\theta$ 
(Eq.~(\ref{eq:p2_exact})). However, in the small-$\alpha$ 
regime where the protocol is valid, $|\alpha_-|^2\ll|\alpha_+|^2$ 
and this effect is negligible~\footnote{the regime where it becomes significant 
is precisely where the Gaussian approximation breaks down and a 
fundamentally different estimator is required anyway, which we leave 
for future work}. The mean momentum $\braket{\hat{p}}\approx 
\alpha\sin(2\theta)$ thus serves as a good estimator of $\theta$ 
within the valid operating range, confirmed numerically in 
Fig.~\ref{fig:phase_est}(a). This construction synthesizes the 
Hamiltonian-imprinting primitive assumed by Liu 
et al.~\cite{liu2016power} from unitary access to $U$ alone together with 
the phase-space instruction set.

This example indicates the use-case
of hybrid QSP in identifying control pulses for Hamiltonian evolution
of oscillators which can be used as primitives to augment DV-only algorithms.
Further technical details are presented in App.~\ref{app:phase-est}.
The homodyne distribution $P(p)$
remains Gaussian (numerically, $\alpha\lesssim 0.1$ keeps deviations
from Gaussianity below $10^{-4}$ in Kullback-Leibler divergence).
At larger $\alpha$ the oscillator-qubit
entanglement generated by Eq.~(\ref{eq:CxU}) produces non-Gaussian
and even bimodal momentum distributions, and a closed-form expression
for $\braket{\hat{p}}$ that is exact for any $\alpha$ is also derived
in App.~\ref{app:phase-est}.  

\paragraph{Precision and overhead.}
We now ask how precisely the protocol determines $\theta$, and at
what cost. The shot-to-shot variance of the homodyne readout is set
by the squeezing parameter $r\geq 0$ of the input oscillator state,
which we summarize by the squeezing factor
$s_0\equiv e^r\in[1,\infty)$: zero squeezing ($s_0=1$) corresponds to
vacuum input, while larger $s_0$ exponentially reduces the noise in
$\hat{p}$. Each individual shot therefore yields only a noisy
estimate of $\braket{\hat{p}}$, and the protocol must be repeated $N$
times to suppress the statistical uncertainty in the inferred value
of $\theta$ by the usual factor of $1/\sqrt{N}$. Standard error
propagation through the Cramér-Rao bound, in the small-$\alpha$
Gaussian regime established above, shows that estimating $\theta$ to
a target precision $\varepsilon$ (that is, the target standard deviation
 in $\theta$) requires
\begin{align}
  Q \;\sim\; \frac{1}{\alpha^2\,s_0^2\,\varepsilon^2}
  \label{eq:Q_scaling}
\end{align}
total applications of $U$, where $Q=2N$ counts each of the $N$ shots
as the two queries to $U$ called by Eq.~(\ref{eq:CxU}).
Equation~(\ref{eq:Q_scaling}) exhibits the standard quantum limit
scaling $Q\propto 1/\varepsilon^2$, with the squeezing factor acting
as a multiplicative resource that reduces the shot count. The full
derivation, together with the closed-form expressions for the mean
$\braket{\hat{p}}$ and the variance $\delta p^2$ on which the bound is
based, is given in App.~\ref{app:phase-est}.

The scaling Eq.~(\ref{eq:Q_scaling}) recovers the ``squeezing as
inverse precision'' behavior of Liu et al.~\cite{liu2016power}, who
reached the same scaling for an assumed Hamiltonian-imprinting
primitive $e^{i\theta\hat{x}\otimes\sigma_\mathrm{y}}$ that is not
natively available if $\theta$ is unknown; the construction
Eq.~(\ref{eq:CxU}) provides this primitive from unitary access to
$U=e^{i\theta\sigma_\mathrm{y}}$ alone. The $1/\alpha^2$ prefactor
reflects a genuine cost of the synthesis: although larger $\alpha$
would give a proportionally larger signal and reduce the shot count,
the Gaussianity required for the Cramér-Rao analysis breaks down
beyond $\alpha\sim 0.1$ (App.~\ref{app:phase-est}), so the bound
Eq.~(\ref{eq:Q_scaling}) only applies in this regime.

For comparison, the simplest qubit-only phase estimation -- applying
$U$ once to $\ket{g}$ and inferring $\theta$ from the empirical
frequency of $\ket{e}$ outcomes -- reaches the same standard quantum
limit scaling $Q_{\mathrm{trivial}}\sim 1/\varepsilon^2$ with no
oscillator ancilla. Our protocol beats this trivial baseline when
the available squeezing satisfies $\alpha\,s_0\gtrsim 1$ at the
optimal operating point ($\theta\approx m\pi/2$ where $m$ is an integer), which in
the analyzed regime $\alpha\lesssim 0.1$ corresponds to a squeezing
factor $s_0\gtrsim 10$; reaching this operating point in practice
requires a coarse prior estimate of $\theta$, available from a small
number of preliminary shots. Heisenberg-scaling alternatives such as
robust phase estimation~\cite{HigginsRPE2007,KimmelRPE2015} trade
circuit depth in $U$ for query count; the constant-depth feature of
Eq.~(\ref{eq:CxU}) -- two applications of $U$ per shot, independent
of $\varepsilon$ -- makes our protocol a natural substitute when
deep coherent circuits in $U$ are limited by qubit decoherence, as
in circuit QED. Whether this constant-depth advantage survives
realistic noise -- qubit dephasing, oscillator photon loss, and
finite-bandwidth pulse imperfections -- is a natural next step that
we leave for future work.

\subsection{Hybrid analytical--numerical optimization}\label{hybrid-opt}

The improved performance of our analytical methods stems from explicitly
exploiting the algebraic structure of the control Hamiltonians, which
allows systematic cancellation of leading-order error terms while
preserving a compact and interpretable pulse structure. One interesting direction is to use our pulse sequences as a seed for numerical optimization methods. This might allow us to mitigate the effect of spurious error sources, for example oscillator non-linearity or finite ancilla lifetime.

\section{Conclusion}\label{sec:conclusion}

The central message of this work is that a single non-Abelian QSP primitive is already remarkably powerful. Using one composite pulse---the Gaussian-Controlled-Rotation (GCR)---we have given fully analytical constructions for problems across hybrid CV-DV control that either previously relied on numerical optimization or had no high-fidelity solution even in the noiseless regime: deterministic preparation of squeezed, cat, GKP, and Fock states; universal control of GKP bosonic qubits including readout, stabilization, and error-corrected gate teleportation; and a phase-estimation primitive synthesized from realistic gates. In each case, the analytical approach matches the performance of state-of-the-art numerical schemes. Crucially, matching numerical optimization is not the essential point: the analytical structure supplies what numerical optimization structurally cannot---mid-circuit error detection at well-defined points and closed-form generalization to other lattices, qudits, and multi-mode codes---turning a collection of separately optimized circuits into a single, transferable building block. 

The primitive itself exploits non-commuting quantum control variables, and we benchmark it analytically and numerically against BB1, the leading univariate composite pulse from classical NMR. GCR achieves comparable performance with a $4.5\times$ reduction in circuit duration, and through that reduction outperforms BB1 in the presence of noise; we verify its correctness using three figures of merit with full numerical confirmation. For state preparation, our analytical squeezing and GKP constructions reach fidelity and circuit duration competitive with state-of-the-art numerics while outperforming semi-analytical approaches in infidelity at comparable duration. The analytical structure further enables mid-circuit ancilla error detection at well-defined intermediate points, a capability inaccessible to purely numerical optimization. With this technique, we prepare a GKP logical Pauli state at fidelity $0.96$ and success probability $0.94$ under experimentally realistic error parameters. 

GCR also provides an intuitive analytical explanation of the GKP stabilization and logical readout protocols originally derived through dissipation engineering and numerical optimization, unifying them as entangling and unentangling operations between a finite-energy GKP code and a qubit ancilla. This insight yields an error-corrected gate teleportation (ECGT) scheme for both single- and two-qubit logical GKP gates, which applies the target logical rotation while simultaneously correcting oscillator displacement errors. Protected against ancilla decay and dephasing by a piecewise construction, the ECGT circuit remains fault-tolerant under biased-noise models---to our knowledge, the most effective approach for non-Clifford operations on GKP qubits to date. 

All of the above applications follow from a single primitive, demonstrating the power of non-Abelian QSP. GCR provides a concrete proof of principle for non-Abelian QSP and motivates the central open problem it leaves: a complete constructive theory characterizing the polynomial classes achievable with non-commuting control variables. Just as quantum singular value transformation generalized QSP and unified essentially all known qubit-only quantum algorithms, we anticipate that such a theory may give rise to a non-Abelian QSVT, potentially unifying hybrid CV-DV quantum algorithms and extending beyond hybrid systems to multi-qubit gate synthesis. Our results provide a foundational step toward that vision.

\section{Code and Data Availability}
The code and data are available via GitHub at Ref.~\cite{singh_shraggynaqsp_simulations_2025}.

\begin{acknowledgments}
SS and SMG were supported by the Army Research Office (ARO), under 
Grant Number W911NF-23-1-0051. The views and conclusions contained in this document are 
those of the authors and should not be interpreted as representing the official policies, 
either expressed or implied, of the Army Research Office (ARO), or the U.S. Government. 
The U.S. Government is authorized to reproduce and distribute reprints for Government purposes 
notwithstanding any copyright notation herein. BR was supported by the National Sciences and Engineering Research Council of Canada (NSERC), the Canada First Research Excellence Fund, as well as the Fonds de Recherche du Québec, Nature et Technologie (FRQNT). External interest disclosure: SMG receives consulting fees and is an equity holder in Quantum Circuits, Inc. SS would also like to thank Xanda Kolesnikow for pointing out some typos upon the arXiv upload of the first version of this manuscript.
\end{acknowledgments}
\appendix
\section{Quantum Phase Space Quadratures}\label{app:phase-space}

Second quantization defines the position and momentum quadratures of a
quantum oscillator via the non-Hermitian operators $\hat{a}$ and
$\hat{a}^\dagger$.  In Wigner units ($\hbar=1$),
\begin{align}
  \hat{x} = \frac{\hat{a}+\hat{a}^\dagger}{2}, \qquad
  \hat{p} = \frac{\hat{a}-\hat{a}^\dagger}{2i},
\end{align}
giving the commutation relation $[\hat{x},\hat{p}]=\frac{i}{2}$, or
equivalently $\hat{p}=-\frac{i}{2}\frac{\partial}{\partial x}$.

The vacuum state $\ket{0}_\mathrm{vac}$ is the minimum-uncertainty state
annihilated by $\hat{a}$:
\begin{align}
  \hat{a}\ket{0}_\mathrm{vac} = 0
  &\implies (\hat{x}+i\hat{p})\ket{0}_\mathrm{vac} = 0 \nonumber\\
  &\implies \left(x + \tfrac{1}{2}\tfrac{\partial}{\partial x}\right)
    e^{-x^2/\sigma} = 0 \nonumber\\
  &\implies \sigma = 1, \qquad
    \braket{x|0}_\mathrm{vac} = \left(\frac{2}{\pi}\right)^{1/4}e^{-x^2}.
\end{align}
This state satisfies $\delta x = \delta p = \frac{1}{2}$ and hence
$\delta x\,\delta p = \frac{1}{4}$, saturating the uncertainty bound.
More generally, states with arbitrary position uncertainty are written
$\psi(x)\propto e^{-x^2/\Delta^2}$, where $\Delta^2 = 4\delta x^2 = 1$
recovers the vacuum.  Following the standard definition
$\hat{a}\ket{\alpha}=\alpha\ket{\alpha}$, the position representation of a
squeezed coherent state is
\begin{align}
  |\braket{x|\alpha_\Delta}|^2
  = \left(\frac{2}{\pi\Delta^2}\right)^{1/4}
    e^{-\frac{2(x-\alpha)^2}{\Delta^2}}.
\end{align}

\subsection{Phase-space instruction set}\label{phase-space}
The hybrid gates $\mathrm{CD}(\beta,\sigma_\phi)$ displace
the oscillator by $\pm\beta$ depending on the qubit eigenstate of
$\sigma_\phi$, where $\beta=\Delta x+i\Delta p$.  These gates have been
realized experimentally via dispersive and sideband
interactions~\cite{leghtas_hardware-efficient_2013,eickbusch2022fast,
fluhmann2019encoding,de2022error}.  In Wigner units, for purely real
$\beta$,
\begin{align}
  \braket{x|\mathrm{CD}(\beta,\sigma_\phi)|0_\Delta,{+\phi}}
  &= \braket{x|e^{-i2\beta\hat{p}\otimes\sigma_\phi}|0_\Delta,{+\phi}}
  \nonumber\\
  &= \left(\frac{2}{\pi}\right)^{1/4}
    e^{-\frac{(x-\beta)^2}{\Delta^2}}\otimes\ket{+\phi},
\end{align}
where $\ket{+\phi}$ is the $+1$ eigenstate of $\sigma_\phi$: the oscillator
wavefunction is displaced along the position axis by $|\beta|$, with the
direction conditioned on the qubit state.
We refer to $\mathrm{CD}$s with purely real (imaginary) $\beta$ as
conditional position displacements (momentum boosts).
\section{Error Analysis of Composite Pulse Sequences in Phase Space}
\label{app:comp_err}

In this appendix, we compute the error analysis for $\mathrm{BB1}(2\theta)$ and $\mathrm{GCR}(2\theta)$. The equations derived here underpin the plots in Sec.~\ref{sec:comp}.

\subsection{Performance metrics}
\label{app:perf_metrics}

We define the metrics used to assess the performance of QSP sequences for extracting a single bit of information from the oscillator via the qubit. For a QSP sequence $U_t$ applied to $\ket{g}\otimes\ket{\pm\alpha_\Delta}$,
\begin{align}
U_t(\ket{g}\otimes\ket{\alpha_\Delta})
  &= \beta_{+g}\ket{\psi_{+g}}\ket{g}+\gamma_{+e}\ket{\phi_{+e}}\ket{e}, \\
U_t(\ket{g}\otimes\ket{-\alpha_\Delta})
  &= \gamma_{-g}\ket{\phi_{-g}}\ket{g}+\beta_{-e}\ket{\psi_{-e}}\ket{e}.
\end{align}
Assuming symmetry ($\beta_{+g}=\beta_{-e}\equiv\beta$, $\gamma_{+e}=\gamma_{-g}\equiv\gamma$), we define two figures of merit. The \textit{failure probability} $P_e(U_t)=|\gamma|^2$ measures the probability of incorrectly identifying the sign of the mean oscillator position. The \textit{hybrid state fidelity},
\begin{align}
F_\mathrm{H} = |\beta|^2|\braket{\alpha|\psi}|^2+|\gamma|^2|\braket{\alpha|\phi}|^2,
\end{align}
quantifies the back action on the oscillator state in the event of a qubit reset. We quote $P_e(U_t)$ and $F_\mathrm{H}(U_t)$ for each QSP sequence used in this work.

\subsection{No QSP correction}
\label{app:no_corr}

Starting from $\ket{\psi_1}=\ket{g}\otimes\ket{\alpha}$ and applying $e^{-i\frac{\pi}{4\alpha}\hat{x}\sigma_\mathrm{x}}$, the overlap with the desired state $\ket{-i}\otimes\ket{\alpha}$ is,
\begin{align}
&\left(\frac{2}{\pi}\right)^{\!\frac{1}{2}}
 \int_{-\infty}^{\infty}dx\;
 \cos\!\left(\frac{\pi}{4\alpha}(x-\alpha)\right)e^{-2(x-\alpha)^2}
 \approx 1-\frac{\pi^2}{128\alpha^2}.
\end{align}
Therefore, the hybrid state fidelity for large $|\alpha|$ is,
\begin{align}
F_\mathrm{H}(\mathrm{no\text{-}QSP})
\approx 1-\frac{\pi^2}{64\alpha^2}, \label{eqn:fidelity}
\end{align}
giving $P_e(\mathrm{no\text{-}QSP})=0.25\chi^2$ since $e^{i\frac{\pi}{4}x\sigma_\mathrm{x}}$ applies no back action in the position basis.

\subsection{Bivariate sequence: \texorpdfstring{$\mathrm{GCR}(2\theta)$}{GCR(2theta)}}
\label{app:error-analysis}

We compute $P_g=1-P_e$ for GCR using $y=(x-\alpha)/\Delta$, $\lambda=-\theta_\mathrm{t}\Delta^2/|\alpha|$, and the key parameter $\chi=|\lambda|/(2\Delta)=\theta_\mathrm{t}\Delta/(2|\alpha|)$. Recall from Sec.~\ref{sec:GCR} that $U_\mathrm{t}^\dagger \mathrm{GCR}(\theta_\mathrm{t})\ket{g}\ket{\alpha_\Delta}=U^\mathrm{Err}\ket{g}\ket{\alpha_\Delta}$, such that 
\begin{align}
    U^\mathrm{Err}= U(\equiv e^{i\frac{\theta_\mathrm{t}}{2|\alpha|}(\hat{x}-\alpha)\sigma_\mathrm{x}})\cdot V(\equiv e^{i\frac{\lambda}{2}\hat{p}\sigma_\mathrm{y}}).   
\end{align}
 In the position basis, the action of $U$ and $V$ is given by the following equations.
\allowdisplaybreaks{
\begin{align}
\braket{x|U|\alpha_\Delta}\ket{g}&=\sum_{m=0}^{\infty}\frac{[i\theta_\mathrm{t}(x-\alpha)\sigma_\mathrm{x}]^m}{2^m|\alpha|^mm!}\alpha_\Delta(x)\ket{g}\\ 
\braket{x|V|\alpha_\Delta}\ket{g}&=\braket{x|\sum_{n=0}^\infty\frac{(i\lambda\hat p\sigma_\mathrm{y})^n}{2^n n!}|\alpha_\Delta}\ket{g}\nonumber\\&=\sum_{n=0}^\infty \Big(-\frac{\lambda\sigma_\mathrm{y}}{4\Delta}\Big)^n \frac{1}{n!}H_n\Big(\frac{x-\alpha}{\Delta}\Big)\alpha_\Delta(x)\ket{g}\label{eq:V}
\end{align}
where $H_n(x)$ denotes the $n^\mathrm{th}$ physicist's Hermite polynomial. We need to compute the distance between our approximate correction $V$ and the exact cancellation operator $U^\dagger$ for the initial state $\ket{g}\otimes\ket{\alpha_\Delta}=\ket{g,\alpha_\Delta}$. 

Expanding $U$ and $V$ in Taylor series and grouping by powers of $m+n$ modulo 4, with $c_{n,m}=(\lambda/\Delta)^{m+n}/(2^n n!\,m!)$, we obtain,
\begin{align}
\braket{x|UV|\alpha_\Delta,g}
&= \sum_{m+n\in 2\mathbb{Z}}(-1)^{\nu_m}c_{n,m}y^m H_n(y)e^{-y^2}\ket{g}
\nonumber\\
&\quad+i\!\sum_{m+n\in 2\mathbb{Z}+1}(-1)^{\mu_m}c_{n,m}y^m H_n(y)e^{-y^2}\ket{e},
\end{align}
where $\nu_m:\bmod(m,4)\ge 2$ and $\mu_m:\bmod(m,4)\in\{0,3\}$. The first-order terms cancel exactly ($ic_{10}H_1(y)=ic_{01}yH_0(y)$). Extracting terms to $\mathcal{O}(\chi^8)$, $\braket{x|UV|\alpha_\Delta,g}=$
\begin{widetext}
\begin{align}
&\mathcal{N}\!\left\{
   1+\chi^2\!\left[y^2-\tfrac{1}{4}\right]
   +\chi^4\!\left[-\tfrac{y^4}{6}-\tfrac{y^2}{4}+\tfrac{1}{32}\right]
   +\chi^6\!\left[-\tfrac{y^6}{90}+\tfrac{y^4}{24}+\tfrac{y^2}{32}-\tfrac{1}{384}\right]
   +\chi^8\!\left[\tfrac{y^8}{2520}+\tfrac{y^6}{360}-\tfrac{y^4}{192}
                  -\tfrac{y^2}{384}+\tfrac{1}{6144}\right]
   +\mathcal{O}(\chi^{10})
   \right\}e^{-y^2}\ket{g} \label{eq:g}\\
&\quad+\mathcal{N}\!\left\{
   i\chi^3\!\left[\tfrac{2y^3}{3}\right]
   -i\chi^5\!\left[\tfrac{y^3}{6}\right]
   +\mathcal{O}(\chi^7)
   \right\}e^{-y^2}\ket{e}, \label{eq:e}
\end{align}
\end{widetext}
where $\mathcal{N}$ is fixed by $\braket{\psi|\psi}=1$.

\medskip\noindent\textit{Success probability.}
Using $dx=\Delta\,dy$ and $\int_{-\infty}^{\infty}dy\,y^{2n}e^{-2y^2}=\sqrt{\pi/2}\,(2n-1)!!/4^n$,
\begin{align}
P_g &= \mathcal{N}^2\Delta\sqrt{\tfrac{\pi}{2}}
       \!\left(1-\tfrac{5\chi^6}{48}+\tfrac{11\chi^8}{768}
       +\mathcal{O}(\chi^{10})\right), \label{eq:succ_norm}\\
\frac{1}{\mathcal{N}^2}
    &= \sqrt{\tfrac{\pi}{2}}\,\Delta
       \!\left(1-\tfrac{29\chi^8}{768}+\mathcal{O}(\chi^{10})\right),
       \label{eq:norm}
\end{align}
giving the failure probability,
\begin{align}
P_e(\mathrm{GCR})
= \frac{5\chi^6/48-5\chi^8/96}{1-29\chi^8/768}
\approx 0.1\chi^6+\mathcal{O}(\chi^8). \label{eq:pe_gcr_app}
\end{align}

\medskip\noindent\textit{Post-selected fidelity.}
The final oscillator state conditioned on $\ket{g}$ is given by Eq.~(\ref{eq:g}). Dividing by the normalization,
\begin{align}
1-F_\mathrm{ps}
= \frac{\chi^4/8-\chi^6/8+\chi^8/64}
       {1-5\chi^6/48+11\chi^8/768}
+\mathcal{O}(\chi^{10}). \label{eq:fps_gcr}
\end{align}

\medskip\noindent\textit{Hybrid fidelity.}
\begin{align}
1-F_\mathrm{H}(\mathrm{GCR})
= \chi^4/8-\chi^6/48+\mathcal{O}(\chi^8). \label{eq:fh_gcr_app}
\end{align}
Both post-selected and reset infidelities scale as $\chi^4/8$ for $\chi\ll 1$.

\subsection{Univariate sequence: \texorpdfstring{$\mathrm{BB1}(2\theta_\mathrm{t})$}{BB1(2theta)}}
\label{app:err_BB1}

We adapt the well-known $\mathrm{BB1}(2\theta_\mathrm{t})$ sequence~\cite{wimperis1994broadband} to quantum control variables, using $y\Delta=x-\alpha$ and $\chi=\theta_\mathrm{t}\Delta/(2|\alpha|)$. Taylor expanding Eq.~(\ref{eq:BB1}) applied to $\ket{\psi}\otimes\ket{g}$ with $\braket{x|\psi}=e^{-(x-\alpha)^2/\Delta^2}$, and simplifying using $\cos(\pi\alpha/|\alpha|)=-1$, $\cos(\pi\alpha/2|\alpha|)=0$, $\sin(\pi\alpha/|\alpha|)=0$, $\sin(\pi\alpha/2|\alpha|)=\alpha/|\alpha|$, and $\sigma_{\phi_1}\sigma_{3\phi_1}\sigma_{\phi_1}=\sigma_{-\phi_1}$, the first- and second-order error terms cancel with the choice $\phi_1=\pm\cos^{-1}(-\theta_\mathrm{t}/2\pi)$, leaving,
\begin{widetext}
\begin{align}
\braket{x|\psi_\mathrm{BB1}}
&= \mathcal{N}\!\left\{
   \left[1+\mathcal{O}(\chi^4)\right]e^{-y^2}\ket{g}
   +\left[-i\chi^3 y^3\!\left(-\frac{\pi^2}{6\theta_\mathrm{t}^2}+\frac{1}{24}
          +i\frac{\pi}{4\theta_\mathrm{t}}\sqrt{1-\frac{\theta_\mathrm{t}^2}{4\pi^2}}
          \right)+\mathcal{O}(\chi^5)\right]e^{-y^2}\ket{e}
   \right\}. \label{eq:BB1_simplified}
\end{align}
\end{widetext}

\noindent\textit{Failure probability.} For $\theta_\mathrm{t}=\pi/4$ (BB1(90)),
\begin{align}
P_e(\mathrm{BB1}) = 1.85\chi^6+\mathcal{O}(\chi^8). \label{eq:pe_bb1_90}
\end{align}
For $\theta_\mathrm{t}=\pi/2$ (BB1(180)), $P_e=0.15\chi^6+\mathcal{O}(\chi^8)$, comparable to GCR. We primarily use BB1(90) in this work.

\medskip\noindent\textit{Reset fidelity.}
The reset fidelity expressions are computed from the $\chi^4$ and $\chi^6$ coefficients in the Taylor expansion (via Mathematica). Since the fourth-order term is purely imaginary in both cases, the infidelity scales as $\chi^6$. For $2\theta_\mathrm{t}=90^\circ$,
\begin{align}
1-F_\mathrm{H}(\mathrm{BB1},90^\circ) = 0.37\chi^6+\mathcal{O}(\chi^8),
\end{align}
and for $2\theta_\mathrm{t}=180^\circ$,
\begin{align}
1-F_\mathrm{H}(\mathrm{BB1},180^\circ) = 15.6\chi^6+\mathcal{O}(\chi^8).
\end{align}
Hence $\mathrm{GCR}(2\theta_\mathrm{t})$ achieves performance on par with $\mathrm{BB1}(2\theta_\mathrm{t})$ in success probability while yielding lower infidelity at significantly reduced circuit depth, as claimed in Sec.~\ref{sec:comp}. Figure~\ref{fig:BB1_180} shows the comparison for $\theta_\mathrm{t}=\pi/2$, contrasting with Fig.~\ref{fig:Correctness}(a) for $\theta_\mathrm{t}=\pi/4$.

\begin{figure}[t]
  \centering
  \includegraphics[width=0.9\linewidth]{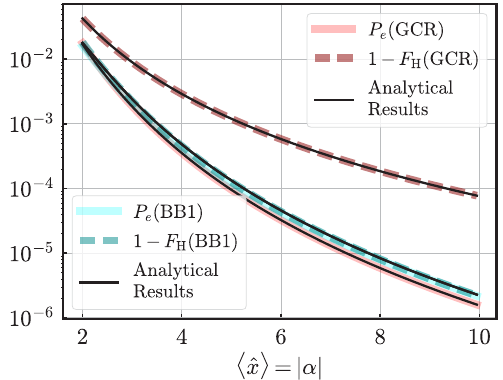}
  \caption{\textbf{Comparison of $\mathrm{GCR}(2\theta_\mathrm{t})$ and
  $\mathrm{BB1}(2\theta_\mathrm{t})$ for $\theta_\mathrm{t}=\pi/2$ and $\braket{\hat p}=0,\ \Delta=1$.}
  The failure probability for $\theta_\mathrm{t}=\pi/2$ is comparable for
  both schemes, in contrast to Fig.~\ref{fig:Correctness}(a)
  where $\theta_\mathrm{t}=\pi/4$ and the prefactors differ by an order
  of magnitude.}
  \label{fig:BB1_180}
\end{figure}

\section{Composition of BB1 and GCR for Modular Readout}
\label{app:bb1gcr}
The goal for this composite pulse, namely BB1(GCR), is to extract bitwise modular position
information~\cite{de2024modular}, in the presence of errors due to Gaussian
uncertainty and a random displacement channel. Starting from
$\ket{g}\otimes\ket{\pm 2\alpha'_\Delta}$ with
$\langle x\rangle/(2|\alpha|)=\alpha'/|\alpha|=m$, the ideal measurement outcome is
$+1$ if $\lfloor m\rfloor\in 2\mathbb{Z}$ and $-1$ if $\lfloor m\rfloor\in 2\mathbb{Z}+1$;
a good readout pulse must therefore produce a response $P(+1)$ that closely
approximates a square wave. To see why BB1(GCR) fails to do so, we will compare the operating-point readout error $P(-1)|_{x=0}$ against the BB1 pulse for which the error comes out to be $6.1\times10^{-3}$.

The Gaussian-Controlled-Rotation (GCR) primitive corrects a
position-dependent qubit rotation error by pairing each
oscillator-position--controlled kick with a compensating
momentum-controlled kick. In isolation, a GCR pulse suppresses the
position-uncertainty error of the bare rotation it replaces.
The resulting concatenated BB1(GCR) sequence is,
\begin{align}
  \mathrm{BB1}(\mathrm{GCR}(\theta))&:
  \mathrm{GCR}_0\!\left(\frac{\theta}{|\alpha|}\hat{x}\right)
  \mathrm{GCR}_{\phi_1}\!\left(\frac{\pi}{|\alpha|}\hat{x}\right)\nonumber\\
  &\quad\times\mathrm{GCR}_{3\phi_1}\!\left(\frac{2\pi}{|\alpha|}\hat{x}\right)
  \mathrm{GCR}_{\phi_1}\!\left(\frac{\pi}{|\alpha|}\hat{x}\right),
  \label{eq:GCR-BB1}
\end{align}
with $\phi_1$ as in Eq.~(\ref{eq:BB1}). We use the reversed BB1
ordering in Eq.~(\ref{eq:GCR-BB1})\footnote{BB1 correction can equivalently be prepended,
appended, or interleaved with the target rotation, without affecting
performance.  The reversed order is chosen to match the pre-correction
structure required by GCR.}, so that the BB1 correction pulses act on the qubit state before the target pulse. In the absence of errors, these pulses intend to rotate qubits to the same state regardless of the value of the mean position $\braket{\hat x}$. The final target rotation by $m\theta$ rotates the qubit to different states depending on $\braket{\hat x}$, hence, yielding modular information about $\braket{x}$. The pre-correction of each GCR pulse should know which direction the qubit is starting in, so by appending the pre-correction we solve this basic problem in the concatenation of BB1 and GCR. Concretely, each GCR pulse in the sequence is given by,
\begin{align}
   \mathrm{GCR}_{\phi}\!\left(\frac{\eta}{|\alpha|}\hat{x}\right)
   = e^{i\frac{\eta}{2|\alpha|}\hat{x}\sigma_{\phi}}
     e^{i\frac{\eta\Delta^2}{2|\alpha|}\hat{p}\sigma_\gamma},
\end{align}
where the pre-correction axis $\sigma_\gamma$ --- the second, momentum-controlled
kick of each GCR unit --- is determined by the ideal qubit state $\ket{\zeta}$
reached after the preceding \emph{error-free} rotation, through
$\sigma_\gamma\ket{\zeta}=i\sigma_\phi\ket{\zeta}$.  For the outermost gate
$\mathrm{GCR}_0$ the ideal qubit state is simply the input $\ket{g}$.

One of the approximations used to prove the correctness of GCR is to satisfy that $e^{i\frac{\eta\Delta^2}{2|\alpha|}\hat p\sigma_\gamma}\ket{\zeta}=e^{-\frac{\eta\Delta^2}{2|\alpha|}\hat p\sigma_\phi}\ket{\zeta}$. When $\hat p$ is written in the position basis, it makes the RHS of this equation an error-canceling unitary of the next pulse, up to first-order in the error (see Sec.~\ref{sec:GCR}). The equality under this approximation exists only when $\sigma_\gamma\ket{\zeta}=i\sigma_\phi\ket{\zeta}$ is true, and while by design of the pulse $\sigma_\gamma=i\sigma_\phi\sigma_\zeta$ can be chosen such that $\sigma_\zeta\ket{\zeta}=\ket{\zeta}$, the qubit state might not be purely $\ket{\zeta}$ due to its residual entanglement with the oscillator (see Fig.~\ref{fig:Correctness}). If there was a way to take the pre-correction axis to be $i\sigma_\phi$ itself, the
resulting ``imaginary'' conditional displacement would cancel the
position-uncertainty error exactly, at every angle, and the readout would be an
almost perfect square wave ($P(-1)|_{x=0}\approx4\times10^{-7}$). This non-unitary pulse where $\sigma_\gamma$ was just replaced by $i\sigma_\phi$ in every GCR pulse is a proof that the other approximation used in the construction of GCR $e^{-\frac{\eta\Delta^2}{2|\alpha|}\hat p\sigma_\phi}\ket{\psi}\equiv e^{-i\frac{\eta}{2|\alpha|}(\hat x-\alpha)\sigma_\phi}\ket{\psi}$ where $\ket{\psi}$ is an oscillator state that is a sum of non-overlapping Gaussian functions in $x$. However, the realizable pre-correction axis choice of $\sigma_\gamma=i\sigma_\phi\sigma_\zeta$,
reproduces this desirable square response only if the initial state of the qubit was completely unentangled from the oscillator. Since BB1's correction
angles run as large as $2\pi$, their higher-order
residuals dominate, and for the $2\pi$ pulse, the GCR pre-correction does more harm than
good. The readout consequently washes out ($P(-1)|_{x=0}\approx0.31$, essentially
random), faring worse than the bare BB1 it was meant to improve. Splitting the pulses cannot
rescue this (sub-$\pi$ sub-rotations break the $x$-agnostic condition, while the only legal split
$2\pi\to\pi+\pi$ reproduces similar operator fidelity). The failure is structural: BB1 requires the
oscillator and ancilla to stay entangled throughout its sequence, whereas GCR needs to start each
pulse from a fresh, unentangled ancilla in a definite state.

\paragraph{Heralded readout.}  In spite of this limitation, it would be interesting to pursue BB1(GCR) for future work. The ideal pulse removes the residual \emph{response} error that limits the deterministic schemes of
Fig.~\ref{fig:bb1-readout}(d). At the no-error point bare BB1 reaches only
$P(-1)|_{x=0}\approx 6\times10^{-3}$. The ideal BB1(GCR)
response is sharp to $\sim\!4\times10^{-7}$ per peak, so a faithful realization would reach the Helstrom-limited optimum value for GKP readout, below any deterministic scheme in Fig.~\ref{fig:bb1-readout}(d). While the performance of the ideal pulse quoted here was realized by using an unphysical $i\sigma_\phi$ followed by enforcing renormalization of the state, the ideal pulse can nevertheless be realized with physical operations \emph{probabilistically}. An extra ancilla can be used to enforce the $i\sigma_\phi$ in the pre-corrections via post-selection. Applied to all four pulses, the operating point readout error for this heralded realization is $\sim\!4\times10^{-7}$. Each herald measures a correction's \emph{ancilla},
not the readout qubit, so the $g/e$ outcome is preserved; heralding therefore costs only yield ($7\%$). The practical choice heralds only the three \emph{correction} pulses
--- the $\pi,2\pi,\pi$ block --- and leaves the final target a \emph{deterministic} GCR, giving the
highest yield ($\sim\!17\%$) with the fewest ancillas. On the $\sim\!17\%$ of shots where all three
heralds succeed the sequence is the full BB1(GCR), reaching $\sim\!5.7\times10^{-4}$. The yield can be increased to $100\%$ using methods from block encoding of quantum imaginary time evolution. The circuit noise on this pulse makes its performance at the no-error point worse than GCR itself~\cite{singh_shraggybb1gcr_simulations_2026}. Hence, we leave further analysis of this direction for future work since this optimization is not directly tied to the framework of non-abelian quantum signal processing.

\section{Squeezing}\label{app:squeezing}

\paragraph{Definitions.}
We consider oscillator states $\psi(x)=e^{-(x-\alpha)^2/4\delta x^2}$ with
position and momentum uncertainties $\delta x=\Delta/2$ and
$\delta p=1/(2\Delta)$.  Squeezing along position acts as
\begin{align}
  \mathrm{S}(\Delta)\ket{\psi}
  = \int_{-\infty}^{\infty}dx\,e^{-x^2/4\delta x^2}\ket{x},
  \quad 4\delta x^2 > 1.
\end{align}
For comparison with Refs.~\cite{eickbusch2022fast,hastrup2021unconditional},
squeezing in dB is defined as $S(r)=10\log_{10}(e^{r/2})$, with
$S_p=S(\log 4\Delta^2)$ and $S_x=S(\log 4/\Delta^2)$, where $r$ is the
bosonic squeezing parameter~\cite{ISA}.  The Fisher information for a
Gaussian state is $F=2/\delta x^2$~\cite{paris2004quantum}; we apply this
to the non-Gaussian states generated here because they remain close to
Gaussian at each step, enabling direct comparison with
Refs.~\cite{hastrup2021unconditional,eickbusch2022fast}.

\paragraph{Circuit duration.}
We report circuit duration rather than gate count because the error and
speed of a conditional displacement depend on displacement amplitude.
Given $H_\mathrm{CD}=\chi(\gamma_0 a^\dagger-\gamma_0^* a)$ with
$\chi/2\pi=50\ \mathrm{kHz}$ and $|\gamma_0|=20$, the duration is
$T_{|\gamma_i|}=|\gamma_i|/(\chi|\gamma_0|)$, lower bounded by
$T_{|\gamma_i|<0.024}=48\ \mathrm{ns}$, including an unconditional
displacement ($24\ \mathrm{ns}$) and a mid-circuit qubit rotation
($24\ \mathrm{ns}$)~\cite{eickbusch2022fast}.

\subsection{Squeezing with GCR}\label{app:squeezing-gcr}

Starting from vacuum ($\Delta=1$, $\beta=0$), after
$\mathrm{CD}(\alpha,\sigma_\mathrm{x})$ the qubit expectation values
conditioned on oscillator position are
\begin{align}
  \langle\sigma_\mathrm{z}\rangle_x &= \mathrm{sech}(4\alpha x/\Delta^2),\\
  \langle\sigma_\mathrm{x}\rangle_x &= \tanh(4\alpha x/\Delta^2),\\
  \langle\sigma_\mathrm{y}\rangle_x &= 0.
\end{align}
When $4\alpha/\Delta^2$ is small, $\langle\sigma_\mathrm{x}\rangle_x$
is approximately linear across the support of $\psi(x)$.  Applying the
corrective rotation $\mathrm{R}_\mathrm{y}(-4\alpha\hat{x}/\Delta^2)
=\mathrm{CD}(i\alpha/\Delta^2,\sigma_\mathrm{y})$ then gives
\begin{align}
  \langle\sigma_\mathrm{z}\rangle_x
    &= \tanh\!\tfrac{4\alpha x}{\Delta^2}\sin\!\tfrac{4\alpha x}{\Delta^2}
      +\mathrm{sech}\!\tfrac{4\alpha x}{\Delta^2}\cos\!\tfrac{4\alpha x}{\Delta^2}
     = 1-\mathcal{O}(x^6),\\
  \langle\sigma_\mathrm{x}\rangle_x
    &= \tanh\!\tfrac{4\alpha x}{\Delta^2}\cos\!\tfrac{4\alpha x}{\Delta^2}
      -\mathrm{sech}\!\tfrac{4\alpha x}{\Delta^2}\sin\!\tfrac{4\alpha x}{\Delta^2}
     = \mathcal{O}(x^3),\label{eq:new_sx}\\
  \langle\sigma_\mathrm{y}\rangle_x &= 0,
\end{align}
confirming disentanglement to $\mathcal{O}(x^3)$.  The resulting oscillator
state is a small even cat approximating a Gaussian with
$\Delta_k^2\approx\Delta_{k-1}^2+\alpha_k^2$.  This mirrors GCR in the
momentum basis, since $\theta/|\alpha|$ is small for large $|\alpha|$ in
Sec.~\ref{sec:GCR}.

\paragraph{Choice of $\alpha_k$.}
For $\langle\sigma_\mathrm{x}\rangle_x$ to remain linear across the FWHM
of $\psi(x)$, the slope $4\alpha_k/\Delta_k^2$ must satisfy
\begin{align}
  \frac{4|\alpha|}{\Delta^2}
  \ll \frac{(2/\pi\Delta^2)^{1/4}}{2\Delta\sqrt{\ln 2}}
  = \frac{0.53}{\Delta^{3/2}},
  \quad\implies\quad |\alpha| \ll 0.13\Delta^{1/2}.
\end{align}
For the accelerated $c=2$ protocol, we fit $\langle\sigma_\mathrm{y}\rangle$
using $|\alpha_0|=0.13$ and $|\alpha_k|=0.06\Delta^2$ for $k\neq 0$,
yielding the optimal correction for the approximately linear slope.  While
not fully analytical, this provides a seed for optimization that converges
significantly faster than schemes where this value is
unconstrained~\cite{hastrup2021unconditional}.

\subsection{Comparison with previous work}\label{app:squeezing-comparison}

Our protocol achieves $8.5\ \mathrm{dB}$ squeezing with infidelity
$\sim 0.003$ in $\sum_i|\alpha_i|\sim 5.7\,\mu\mathrm{s}$.  The two protocols operate in complementary regimes:
Ref.~\cite{hastrup2021unconditional} begins with a large conditional
displacement to prepare a large odd cat and displaces toward vacuum,
exploiting the $S_x\neq -S_p$ regime to achieve faster convergence at the
cost of numerical optimization for disentanglement and mild non-Gaussian
distortion near the origin.  Our protocol builds up squeezing incrementally
from vacuum, keeping the state close to Gaussian at every step and enabling
analytical disentanglement via GCR, at the cost of being restricted to the
$S_x=-S_p$ regime.  At $S_p=8.5\ \mathrm{dB}$,
Ref.~\cite{hastrup2021unconditional} reports infidelity
$\sim\mathcal{O}(10^{-2})$ vs.\ our $\sim\mathcal{O}(10^{-3})$ in
comparable circuit duration.  Our scheme is also on par with numerically
optimized schemes~\cite{eickbusch2022fast}, as shown in
Fig.~\ref{fig:numerical_squeezing}.

\begin{figure}
  \centering
  \includegraphics[width=\linewidth]{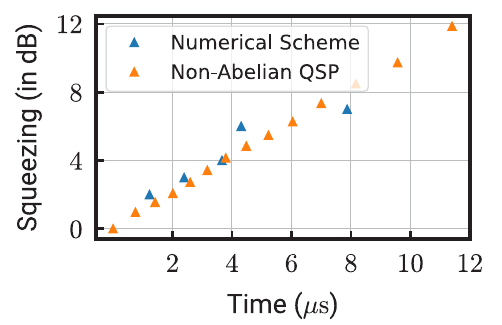}
  \caption{Squeezing of vacuum with fidelity $\mathcal{F}>0.99$ using
  non-Abelian QSP (Sec.~\ref{sec:squeezing}) and the numerically optimal
  scheme of Ref.~\cite{eickbusch2022fast}.}
  \label{fig:numerical_squeezing}
\end{figure}

\subsection{Comparison with Trotterization}\label{app:squeezing-trotter}

A Trotterization-based squeezing sequence using the BCH formula gives
\begin{align}
  &\bigl(e^{i\epsilon\hat{x}\sigma_\mathrm{x}/N}
          e^{i\epsilon\hat{p}\sigma_\mathrm{y}/N}
          e^{-i\epsilon\hat{x}\sigma_\mathrm{x}/N}
          e^{-i\epsilon\hat{p}\sigma_\mathrm{y}/N}\bigr)^N\ket{0,g}
  \nonumber\\
  &= e^{i\epsilon^2(\hat{x}\hat{p}+\hat{p}\hat{x})\sigma_\mathrm{z}}
     \ket{0,g}+\mathcal{O}(\epsilon^{4N}),
\end{align}
which generates the squeezing Hamiltonian $\hat{x}\hat{p}+\hat{p}\hat{x}$.
For a state with variance $\Delta^2$, the analogous sequence uses
$\epsilon/\Delta^2$ and $\epsilon$ for the two CDs respectively, yielding
squeezing proportional to $\epsilon^2/\Delta^4$ per step.  Effective
qubit disentanglement requires both increasing $N$ and decreasing $\epsilon$
as $\Delta$ grows, significantly increasing circuit duration.  As noted in
the main text, this makes Trotterization substantially slower than our
protocol at comparable fidelity: for $8.52\ \mathrm{dB}$ squeezing,
Trotterization requires $40.67\,\mu\mathrm{s}$ vs.\ $5.8\,\mu\mathrm{s}$
for our scheme.
\section{Cat State Preparation}\label{app:cat-prep}

We compute fidelities of cat states against the output of the preparation
circuits in Fig.~\ref{fig:Cat_states}(a) (large cats) and
Fig.~\ref{fig:squeezing}(a) (small cats).  In CV architectures, cat states
are superpositions of coherent states located diametrically opposite in
phase space,
\begin{align}
\ket{C_{\pm\alpha}}
  &\propto (\mathrm{D}(\alpha)\pm\mathrm{D}(-\alpha))\ket{0}_\mathrm{vac}
   \approx \frac{\ket{\alpha}\pm\ket{-\alpha}}{\sqrt{2}},\\
\psi(x)
  &= \braket{x|C_{\pm\alpha}}
   = \left(\frac{2}{\pi}\right)^{1/4}
     \frac{e^{-(x-\alpha)^2}\pm e^{-(x+\alpha)^2}}{\sqrt{2}}.
\end{align}

\paragraph{Non-deterministic preparation.}
Entangling the cavity vacuum with the qubit in $\ket{\pm}$ via
$\mathrm{CD}(\alpha,\sigma_\mathrm{x})$ gives
\begin{align}
\ket{\psi_1}
  = \mathrm{CD}(\alpha,\sigma_\mathrm{x})\ket{0}\ket{\pm}
  \propto \ket{\alpha}\ket{+i}\pm\ket{-\alpha}\ket{-i}.
  \label{eqn:entangled}
\end{align}
A subsequent $\mathrm{R}_0(-\pi/2)$ rotation yields
\begin{align}
\ket{\psi_2}
  \propto (\ket{\alpha}\pm\ket{-\alpha})\ket{g}
          - i(\ket{\alpha}\mp\ket{-\alpha})\ket{e}.
\end{align}
Measuring the qubit projects the cavity onto an even or odd cat state, each
with probability $\frac{1}{2}$.

\paragraph{Deterministic preparation.}
To fully disentangle the qubit from the cavity, we apply a momentum boost
$e^{i\beta\hat{x}\sigma_\mathrm{z}}$, which implements a position-dependent
$\sigma_\mathrm{z}$-rotation by $-2\beta\hat{x}$.  Using
$\mathrm{R}_\phi(\theta)=\mathrm{R}_{\pi/2-\phi}(-\pi/2)\,
\mathrm{R}_\mathrm{z}(\theta)\,\mathrm{R}_{\pi/2-\phi}(\pi/2)$,
\begin{align}
\ket{\psi_4}
  = \mathrm{R}_0(-\pi/2)\,e^{i\beta\hat{x}\sigma_\mathrm{z}}\ket{\psi_2}.
  \label{eqn:theta}
\end{align}
Setting $\beta=-\pi/(4|\alpha|)$ aligns the qubit polarization with the
cavity peaks at $x=\pm\alpha$, as shown in Fig.~\ref{fig:Cat_states}(a).

\subsection{Fidelity without QSP correction}\label{app:cat_I}

Writing $\ket{\psi_4}$ in the position basis and computing the overlap with
the target $\ket{C_{+\alpha}}\ket{+}$, neglecting inter-Gaussian overlap
and using the symmetry
$\int e^{-2(x-\alpha)^2}\cos(\beta(x-\alpha))\,dx
=\int e^{-2(x+\alpha)^2}\cos(\beta(x+\alpha))\,dx$,
the integral reduces to
\begin{align}
\braket{\psi_\mathrm{cat}|\psi_4}
  &\approx \left(\frac{2}{\pi}\right)^{\frac{1}{2}}
  \int_{-\infty}^{\infty}dx\;
  e^{-2(x-\alpha)^2}\!\left(1 - \frac{(\beta x - \pi/4)^2}{2}\right)\\
  &= 1 - \frac{\pi^2}{128\alpha^2}.
\end{align}
The fidelity for large cats is therefore
\begin{align}
\mathcal{F} \approx 1 - \frac{\pi^2}{64\alpha^2}.\label{eqn:cat_fidelity}
\end{align}
This matches the no-QSP-correction case for rotation gadgets
(App.~\ref{app:no_corr}), confirming that GCR and BB1 improvements carry
over directly to cat state preparation.

\subsection{Deterministic odd cat preparation}\label{app:odd-cat}

The unentangling gadget for the odd cat acts as
\begin{align}
  &\mathcal{U}(\pi/2,|\alpha|,1)\,
    e^{-i2\alpha\hat{p}\sigma_\mathrm{x}}
    (\ket{0}_\mathrm{vac}\otimes\ket{g})
  \nonumber\\
  &\quad= e^{-i\frac{\pi}{4|\alpha|}\hat{p}\sigma_\mathrm{x}}
          e^{-i\frac{\pi}{4|\alpha|}\hat{x}\sigma_\mathrm{y}}
    (\ket{\alpha}\ket{+}+\ket{-\alpha}\ket{-})
    \label{eq:cat-stated}\\
  &\quad\approx (\ket{\alpha}-\ket{-\alpha})\otimes\ket{e}
    +\mathcal{O}(\chi^2)\ket{\psi'}\otimes\ket{e}
    +\mathcal{O}(\chi^3)\ket{\psi''}\otimes\ket{g},
\end{align}
targeting $\ket{e}$ with maximum probability.  Cat fidelity improves with
increasing $|\alpha|$ and decreasing $\Delta$, consistent with
$\chi=\theta\Delta/2|\alpha|$ (Sec.~\ref{sec:comp}).

\subsection{The problem with small cat states}\label{app:small_cats}

For small cats ($|\alpha|^2\lesssim 4$), the overlap 
$\braket{\alpha|-\alpha}\not\to 0$ invalidates 
Eq.~(\ref{eqn:cat_fidelity}). The finite-overlap correction enters the 
fidelity integral as an additional term proportional to 
$e^{-2\alpha^2}$, representing the contribution from the cross term 
between the two Gaussian peaks. This correction adds to the even cat 
fidelity and subtracts from the odd cat fidelity, making 
$\mathcal{F}_\mathrm{odd}<\mathcal{F}_\mathrm{even}$ with the gap growing 
exponentially as $\alpha$ decreases. For $\beta=-2\alpha$, including the 
normalization $\mathcal{N}=\bigl(\frac{1}{2\pi}\bigr)^{1/4}/
\sqrt{1\pm e^{-2\alpha^2}}$, the fidelities are
\begin{align}
\mathcal{F}_\mathrm{even}
  &\approx\Biggl|
    \frac{1-\frac{\alpha^2}{2}-2\alpha^4+\frac{\pi\alpha^2}{2}
          -\frac{\pi^2}{32}}{\sqrt{1+e^{-2\alpha^2}}}
   +\frac{e^{-2\alpha^2}\bigl(1-\frac{\alpha^2}{2}
          -\frac{\pi^2}{32}\bigr)}{\sqrt{1+e^{-2\alpha^2}}}
  \Biggr|^2,\label{Fidelity_small_even}\\
\mathcal{F}_\mathrm{odd}
  &\approx\Biggl|
    \frac{1-\frac{\alpha^2}{2}-2\alpha^4+\frac{\pi\alpha^2}{2}
          +\frac{\pi^2}{32}}{\sqrt{1-e^{-2\alpha^2}}}
   -\frac{e^{-2\alpha^2}\bigl(1-\frac{\alpha^2}{2}
          -\frac{\pi^2}{32}\bigr)}{\sqrt{1-e^{-2\alpha^2}}}
  \Biggr|^2.\label{Fidelity_small_odd}
\end{align}
This asymmetry reappears in Fock-state preparation — projecting onto the 
even (odd) cat corresponds to $\ket{0}\to\ket{0}$ ($\ket{0}\to\ket{1}$) 
for small $\alpha$ — and is addressed in Sec.~\ref{ssec:universal} via 
the amplification gadget.
\section{GKP Logical Pauli States}\label{app:GKP-prep}

We derive the numerical circuit presented in Sec.~\ref{ssec:GKP-States}
for GKP state preparation. The required states described by Eq.~
(\ref{eq:GKP-bin-logical}) are superpositions of
finite-energy basis states $\ket{\alpha}_\Delta$ at positions
$\alpha=m\sqrt{2\pi}$, with $m\in 2\mathbb{Z}$ for
$\ket{0}_\mathrm{GKP},\ket{+}_\mathrm{GKP},\ket{+i}_\mathrm{GKP}$ and
$m\in 2\mathbb{Z}+1$ for
$\ket{1}_\mathrm{GKP},\ket{-}_\mathrm{GKP},\ket{-i}_\mathrm{GKP}$.
These superpositions are built using repeated applications of the cat-state
transfer circuit $\mathcal{C}$, which generates superposition coefficients
following a binomial distribution.  Below we quantify the relationship
between state fidelity and circuit depth, describe the unentanglement
procedure, and explain the role of the appended stabilization step.

\paragraph{State fidelity vs.\ circuit depth.}
The superposition coefficients generated by $\mathcal{C}_k$
(Fig.~\ref{fig:GKP-prep}(a)) follow Pascal's triangle
(Fig.~\ref{fig:pascal}): after $N$ cat-state transfer steps, the
$m^\text{th}$ peak has amplitude $\sqrt{\binom{N}{m}/2^N}$.  The target
GKP state requires amplitudes proportional to $k\,e^{-\pi m^2\Delta^2/2}$
at positions $m\sqrt{\pi}$.  For a given $\Delta$ and $Z_\mathrm{GKP}$
codeword $\mu$, the optimal $N$ satisfies
\begin{align}
  \sqrt{\binom{N}{m+\mu}\Big/2^N}
  = \exp\!\left(-\pi\frac{(2m+\mu-N/2)^2\Delta^2}{4}\right),
  \quad m\in\mathbb{Z}.
\end{align}
Applying Stirling's approximation and setting $x=m/N$ reduces this to the
transcendental equation
\begin{align}
  \frac{x^{-x}(1-x)^{-(1-x)}}{2} = e^{-2N\pi\Delta^2(x-0.5)^2},
\end{align}
solved numerically via Newton--Raphson iteration.  The overlap is maximized
at $N\Delta^2\approx 0.32$.  Table~\ref{tab:GKP-prep-circuit-depth} lists
the optimal $N$ for several values of $\Delta$ achieving fidelity
$\mathcal{F}\ge 0.98$.

\begin{figure}
  \centering
  \includegraphics[width=0.5\textwidth]{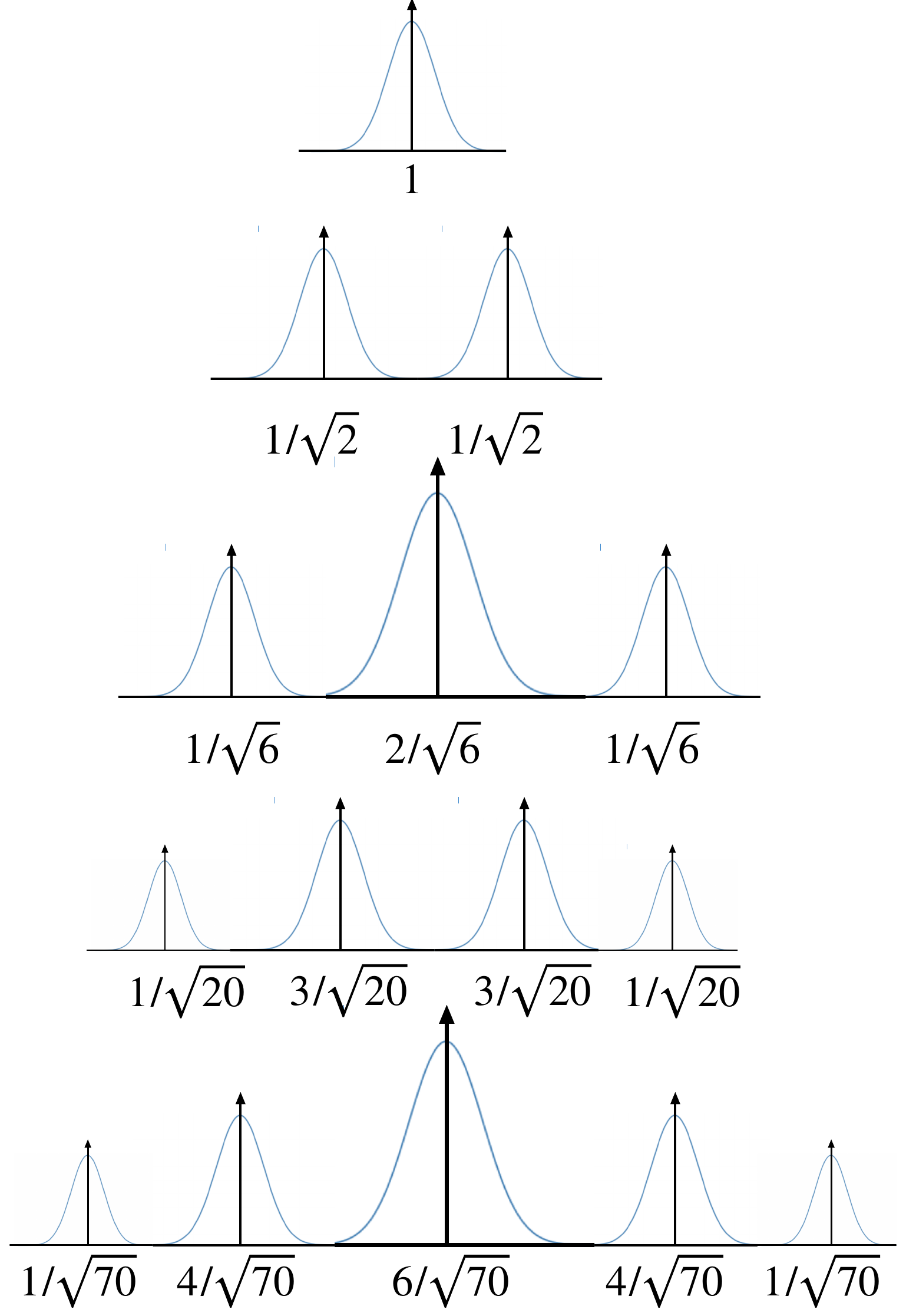}
  \caption{Probability amplitudes of the finite-energy basis states generated by repeated cat-state transfer circuits, arranged according to Pascal’s triangle. This structure defines the state prepared by the circuit in Fig.~\ref{fig:GKP-prep}. The numerators denote the binomial coefficients associated with each peak, while the denominators arise from the normalization of the corresponding binomial distribution.}
  \label{fig:pascal}
\end{figure}

\begin{table}[htb]
  \centering
  \begin{tabular}{||c|c|c|c||}
  \hline\hline
  $\Delta$ & Squeezing (dB) & $N$ & $1-\mathcal{F}$ \\
  \hline
  $0.1$ & $20$    & $31$ & $\mathcal{O}(10^{-5})$ \\
  $0.2$ & $14$    & $7$  & $\mathcal{O}(10^{-3})$ \\
  $0.3$ & $10.45$ & $3$  & $\mathcal{O}(10^{-2})$ \\
  $0.4$ & $7.95$  & $1$  & $\mathcal{O}(10^{-2})$ \\
  \hline\hline
  \end{tabular}
  \caption{Optimal number of large conditional displacements ($\sqrt{\pi}$)
  required to prepare $\ket{0}_\mathrm{GKP}$ at various squeezing levels,
  starting from $e^{-x^2/\Delta^2}$, obtained from $N\Delta^2=0.32$.}
  \label{tab:GKP-prep-circuit-depth}
\end{table}

\paragraph{Unentanglement.}
For $k=2$, the optimal unentanglement angle is not immediately obvious
(see Fig.~\ref{fig:GKP-prep}(a)).  Defining the fidelity as a function of
the conditional displacement amplitude $a$,
\begin{align}
  F &= \frac{\bigl[\frac{\pi}{4}-a\sqrt{\pi}x\bigr]^2
       +\sum_{i\in 2\mathbb{Z},\,2\le i\le x}
        \bigl[\binom{k}{i/2}a(x-i)\bigr]^2}
       {1+\sum_{i\in 2\mathbb{Z},\,2\le i\le x}\binom{k}{i/2}^2},
\end{align}
where $x=(k+1)/2+1$ for $k\in 2\mathbb{Z}+1$ and $x=k/2+1$ for
$k\in 2\mathbb{Z}$, the optimal angle minimizes $1-F$ subject to
$a\le\sqrt{\pi}/4k$.  The central peak is excluded for even $k$ since it
contributes no qubit rotation.  One can verify that the minimum is achieved
at $a=\sqrt{\pi}/4k$ for $k\le 3$.

\paragraph{Appending SBS.}
The states generated by $\mathcal{C}_k$ alone lack the correct coefficients
for the outermost peaks — visible as the absence of fringes in those peaks
in Fig.~\ref{fig:GKP-prep}(a).  Appending a single round of SBS corrects
this.  The combined scheme outperforms pure SBS preparation for two reasons:
the faster $\mathcal{C}_k$ circuits bring the state close to the GKP
codespace before SBS is invoked, avoiding its slow convergence
(Fig.~\ref{fig:GKP-prep}(b)); and since no ancilla measurement outcome is
relied upon during the $\mathcal{C}_k$ steps, mid-circuit error detection
can be used throughout.

\paragraph{State preparation from vacuum.}
Repeating the scheme with $\Delta=1$ (vacuum input) produces a
momentum-squeezed state after $\mathcal{C}_4$, whose finite-energy
parameter $\delta p$ is then used to repeat the protocol in the momentum
quadrature.  This yields a magic state with fidelity $0.85$ and success
probability $0.90$; the reduced fidelity arises because the unentangling
gadget achieves only $0.94$ success probability when generating a squeezed
cat from squeezed vacuum rather than from a state prepared by the squeezing
gadget.  This circuit does not require an appended SBS round, but its
hybrid fidelity is lower than that of the circuit in Fig.~\ref{fig:GKP-prep}(b),
and it is included here for completeness.

\paragraph{Arbitrary GKP states.}
Universal state transfer requires preparation of arbitrary GKP codewords
beyond the logical Pauli eigenstates.  Starting from $\ket{0}_\mathrm{GKP}$
in the cavity and the qubit in $a\ket{g}+b\ket{e}$, a finite-energy logical
$Z_\mathrm{GKP}$ operation conditioned on the qubit,
$e^{i\sqrt{\pi/2}\,\hat{x}\sigma_\mathrm{z}}$, prepares the hybrid state
\begin{align}
  \ket{\psi} = \mathrm{D}(i\sqrt{\pi}/2\sqrt{2})
               \bigl[a\ket{0' g}+b\ket{1' e}\bigr],
\end{align}
where $\ket{0'}=\mathrm{D}(-i\sqrt{\pi}/\sqrt{2})\ket{0}$ and
$\ket{1'}=\mathrm{D}(i\sqrt{\pi}/\sqrt{2})\ket{0}$ are off-centered GKP
states.  The off-centered states $\ket{0'}$ and $\ket{1'}$ are not in the GKP 
codespace, so a few rounds of stabilization are required to project back, 
giving $a\ket{0\,g}+b\ket{1\,e}$, after which GCR unentangles the qubit 
to yield $[a\ket{0}_\mathrm{GKP}+b\ket{1}_\mathrm{GKP}]\otimes\ket{g}$.
This construction performs no error correction during the displacement and
unentanglement steps; the error-corrected gate teleportation scheme in
Sec.~\ref{sec:GKP} addresses this limitation.

\section{Error-Corrected Gate Teleportation}\label{app:c-pauli}

We verify the error-corrected $\mathrm{CX}_\mathrm{GKP}$ teleportation
scheme described in Sec.~\ref{ssec:piecewise-teleportation} by tracking
the joint oscillator-qubit state through each step.  The setup consists
of two GKP-encoded oscillators, each stabilized by a dedicated ancilla
qubit.  Defining
\begin{align}
  \ket{\psi_1}_\mathrm{GKP} &= a\ket{0_1}_\mathrm{GKP}+b\ket{1_1}_\mathrm{GKP},\\
  \ket{\psi_2}_\mathrm{GKP} &= c\ket{+_2}_\mathrm{GKP}+d\ket{-_2}_\mathrm{GKP},
\end{align}
the protocol proceeds as follows.

\textit{Step 1: Entanglement.}
Applying the $x$-entangling gadget $\mathcal{E}_{\hat{x}_1}$ to the
control and the $p$-entangling gadget $\mathcal{E}_{\hat{p}_2}$ to the
target gives
\begin{align}
  &\mathcal{E}_{\hat{x}_1}\mathcal{E}_{\hat{p}_2}
   \bigl[\ket{\psi_1}_\mathrm{GKP}\otimes\ket{g_1}\bigr]
   \bigl[\ket{\psi_2}_\mathrm{GKP}\otimes\ket{g_2}\bigr]
   \nonumber\\
  &= \bigl[(a\ket{0_1}_\mathrm{GKP}\ket{g_1}
            -b\ket{1_1}_\mathrm{GKP}\ket{e_1})\bigr]
     \bigl[(c\ket{+_2}_\mathrm{GKP}\ket{g_2}
            -d\ket{-_2}_\mathrm{GKP}\ket{e_2})\bigr].
\end{align}

\textit{Step 2: Qubit CZ.}
Applying $C_1Z_2$ between the two ancilla qubits yields
\begin{align}
  &a\ket{0_1}_\mathrm{GKP}\ket{g_1}
   \bigl[c\ket{+_2}_\mathrm{GKP}\ket{g_2}
         -d\ket{-_2}_\mathrm{GKP}\ket{e_2}\bigr]
  \nonumber\\
  &-b\ket{1_1}_\mathrm{GKP}\ket{e_1}
   \bigl[c\ket{+_2}_\mathrm{GKP}\ket{g_2}
         +d\ket{-_2}_\mathrm{GKP}\ket{e_2}\bigr].
\end{align}

\textit{Step 3: Unentanglement.}
Applying $\mathcal{E}_{\hat{x}_1}\mathcal{E}_{\hat{p}_2}$ again
disentangles both qubits, leaving
\begin{align}
  &\bigl[a\ket{0_1}_\mathrm{GKP}(c\ket{+_2}_\mathrm{GKP}
         -d\ket{-_2}_\mathrm{GKP})
  \nonumber\\
  &-b\ket{1_1}_\mathrm{GKP}(c\ket{+_2}_\mathrm{GKP}
    +d\ket{-_2}_\mathrm{GKP})\bigr]
   \otimes\ket{g_1}\ket{g_2},
\end{align}
which is the $\mathrm{CX}_\mathrm{GKP}$ image of
$\ket{\psi_1}\otimes\ket{\psi_2}$ with both ancillae returned to $\ket{g}$
and ready for the next stabilization or gate cycle.

As a concrete check, initialising the control in
$\ket{-}_\mathrm{GKP}$ ($a=1,b=-1$) and the target in
$\ket{1}_\mathrm{GKP}$ ($c=1,d=-1$) produces the GKP Bell pair
$\ket{0_10_2}_\mathrm{GKP}+\ket{1_11_2}_\mathrm{GKP}$, with success
probability $0.9987$ and gate fidelity $99.92\%$.  This scheme provides no
protection against ancilla errors within a single teleportation step;
fault-tolerance is restored by the two-qubit Pauli rotation construction
$P_iP_j(\theta)$ described in Sec.~\ref{sec:GKP}.

\section{Finite-Energy SUM Gate}\label{app:finite-SUM}

Logical gates for GKP qubits are obtained via the non-unitary conjugation
$\hat{E}_\Delta A\hat{E}_\Delta^{-1}$, where $A$ is the corresponding
infinite-energy gate and $\hat{E}_\Delta=e^{-\Delta^2\hat{n}}$ is the
envelope operator (Sec.~\ref{sec:GKP}).  For the entangling gate
$\mathrm{CX}_\mathrm{GKP}$ with $A=e^{i2\hat{x}\otimes\hat{p}}$,
\begin{align}
  \hat{E}_\Delta\, e^{i2\hat{x}\otimes\hat{p}}\,\hat{E}_\Delta^{-1}
  &\approx e^{i(2\hat{x}_1\hat{p}_2
              +2\Delta^4\hat{p}_1\hat{x}_2
              -i2\Delta^2(\hat{x}_1\hat{x}_2
              -2\hat{p}_1\hat{p}_2))},
\end{align}
valid to leading order in $\Delta$ using $\cosh\Delta^2\approx 1$ and
$\sinh\Delta^2\approx\Delta^2$.  This can be realized with an auxiliary
qubit by defining $\mathrm{S}\equiv e^{-i\Delta^2(\hat{x}_1\hat{x}_2
-\hat{p}_1\hat{p}_2)\sigma_\mathrm{y}}$ and $\mathrm{B}\equiv
e^{i2(\hat{x}_1\hat{p}_2-\Delta^4\hat{p}_1\hat{x}_2)\sigma_\mathrm{x}}$,
so that
\begin{align}
  \mathrm{CX}_\mathrm{GKP}
  \approx \mathrm{S}\,\mathrm{B}\,\mathrm{S}\,\ket{\psi}_\mathrm{GKP}\ket{0},
\end{align}
implementable via dissipation-based methods with Trotterization~\cite{royer2020stabilization,rojkov2023two}
or via GCR-type correction using $\sigma_\mathrm{x}\ket{g}=-i\sigma_\mathrm{y}\ket{g}$.

\paragraph{Gate S: echoed conditional two-mode squeezing.}
Using the Baker--Campbell--Hausdorff expansion,
$\mathrm{TMS}(\alpha,\pi)\,a\,\mathrm{TMS}^\dagger(\alpha,\pi)
= a\cosh\alpha + b^\dagger\sinh\alpha$,
conjugating the dispersive interaction $e^{-i\chi t_\mathrm{S}a^\dagger a\sigma_\mathrm{z}}$
by $\mathrm{TMS}(\alpha,\pi)$ gives
\begin{align}
&\mathrm{TMS}(\alpha,\pi)\,e^{-i\chi t_\mathrm{S}a^\dagger a\sigma_\mathrm{z}}\,
  \mathrm{TMS}^\dagger(\alpha,\pi) \nonumber\\
&= \exp\!\Bigl[-i\chi t_\mathrm{S}\Bigl(
    \cosh^2\!\alpha\,a^\dagger a
   +\sinh^2\!\alpha\,bb^\dagger
   \nonumber\\&\quad+\tfrac{1}{2}\sinh(2\alpha)(a^\dagger b^\dagger+ab)
  \Bigr)\sigma_\mathrm{z}\Bigr].
  \label{eq:TMS-rot}
\end{align}
The first two terms are even in $\alpha$; the last is odd.  Echoing with a
qubit $\sigma_\mathrm{x}$ between $\pm\alpha$ therefore cancels the first
two terms, leaving
\begin{align}
  &\mathrm{TMS}(\alpha,\pi)\,e^{-i\chi t_\mathrm{S}a^\dagger a\sigma_\mathrm{z}}\,
    \mathrm{TMS}^\dagger(\alpha,\pi)\cdot\sigma_\mathrm{x}\nonumber\\
  &\cdot\mathrm{TMS}(-\alpha,\pi)\,e^{-i\chi t_\mathrm{S}a^\dagger a\sigma_\mathrm{z}}\,
    \mathrm{TMS}^\dagger(-\alpha,\pi)\nonumber\\
  &\approx e^{-i\chi t_\mathrm{S}\sinh(2\alpha)(a^\dagger b^\dagger+ab)\sigma_\mathrm{z}}
   = e^{-i\chi t_\mathrm{S}\sinh(2\alpha)(\hat{x}_1\hat{x}_2-\hat{p}_1\hat{p}_2)\sigma_\mathrm{z}}.
  \label{eq:Echoed-TMS}
\end{align}
Rotating the qubit Bloch sphere with $\mathrm{R}_\mathrm{x}(\pi/2)$ converts
this to $\mathrm{S}$, with gate time
$t_\mathrm{S}=\Delta^2/(\chi\sinh 2\alpha)$.  Since $\sinh 2\alpha$ is
unbounded, arbitrarily fast gates are achievable at large $\alpha$ even in
the weak dispersive regime.  The two-mode squeezed frame at large $\alpha$
can be reached via single-mode squeezing and beam splitter operations using
the Bloch--Messiah decomposition~\cite{ISA}.

\paragraph{Gate B: echoed conditional beam splitter.}
An analogous echo construction using $\mathrm{BS}(\alpha,\pi)$ gives
\begin{align}
  &\mathrm{BS}(\alpha,\pi)\,e^{-i\chi t_\mathrm{B}a^\dagger a\sigma_\mathrm{z}}\,
    \mathrm{BS}^\dagger(\alpha,\pi)\cdot\sigma_\mathrm{x}\nonumber\\
  &\cdot\mathrm{BS}(-\alpha,\pi)\,e^{-i\chi t_\mathrm{B}a^\dagger a\sigma_\mathrm{z}}\,
    \mathrm{BS}^\dagger(-\alpha,\pi)\nonumber\\
  &\approx e^{-i\chi t_\mathrm{B}\sin\alpha(\hat{x}_1\hat{p}_2
             +\hat{p}_1\hat{x}_2)\sigma_\mathrm{z}}.
  \label{eq:Echoed-BS}
\end{align}
To extract $\mathrm{B}$ with the $\Delta^4$ correction, we apply
single-mode squeezing $\mathrm{S}_a(r)$ with $r=\frac{1}{2}\ln\Delta^{-4}$
to mode 1, rescaling $\hat{x}_1\to e^r\hat{x}_1$ and
$\hat{p}_1\to e^{-r}\hat{p}_1$.  This gives gate time
\begin{align}
  t_\mathrm{B} = \frac{2\Delta^2}{\chi\sin\alpha},
\end{align}
which decreases with $\Delta$, favoring the low-$\Delta$ regime relevant
for GKP states~\cite{eickbusch2022fast}.

\paragraph{Gate times.}
Although $t_\mathrm{S}$ and $t_\mathrm{B}$ enter asymmetrically in 
$\mathrm{CX}_\mathrm{GKP}$ and $\mathrm{CZ}_\mathrm{GKP}$, the total 
circuit duration $t_\mathrm{CX/CZ}=2t_\mathrm{S}+t_\mathrm{B}$ is the 
same for both gates since the roles of $t_\mathrm{S}$ and $t_\mathrm{B}$ 
are simply exchanged:
\begin{align}
  t_\mathrm{S} = \frac{\Delta^2}{\chi\sinh 2\alpha}, \qquad
  t_\mathrm{B} \ge \frac{2\Delta^2}{\chi}
  \quad (\mathrm{CX}_\mathrm{GKP}),
\end{align}
and symmetrically with $t_\mathrm{S}\leftrightarrow t_\mathrm{B}$ for
$\mathrm{CZ}_\mathrm{GKP}$. Both gates benefit from the $\Delta^2$ 
suppression of gate time, making this construction particularly 
advantageous in the low-$\Delta$ regime relevant for GKP states. This 
construction is the two-mode extension of the echoed conditional 
displacement introduced in Ref.~\cite{ISA}.

\section{Rotationally Symmetric States}
\subsection{Four-Legged Cat State Preparation}\label{app:4-legged-cat}

We derive the unentanglement conditions for the four-legged cat state
preparation circuit of Sec.~\ref{ssec:universal}.  Starting from an even
cat state $\ket{C_{i\beta}}$ with $\beta\in\mathbb{R}$,
\begin{align}
  &\mathcal{U}(\theta,|\alpha|,1)\,
   e^{-i2\alpha\hat{p}\sigma_\mathrm{x}}\ket{C_{i\beta}}\ket{g}
   \nonumber\\
  &= \mathcal{U}(\theta,|\alpha|,1)
   \bigl[e^{-i2\alpha\hat{p}}\ket{C_{i\beta}}\ket{+}
        +e^{+i2\alpha\hat{p}}\ket{C_{i\beta}}\ket{-}\bigr],
\end{align}
where $\mathcal{U}\equiv\mathrm{GCR}$ or $\mathrm{BB1}$ and $\theta=\pi/2$.

\paragraph{GCR.}
The GCR unentanglement gadget acts as
\begin{align}
  &\mathcal{U}(\theta,|\alpha|,1)\,
   e^{-i2\alpha\hat{p}\sigma_\mathrm{x}}\ket{C_{i\beta}}\ket{g}
   \nonumber\\
  &\approx\bigl[e^{i\alpha\frac{\pi}{4\beta}\sigma_\mathrm{x}}
   e^{-i2\alpha\hat{p}}\ket{C_{i\beta}}
   +e^{-i\alpha\frac{\pi}{4\beta}\sigma_\mathrm{x}}
   e^{i2\alpha\hat{p}}\ket{C_{i\beta}}\bigr]\ket{g}.
\end{align}
Unentanglement succeeds when $\pi\alpha/4\beta=m\pi/2$, giving
$\alpha/\beta=2m$ for $m\in\mathbb{Z}$.  The resulting state is
\begin{equation}
  \ket{4\mathcal{C}}\propto
  (\ket{2\beta+i\beta}-\ket{2\beta-i\beta})
  -(\ket{-2\beta+i\beta}+\ket{-2\beta-i\beta}),
\end{equation}
a rectangular cat with even aspect ratio.  Squeezing this state to fix
the aspect ratio would distort the individual blobs and is not recommended.

\paragraph{BB1.}
Since all operations in $\mathcal{U}\equiv\mathrm{BB1}$ are controlled
momentum boosts, the unentanglement proceeds as in the two-legged cat case,
up to a local phase on two of the four blobs:
\begin{align}
  &\mathcal{U}(\theta,|\alpha|,1)\,
   e^{-i2\alpha\hat{p}\sigma_\mathrm{x}}\ket{C_{i\beta}}\ket{g}
   \nonumber\\
  &\approx\bigl[\ket{\alpha+i\beta}
   +e^{i\alpha\beta}\ket{-\alpha+i\beta}
   +e^{-i\alpha\beta}\ket{\alpha-i\beta}
   +\ket{-\alpha-i\beta}\bigr]\ket{g},
\end{align}
with the phase vanishing when $\alpha^2=2\pi$.  BB1 therefore prepares
square four-legged cats without geometric restriction.
\subsection{Fock States}\label{app:fock-states}

Fock states can be approximated as rotationally symmetric superpositions of
coherent states,
\begin{align}
  \ket{\psi_n} &= \frac{1}{\mathcal{N}}\sum_{j=0}^{m-1}
    e^{i\frac{2\pi n}{m}j}\ket{\alpha e^{-i\frac{2\pi j}{m}}},
    \label{eq:fock}\\
  F_n &= |\braket{n|\psi_n}|^2
       = m^2\frac{\alpha^{2n}}{n!}\frac{e^{-\alpha^2}}{\mathcal{N}^2},
    \quad\alpha\in\mathbb{R},
    \label{eq:prob_fock}
\end{align}
where fidelity improves with increasing $m$ beyond the minimum $m_n$
required for each $\ket{n}$.  Preparation can be achieved via a
Trotterized circuit for the anti-Jaynes--Cummings (AJC) Hamiltonian,
\begin{align}
  \mathrm{AJC} = \hat{a}\sigma_- + \hat{a}^\dagger\sigma_+
               = 2(\hat{x}\sigma_\mathrm{x}-\hat{p}\sigma_\mathrm{y}),
\end{align}
whose right-hand side is approximated by a conditional displacement and a
conditional momentum boost.  Below we describe the Law--Eberly protocol
and derive analytical preparation circuits for $N=\{1,2,3\}$ conditional
displacements, matching the fidelities of the numerical
scheme~\cite{eickbusch2022fast}.

\subsubsection{Law--Eberly protocol}\label{app:law-eberly}

A conditional displacement $e^{i(\gamma\hat{x}+\beta\hat{p})\otimes\sigma_\phi}$
with $r^2=\gamma^2+\beta^2$ decomposes into JC and AJC interactions as
\begin{align}
  \mathrm{CD} &= \exp\!\Bigl[i\tfrac{r}{2}\bigl(
    \underbrace{e^{i(\theta+\phi)}\hat{a}\sigma_- +
                e^{-i(\theta+\phi)}\hat{a}^\dagger\sigma_+}_\text{AJC}
   \nonumber\\&\quad\quad\quad+\underbrace{e^{i(\theta-\phi)}\hat{a}\sigma_+ +
                e^{-i(\theta-\phi)}\hat{a}^\dagger\sigma_-}_\text{JC}
    \bigr)\Bigr].
\end{align}
Acting on $\ket{n}\ket{g}$ with the AJC evolution alone gives
\begin{align}
  e^{i\alpha\,\mathrm{AJC}}\ket{n}\ket{g}
  &= \cos\!\bigl(2\alpha\sqrt{n+1}\bigr)\ket{n}\ket{g}
    \nonumber\\&\quad\quad+e^{-i(\theta+\phi)}\sin\!\bigl(2\alpha\sqrt{n+1}\bigr)
     \ket{n+1}\ket{e},
  \label{eq:law-eberly}
\end{align}
which transfers a single photon to the oscillator at
$\alpha=\pi/(4\sqrt{n+1})$.  Since $\mathrm{CD}$ also contains the JC
term, we alternate between $\theta-\phi=-\pi/2$ and $\theta-\phi=+\pi/2$
to cancel it, realizing the Trotterization
$(e^{i\alpha\hat{x}\sigma_\mathrm{y}/N}e^{i\alpha\hat{p}\sigma_\mathrm{x}/N})^N$.
Given pure AJC or JC evolution, the Law--Eberly
protocol~\cite{law1996arbitrary} achieves universal Fock state preparation
from vacuum via sequential single-photon addition.  The Trotterization-based
scheme has been shown to improve upon the numerical scheme~\cite{eickbusch2022fast}
in circuit duration, though the numerical scheme remains optimal in circuit
depth~\cite{ISA}.  Our analytical scheme in Sec.~\ref{ssec:universal} is
the only one matching the numerical circuit depth at comparable fidelity.

\subsubsection{Optimal protocols for $N=\{1,2,3\}$}\label{app:fock_circuit}

We focus on $\ket{n=1}$, noting that Eq.~(\ref{eq:fock}) represents it as
a small odd cat state.  As discussed in App.~\ref{app:small_cats}, the
probability of projecting onto a small odd cat is lower than for an even
cat, so there is an optimal $\alpha$ balancing fidelity with $\ket{1}$
against success probability.

\paragraph{$N=1$.}
Starting from $\ket{0,g}$,
\begin{align}
  \ket{0,g}
  \xrightarrow{\mathrm{CD}(\alpha,\sigma_\mathrm{y})}
  i\mathcal{N}_\mathrm{odd}(\ket{\alpha}-\ket{-\alpha})\ket{e}
  +\mathcal{N}_\mathrm{even}(\ket{\alpha}+\ket{-\alpha})\ket{g}.
\end{align}
The optimal $\alpha=\pi/4$ maximizes the product of the fidelity with
$\ket{1}$ (Eq.~(\ref{eq:prob_fock})) and the relative success probability
$|\mathcal{N}_\mathrm{odd}|^2/|\mathcal{N}_\mathrm{even}|^2$.  This
coincides with the JC parameter for $\ket{0,g}\to\ket{1,e}$ achieved with
a single conditional displacement.

\paragraph{$N=2$.}
The first-order Trotterized AJC circuit gives
$\ket{0,g}\to\ket{1,e}$ at $\alpha=\pi/4$, with hybrid fidelity $0.70$
— slightly below the numerically optimized result.  A better strategy
appends $e^{i\frac{\pi}{4\alpha}\hat{x}\sigma_\mathrm{x}}$ instead, which
is the unentanglement sequence for large cat states without QSP correction
(Sec.~\ref{ssec:Cat_States}, App.~\ref{app:cat_I}).  Since $|\alpha|<2$,
QSP correction does not apply (Fig.~\ref{fig:Correctness}), but the
momentum-basis back action is here favorable: the enhanced non-commutativity
of displacement and momentum boost at small amplitudes creates a state whose
position and momentum peaks nearly coincide, yielding a near-circular Wigner
distribution close to $\ket{1}$.  The resulting hybrid fidelity is $0.84$,
matching the numerical scheme (Fig.~\ref{fig:fock1}).

\paragraph{$N=3$.}
The $N=2$ circuit has effectively transferred the unentanglement problem 
from the position to the momentum quadrature, creating a large cat of size 
$\beta_1$ along $\hat{p}$. The third gate must therefore unentangle the 
qubit in the momentum basis, requiring 
$\langle\sigma_\mathrm{x}\rangle_\mathrm{new}=0$, where
\begin{align}
\langle\sigma_\mathrm{x}\rangle_\mathrm{new}
  = \langle\sigma_\mathrm{x}\rangle_\mathrm{old}\cos(4\alpha_2 p)
   -\langle\sigma_\mathrm{z}\rangle_\mathrm{old}\sin(4\alpha_2 p),
  \label{eq:sx_new}
\end{align}
giving $\tan(4\alpha_2 p)=\langle\sigma_\mathrm{x}\rangle_\mathrm{old}/
\langle\sigma_\mathrm{z}\rangle_\mathrm{old}=\sinh(4\beta_1 p)$.
Imposing this condition at the momentum wavefunction peak $p=\alpha_1$
(Fig.~\ref{fig:fock1}),
\begin{align}
  \alpha_2 = \frac{1}{4\alpha_1}\tan^{-1}\!\bigl(\sinh(4\beta_1\alpha_1)\bigr)
           = \frac{1}{\pi}\tan^{-1}\!\bigl(\sinh(\pi/2)\bigr).
\end{align}
This $N=3$ sequence has higher operator fidelity to the AJC Hamiltonian
evolution than the symmetric second-order Trotterized circuit in
Eq.~(\ref{eq:CD_circuit}).
\section{Constructing the Hybrid Unitary for Phase Estimation}
\label{app:phase-est}
We derive the explicit form of $C_xU$ in Eq.~(\ref{eq:CxU}) for
$U=e^{i\theta\sigma_\mathrm{y}}$. We work throughout in the convention
$[\hat{x},\hat{p}]=i$ adopted in Sec.~\ref{sssec:quantum-phase-estimation}, and view the four-pulse
synthesis as constructing -- from unitary access to $U$ together with
the phase-space instruction set -- the hybrid Hamiltonian-imprinting
primitive that Liu et al.~\cite{liu2016power} take as given. The
derivation proceeds in four steps. We first reduce the four-factor
product $C_xU$ to a single qubit rotation whose axis and angle depend
on $\hat{x}$ and $\theta$. We then identify, in the small-$\alpha$
limit, the part of this rotation that produces a measurable momentum
displacement of the oscillator. We then go beyond the small-$\alpha$
approximation and obtain a closed-form expression for $\braket{\hat{p}}$
that is exact for any $\alpha$ when the oscillator is initialised in
the squeezed vacuum. Finally, we compute the homodyne variance and
use the Cramér-Rao bound to characterize the precision of the
resulting phase estimate.

\paragraph{Reducing $C_xU$ to a single rotation.}
Each two-qubit rotation can be represented as a quaternion with basis
elements $\{I,\sigma_\mathrm{x},\sigma_\mathrm{y},\sigma_\mathrm{z}\}$,
and the product formula for quaternions (equivalently, the standard
composition rule for qubit rotations) gives the rotation axis and
angle of the product of two rotations directly. Using
$\sqrt{X}^{\dagger}\,e^{i\theta\sigma_\mathrm{y}}\,\sqrt{X}
=e^{i\theta\sigma_\mathrm{z}}$, we first simplify Eq.~(\ref{eq:CxU})
as
\begin{align}
  C_xU
  &= e^{i\alpha\hat{x}\otimes\sigma_\mathrm{x}}\,
     e^{i\theta\sigma_\mathrm{z}}\,
     e^{-i\alpha\hat{x}\otimes\sigma_\mathrm{x}}\,
     e^{-i\theta\sigma_\mathrm{z}}.
\end{align}
Setting $\hat{n}\cdot\vec{\sigma}=\sigma_\mathrm{x}$ and
$\hat{m}\cdot\vec{\sigma}=\sigma_\mathrm{z}$, so that
$\hat{n}\cdot\hat{m}=0$ and
$(\hat{n}\times\hat{m})\cdot\vec{\sigma}=-\sigma_\mathrm{y}$, the
composition rule
$(\hat{n}\cdot\vec{\sigma})(\hat{m}\cdot\vec{\sigma})
=(\hat{n}\cdot\hat{m})I+i(\hat{n}\times\hat{m})\cdot\vec{\sigma}$
gives the intermediate rotations
\begin{align}
  \cos\gamma &= \cos(\alpha\hat{x})\cos\theta,\\
  \sin\gamma\,\hat{k}\cdot\vec{\sigma}
    &= \sin(\alpha\hat{x})\cos\theta\,\sigma_\mathrm{x} \nonumber\\
    &\quad +\cos(\alpha\hat{x})\sin\theta\,\sigma_\mathrm{z}
       +\sin(\alpha\hat{x})\sin\theta\,\sigma_\mathrm{y},\\
  \sin\gamma\,\hat{k}^\prime\cdot\vec{\sigma}
    &= -\sin(\alpha\hat{x})\cos\theta\,\sigma_\mathrm{x} \nonumber\\
    &\quad -\cos(\alpha\hat{x})\sin\theta\,\sigma_\mathrm{z}
       +\sin(\alpha\hat{x})\sin\theta\,\sigma_\mathrm{y},
\end{align}
with $\gamma^\prime=\gamma$. Composing
$e^{i\gamma\hat{k}\cdot\vec{\sigma}}$ and
$e^{i\gamma\hat{k}^\prime\cdot\vec{\sigma}}$ via the same rule yields
$C_xU=e^{i\hat{g}\,\hat{n}^\prime\cdot\vec{\sigma}}$ with
\begin{align}
  \cos\hat{g}
    &= 1 - 2\sin^2(\alpha\hat{x})\sin^2\theta,
    \label{eq:start_phase_est}\\
  \sin\hat{g}\,\hat{n}^\prime\cdot\vec{\sigma}
    &= \sin(2\alpha\hat{x})\sin^2\theta\,\sigma_\mathrm{x} \nonumber\\
    &\quad +\tfrac{1}{2}\sin(2\alpha\hat{x})\sin(2\theta)\,\sigma_\mathrm{y} \nonumber\\
    &\quad -\sin^2(\alpha\hat{x})\sin(2\theta)\,\sigma_\mathrm{z}.
       \label{eq:end_phase_est}
\end{align}

\paragraph{Leading-order momentum displacement.}
In the limit $\alpha\hat{x}\ll 1$, the rotation generator
$\hat{g}\,\hat{n}^\prime\cdot\vec{\sigma}$ reduces at leading order to
\begin{align}
  \hat{g}\,\hat{n}_\mathrm{x}^\prime &= 2\alpha\hat{x}\sin^2\theta, \\
  \hat{g}\,\hat{n}_\mathrm{y}^\prime &= \alpha\hat{x}\sin(2\theta), \\
  \hat{g}\,\hat{n}_\mathrm{z}^\prime &= 0,
\end{align}
where the $\sigma_\mathrm{z}$ component is suppressed at
$O((\alpha\hat{x})^2)$ and so does not contribute to leading order.
When the qubit is initialised in a $\sigma_\mathrm{y}$ eigenstate,
only the $\sigma_\mathrm{y}$ component of the generator acts as a
c-number on the qubit; it then simply imprints an $\hat{x}$-dependent
phase $e^{i\alpha\hat{x}\sin(2\theta)}$ on the oscillator -- precisely
a momentum-displacement operator. The $\sigma_\mathrm{x}$ piece, by
contrast, anticommutes with $\sigma_\mathrm{y}$ and flips the qubit
out of $\ket{+y}$; it entangles the qubit with the oscillator in a
manner that, after tracing over the qubit, produces no net mean
momentum displacement at leading order. Accordingly, at leading order
in $\alpha\hat{x}$ and with the qubit prepared and measured in the
$\sigma_\mathrm{y}$ basis, the mean oscillator momentum after $C_xU$
is
\begin{align}
  \braket{\hat{p}} \approx \alpha\sin(2\theta),
  \label{eq:p_leading}
\end{align}
from which $\theta$ can be extracted directly.

\paragraph{Exact result for vacuum input.}
We can derive a closed-form expression for $\braket{\hat{p}}$ that is
exact for any $\alpha$ when the oscillator is in the squeezed vacuum.
The action of $C_xU$ on $\ket{+y}$ at fixed oscillator coordinate
$x$ is
$C_xU\ket{x,+y}=\ket{x}\,[\,\alpha_+(x)\ket{+y}+\alpha_-(x)\ket{-y}\,]$,
where the qubit amplitudes
\begin{align}
  \alpha_+(x) &= \cos\hat{g}(x) + i\,\hat{g}\,\hat{n}_\mathrm{y}^\prime(x), \nonumber\\
  \alpha_-(x) &= -\hat{g}\,\hat{n}_\mathrm{x}^\prime(x)
                 + i\,\hat{g}\,\hat{n}_\mathrm{z}^\prime(x),
\end{align}
follow from acting Eqs.~(\ref{eq:start_phase_est})--(\ref{eq:end_phase_est})
on $\ket{+y}$ using $\sigma_\mathrm{x}\ket{+y}=i\ket{-y}$,
$\sigma_\mathrm{y}\ket{+y}=\ket{+y}$, and
$\sigma_\mathrm{z}\ket{+y}=\ket{-y}$. The mean momentum after the
protocol is then
\begin{align}
  \braket{\hat{p}} = \int dx\,|\phi_0(x)|^2\,
    \mathrm{Im}\!\left[\alpha_+^*\,\partial_x\alpha_+
                       +\alpha_-^*\,\partial_x\alpha_-\right],
\end{align}
where $\phi_0(x)$ is the squeezed-vacuum wavefunction with squeezing
parameter $r$ in the momentum quadrature. The integrand simplifies to
\begin{align}
  \mathrm{Im}[\cdots] = \alpha\sin(2\theta)\bigl[
     \cos(2\theta)\cos(2\alpha x) + 2\sin^2\theta\bigr],
\end{align}
and the Gaussian integral
$\int dx\,|\phi_0(x)|^2\cos(2\alpha x) = e^{-\alpha^2 e^{2r}}$ then
gives
\begin{align}
  \boxed{\;\braket{\hat{p}} = \alpha\sin(2\theta)
    - \frac{\alpha}{2}\bigl(1-e^{-\alpha^2 e^{2r}}\bigr)\sin(4\theta)\;}
  \label{eq:p_exact}
\end{align}
exactly, for squeezed-vacuum input and any $\alpha$.
Equation~(\ref{eq:p_exact}) reduces to the leading-order result
Eq.~(\ref{eq:p_leading}) at small $\alpha$ -- where
$1-e^{-\alpha^2 e^{2r}}=\alpha^2 e^{2r}+\cdots$ makes the correction
$O(\alpha^3)$ -- and remains valid arbitrarily far from that limit.

Two features of Eq.~(\ref{eq:p_exact}) are worth highlighting. First,
only two harmonics of $\theta$ appear: the desired $\sin(2\theta)$
signal and a $\sin(4\theta)$ correction. Second, the correction
vanishes at $\theta=\pi/4$ because $\sin(4\theta)=0$ there: the
protocol returns $\braket{\hat{p}}=\alpha$ \emph{exactly} for any
$\alpha$ when $\theta=\pi/4$, independently of squeezing.

\paragraph{Variance and precision analysis.}
The variance of the homodyne readout admits an equally clean form.
Using the same wavefunction decomposition that gave
Eq.~(\ref{eq:p_exact}), the second moment of $\hat{p}$ for
squeezed-vacuum input is
\begin{align}
  \braket{\hat{p}^2} = \frac{e^{-2r}}{2} + 4\alpha^2\sin^2\theta,
  \label{eq:p2_exact}
\end{align}
exact for all $\alpha$ and $r$. The first term is the squeezed-vacuum
$\hat{p}$-quadrature variance; the second term is an excess noise
contribution that arises from the oscillator-qubit entanglement
generated by $C_xU$. Combining Eqs.~(\ref{eq:p_exact}) and
(\ref{eq:p2_exact}) gives the homodyne variance
$\delta p^2 = \braket{\hat{p}^2}-\braket{\hat{p}}^2$, which
at leading order in $\alpha$ simplifies to ($\braket{p^2}$ converts $\sin^2\theta$ to $\sin^4\theta$)
\begin{align}
  \delta p^2 = \frac{e^{-2r}}{2}
    + 4\alpha^2\sin^4\theta + O(\alpha^4),
\end{align}
and is dominated by the input squeezed-vacuum variance $e^{-2r}/2$ in
the small-$\alpha$ regime.

In this small-$\alpha$ regime, the homodyne distribution $P(p)$ is
Gaussian; deviations from a Gaussian fit have Kullback-Leibler
divergence $\lesssim 10^{-4}$ at $\alpha\lesssim 0.1$ (numerically
verified). At larger $\alpha$, the oscillator-qubit entanglement
generates non-Gaussian and even bimodal distributions whose precision
analysis would require maximum-likelihood estimation from the full
$P(p)$ shape; this regime represents an interesting target for future
analysis of non-Gaussian phase-estimation protocols, and we restrict
our analysis here to small $\alpha$ where the distribution is
Gaussian.

The Cramér-Rao bound on estimating $\theta$ from $N$ homodyne shots is
\begin{align}
  \sigma_\theta \;\ge\; \frac{\delta p}
                              {\sqrt{N}\,|\partial g/\partial\theta|}
       \;\xrightarrow{\,\alpha\to 0\,}\;
       \frac{e^{-r}}{2\sqrt{2N}\,\alpha\,|\cos(2\theta)|},
  \label{eq:CRB_small_alpha}
\end{align}
where $g(\theta;\alpha,r)$ is the right-hand side of
Eq.~(\ref{eq:p_exact}) and
$\partial g/\partial\theta=2\alpha\cos(2\theta)+O(\alpha^3)$.
Equivalently, reaching a target precision
$\sigma_\theta=\varepsilon$ requires
\begin{align}
  N \;\sim\; \frac{e^{-2r}}{8\,\alpha^2\,\varepsilon^2\cos^2(2\theta)}
  \label{eq:shots_small_alpha}
\end{align}
homodyne measurements, each consuming two queries to $U$ (one $U$ and
one $\sqrt{X}\,U\,\sqrt{X}^\dagger$). Note that the leading-order
slope $\partial g/\partial\theta=2\alpha\cos(2\theta)$ vanishes at
$\theta=\pi/4$, so although the mean response Eq.~(\ref{eq:p_exact})
is exact at this point, the estimation precision is correspondingly
poor; the optimal operating point is $\theta\approx 0$ or $\pi/2$
where $|\cos(2\theta)|=1$. Equation~(\ref{eq:shots_small_alpha}) sets
the per-shot precision underlying the main-text scaling
Eq.~(\ref{eq:Q_scaling}); the resulting comparison to Liu et al.\ and
to the qubit-only baseline is given alongside Fig.~\ref{fig:phase_est}
in the main text.
\bibliographystyle{elsarticle-num} 
\bibliography{CV_QSP_cited.bib}

@misc{fong2025engineeringnongaussianbosonicgates,
      title={Engineering Non-Gaussian Bosonic Gates through Quantum Signal Processing}, 
      author={Pak-Tik Fong and Hoi-Kwan Lau},
      year={2025},
      eprint={2508.20261},
      archivePrefix={arXiv},
      primaryClass={quant-ph},
      url={https://arxiv.org/abs/2508.20261}, 
}

@article{ChuangRobust,
	title        = {Arbitrarily accurate composite pulse sequences},
	author       = {Brown, Kenneth R. and Harrow, Aram W. and Chuang, Isaac L.},
	year         = 2004,
	month        = {Nov},
	journal      = {Phys. Rev. A},
	publisher    = {American Physical Society},
	volume       = 70,
	pages        = {052318},
	doi          = {10.1103/PhysRevA.70.052318},
	url          = {https://link.aps.org/doi/10.1103/PhysRevA.70.052318},
	issue        = 5,
	numpages     = 4
}

@article{ChuangRobustErratum,
	title        = {Erratum: Arbitrarily accurate composite pulse sequences [Phys. Rev. A 70, 052318 (2004)]},
	author       = {Brown, Kenneth R. and Harrow, Aram W. and Chuang, Isaac L.},
	year         = 2005,
	month        = {Sep},
	journal      = {Phys. Rev. A},
	publisher    = {American Physical Society},
	volume       = 72,
	pages        = {039905(E)},
	doi          = {10.1103/PhysRevA.72.039905},
	url          = {https://link.aps.org/doi/10.1103/PhysRevA.72.039905},
	issue        = 3,
	numpages     = 1
}

@article{WhaleyRobust,
	title        = {Efficient phase-factor evaluation in quantum signal processing},
	author       = {Dong, Yulong and Meng, Xiang and Whaley, K. Birgitta and Lin, Lin},
	year         = 2021,
	month        = {Apr},
	journal      = {Phys. Rev. A},
	publisher    = {American Physical Society},
	volume       = 103,
	pages        = {042419},
	doi          = {10.1103/PhysRevA.103.042419},
	url          = {https://link.aps.org/doi/10.1103/PhysRevA.103.042419},
	issue        = 4,
	numpages     = 22
}

@article{M-S-gate1,
  title = {Multiparticle Entanglement of Hot Trapped Ions},
  author = {M\o{}lmer, Klaus and S\o{}rensen, Anders},
  journal = {Phys. Rev. Lett.},
  volume = {82},
  issue = {9},
  pages = {1835--1838},
  numpages = {0},
  year = {1999},
  month = {Mar},
  publisher = {American Physical Society},
  doi = {10.1103/PhysRevLett.82.1835},
  url = {https://link.aps.org/doi/10.1103/PhysRevLett.82.1835}
}

@article{M-S-gate2,
  title = {Quantum Computation with Ions in Thermal Motion},
  author = {S\o{}rensen, Anders and M\o{}lmer, Klaus},
  journal = {Phys. Rev. Lett.},
  volume = {82},
  issue = {9},
  pages = {1971--1974},
  numpages = {0},
  year = {1999},
  month = {Mar},
  publisher = {American Physical Society},
  doi = {10.1103/PhysRevLett.82.1971},
  url = {https://link.aps.org/doi/10.1103/PhysRevLett.82.1971}
}

@article{PhysRevA.97.022341,
  title = {Generating grid states from Schr\"odinger-cat states without postselection},
  author = {Weigand, Daniel J. and Terhal, Barbara M.},
  journal = {Phys. Rev. A},
  volume = {97},
  issue = {2},
  pages = {022341},
  numpages = {13},
  year = {2018},
  month = {Feb},
  publisher = {American Physical Society},
  doi = {10.1103/PhysRevA.97.022341},
  url = {https://link.aps.org/doi/10.1103/PhysRevA.97.022341}
}

@article{LukeBellQSVT,
  title = {Co-designing Spectral Transformation Oracles with Hybrid Oscillator-Qubit Quantum Processors: From Algorithms to Compilation},
  author = {Bell, Luke and Wang, Yan and Smith, Kevin C. and Liu, Yuan and Dumitrescu, Eugene and Girvin, S.M.},
  journal = {PRX Quantum},
  volume = {6},
  issue = {4},
  pages = {040359},
  numpages = {27},
  year = {2025},
  month = {Dec},
  publisher = {American Physical Society},
  doi = {10.1103/1496-tlmm},
  url = {https://link.aps.org/doi/10.1103/1496-tlmm}
}

@misc{brenner2024factoring,
	title = {Factoring an integer with three oscillators and a qubit},
	url = {http://arxiv.org/abs/2412.13164},
	doi = {},
	urldate = {2025-03-25},
	publisher = {arXiv},
	author = {Brenner, Lukas and Caha, Libor and Coiteux-Roy, Xavier and Koenig, Robert},
	month = dec,
	year = {2024},
	note = {arXiv:2412.13164 [quant-ph]},
}

@article{leghtas_hardware-efficient_2013,
	title = {Hardware-{Efficient} {Autonomous} {Quantum} {Memory} {Protection}},
	volume = {111},
	url = {https://link.aps.org/doi/10.1103/PhysRevLett.111.120501},
	doi = {},
	number = {12},
	urldate = {2025-04-28},
	journal = {Physical Review Letters},
	author = {Leghtas, Zaki and Kirchmair, Gerhard and Vlastakis, Brian and Schoelkopf, Robert J. and Devoret, Michel H. and Mirrahimi, Mazyar},
	month = sep,
	year = {2013},
	note = {Publisher: American Physical Society},
	pages = {120501},
}

@article{bruzewicz2019trapped,
	title = {Trapped-ion quantum computing: {Progress} and challenges},
	volume = {6},
	issn = {1931-9401},
	shorttitle = {Trapped-ion quantum computing},
	url = {https://doi.org/10.1063/1.5088164},
	doi = {},
	number = {2},
	urldate = {2025-03-25},
	journal = {Applied Physics Reviews},
	author = {Bruzewicz, Colin D. and Chiaverini, John and McConnell, Robert and Sage, Jeremy M.},
	month = may,
	year = {2019},
	pages = {021314},
}

@article{ReinholdErrorCorrectedGates,
	title = {Error-corrected gates on an encoded qubit},
	volume = {16},
	copyright = {2020 The Author(s), under exclusive licence to Springer Nature Limited},
	issn = {1745-2481},
	url = {https://www.nature.com/articles/s41567-020-0931-8},
	doi = {},
	language = {en},
	number = {8},
	urldate = {2025-03-25},
	journal = {Nature Physics},
	author = {Reinhold, Philip and Rosenblum, Serge and Ma, Wen-Long and Frunzio, Luigi and Jiang, Liang and Schoelkopf, Robert J.},
	month = aug,
	year = {2020},
	note = {Publisher: Nature Publishing Group},
	pages = {822--826},
}

@article{braunstein2005quantum,
	title = {Quantum information with continuous variables},
	volume = {77},
	url = {https://link.aps.org/doi/10.1103/RevModPhys.77.513},
	doi = {},
	number = {2},
	urldate = {2025-03-25},
	journal = {Reviews of Modern Physics},
	author = {Braunstein, Samuel L. and van Loock, Peter},
	month = jun,
	year = {2005},
	note = {Publisher: American Physical Society},
	pages = {513--577},
}

@article{liu2016power,
	title = {Power of one qumode for quantum computation},
	volume = {93},
	url = {https://link.aps.org/doi/10.1103/PhysRevA.93.052304},
	doi = {},
	number = {5},
	urldate = {2025-03-25},
	journal = {Physical Review A},
	author = {Liu, Nana and Thompson, Jayne and Weedbrook, Christian and Lloyd, Seth and Vedral, Vlatko and Gu, Mile and Modi, Kavan},
	month = may,
	year = {2016},
	note = {Publisher: American Physical Society},
	pages = {052304},
}

@article{ISA,
	 title = {Hybrid Oscillator-Qubit Quantum Processors: Instruction Set Architectures, Abstract Machine Models, and Applications},
  author = {Liu, Yuan and Singh, Shraddha and Smith, Kevin C. and Crane, Eleanor and Martyn, John M. and Eickbusch, Alec and Schuckert, Alexander and Li, Richard D. and Sinanan-Singh, Jasmine and Soley, Micheline B. and Tsunoda, Takahiro and Chuang, Isaac L. and Wiebe, Nathan and Girvin, Steven M.},
  journal = {PRX Quantum},
  volume = {7},
  issue = {1},
  pages = {010201},
  numpages = {166},
  year = {2026},
  month = {Jan},
  publisher = {American Physical Society},
  doi = {10.1103/4rf7-9tfx},
  url = {https://link.aps.org/doi/10.1103/4rf7-9tfx}
}

@article{xu2024fault,
	title = {Fault-{Tolerant} {Operation} of {Bosonic} {Qubits} with {Discrete}-{Variable} {Ancillae}},
	volume = {14},
	url = {https://link.aps.org/doi/10.1103/PhysRevX.14.031016},
	doi = {},
	number = {3},
	urldate = {2025-03-25},
	journal = {Physical Review X},
	author = {Xu, Qian and Zeng, Pei and Xu, Daohong and Jiang, Liang},
	month = jul,
	year = {2024},
	note = {Publisher: American Physical Society},
	pages = {031016},
}

@article{low2016methodology,
	title = {The methodology of resonant equiangular composite quantum gates},
	volume = {6},
	issn = {2160-3308},
	url = {http://arxiv.org/abs/1603.03996},
	doi = {},
	number = {4},
	urldate = {2025-03-25},
	journal = {Physical Review X},
	author = {Low, Guang Hao and Yoder, Theodore J. and Chuang, Isaac L.},
	month = dec,
	year = {2016},
	note = {arXiv:1603.03996 [quant-ph]},
	pages = {041067},
}

@phdthesis{low2017quantum,
	type = {Thesis},
	title = {Quantum signal processing by single-qubit dynamics},
	copyright = {MIT theses are protected by copyright. They may be viewed, downloaded, or printed from this source but further reproduction or distribution in any format is prohibited without written permission.},
	url = {https://dspace.mit.edu/handle/1721.1/115025},
	language = {eng},
	urldate = {2025-03-25},
	school = {Massachusetts Institute of Technology},
	author = {Low, Guang Hao},
	year = {2017},
	note = {Accepted: 2018-04-27T18:10:33Z},
}

@article{eickbusch2022fast,
	title = {Fast universal control of an oscillator with weak dispersive coupling to a qubit},
	volume = {18},
	copyright = {2022 The Author(s), under exclusive licence to Springer Nature Limited},
	issn = {1745-2481},
	url = {https://www.nature.com/articles/s41567-022-01776-9},
	doi = {},
	language = {en},
	number = {12},
	urldate = {2025-03-25},
	journal = {Nature Physics},
	author = {Eickbusch, Alec and Sivak, Volodymyr and Ding, Andy Z. and Elder, Salvatore S. and Jha, Shantanu R. and Venkatraman, Jayameenakshi and Royer, Baptiste and Girvin, S. M. and Schoelkopf, Robert J. and Devoret, Michel H.},
	month = dec,
	year = {2022},
	note = {Publisher: Nature Publishing Group},
	pages = {1464--1469},
}

@article{gottesman2001encoding,
	title = {Encoding a qubit in an oscillator},
	volume = {64},
	url = {https://link.aps.org/doi/10.1103/PhysRevA.64.012310},
	doi = {},
	number = {1},
	urldate = {2025-03-25},
	journal = {Physical Review A},
	author = {Gottesman, Daniel and Kitaev, Alexei and Preskill, John},
	month = jun,
	year = {2001},
	note = {Publisher: American Physical Society},
	pages = {012310},
}

@article{campagne2020quantum,
	title = {Quantum error correction of a qubit encoded in grid states of an oscillator},
	volume = {584},
	copyright = {2020 The Author(s), under exclusive licence to Springer Nature Limited},
	issn = {1476-4687},
	url = {https://www.nature.com/articles/s41586-020-2603-3},
	doi = {},
	language = {en},
	number = {7821},
	urldate = {2025-03-25},
	journal = {Nature},
	author = {Campagne-Ibarcq, P. and Eickbusch, A. and Touzard, S. and Zalys-Geller, E. and Frattini, N. E. and Sivak, V. V. and Reinhold, P. and Puri, S. and Shankar, S. and Schoelkopf, R. J. and Frunzio, L. and Mirrahimi, M. and Devoret, M. H.},
	month = aug,
	year = {2020},
	note = {Publisher: Nature Publishing Group},
	pages = {368--372},
}

@article{hastrup2021improved,
	title = {Improved readout of qubit-coupled {Gottesman}–{Kitaev}–{Preskill} states},
	volume = {6},
	issn = {2058-9565},
	url = {https://dx.doi.org/10.1088/2058-9565/ac070d},
	doi = {},
	language = {en},
	number = {3},
	urldate = {2025-03-25},
	journal = {Quantum Science and Technology},
	author = {Hastrup, Jacob and Andersen, Ulrik Lund},
	month = jun,
	year = {2021},
	note = {Publisher: IOP Publishing},
	pages = {035016},
}

@article{rossi2022multivariable,
	title = {Multivariable quantum signal processing ({M}-{QSP}): prophecies of the two-headed oracle},
	volume = {6},
	shorttitle = {Multivariable quantum signal processing ({M}-{QSP})},
	url = {https://quantum-journal.org/papers/q-2022-09-20-811/},
	doi = {},
	language = {en-GB},
	urldate = {2025-03-25},
	journal = {Quantum},
	author = {Rossi, Zane M. and Chuang, Isaac L.},
	month = sep,
	year = {2022},
	note = {Publisher: Verein zur Förderung des Open Access Publizierens in den Quantenwissenschaften},
	pages = {811},
}

@article{hastrup2021measurement,
	title = {Measurement-free preparation of grid states},
	volume = {7},
	copyright = {2021 The Author(s)},
	issn = {2056-6387},
	url = {https://www.nature.com/articles/s41534-020-00353-3},
	doi = {},
	language = {en},
	number = {1},
	urldate = {2025-03-25},
	journal = {npj Quantum Information},
	author = {Hastrup, Jacob and Park, Kimin and Brask, Jonatan Bohr and Filip, Radim and Andersen, Ulrik Lund},
	month = jan,
	year = {2021},
	note = {Publisher: Nature Publishing Group},
	pages = {1--8},
}

@article{rojkov2023two,
	title = {Two-{Qubit} {Operations} for {Finite}-{Energy} {Gottesman}-{Kitaev}-{Preskill} {Encodings}},
	volume = {133},
	url = {https://link.aps.org/doi/10.1103/PhysRevLett.133.100601},
	doi = {},
	number = {10},
	urldate = {2025-03-25},
	journal = {Physical Review Letters},
	author = {Rojkov, Ivan and Röggla, Paul Moser and Wagener, Martin and Fontboté-Schmidt, Moritz and Welte, Stephan and Home, Jonathan and Reiter, Florentin},
	month = sep,
	year = {2024},
	note = {Publisher: American Physical Society},
	pages = {100601},
}

@article{hastrup2021unsuitability,
	title = {Unsuitability of cubic phase gates for non-{Clifford} operations on {Gottesman}-{Kitaev}-{Preskill} states},
	volume = {103},
	url = {https://link.aps.org/doi/10.1103/PhysRevA.103.032409},
	doi = {},
	number = {3},
	urldate = {2025-03-25},
	journal = {Physical Review A},
	author = {Hastrup, Jacob and Larsen, Mikkel V. and Neergaard-Nielsen, Jonas S. and Menicucci, Nicolas C. and Andersen, Ulrik L.},
	month = mar,
	year = {2021},
	note = {Publisher: American Physical Society},
	pages = {032409},
}

@article{low2017optimal,
	title = {Optimal {Hamiltonian} {Simulation} by {Quantum} {Signal} {Processing}},
	volume = {118},
	url = {https://link.aps.org/doi/10.1103/PhysRevLett.118.010501},
	doi = {},
	number = {1},
	urldate = {2025-03-25},
	journal = {Physical Review Letters},
	author = {Low, Guang Hao and Chuang, Isaac L.},
	month = jan,
	year = {2017},
	note = {Publisher: American Physical Society},
	pages = {010501},
}

@article{martyn2021grand,
	title = {Grand {Unification} of {Quantum} {Algorithms}},
	volume = {2},
	url = {https://link.aps.org/doi/10.1103/PRXQuantum.2.040203},
	doi = {},
	number = {4},
	urldate = {2025-03-25},
	journal = {PRX Quantum},
	author = {Martyn, John M. and Rossi, Zane M. and Tan, Andrew K. and Chuang, Isaac L.},
	month = dec,
	year = {2021},
	note = {Publisher: American Physical Society},
	pages = {040203},
}

@article{motlagh2023generalized,
	title = {Generalized {Quantum} {Signal} {Processing}},
	volume = {5},
	url = {https://link.aps.org/doi/10.1103/PRXQuantum.5.020368},
	doi = {},
	number = {2},
	urldate = {2025-03-25},
	journal = {PRX Quantum},
	author = {Motlagh, Danial and Wiebe, Nathan},
	month = jun,
	year = {2024},
	note = {Publisher: American Physical Society},
	pages = {020368},
}

@article{law1996arbitrary,
	title = {Arbitrary {Control} of a {Quantum} {Electromagnetic} {Field}},
	volume = {76},
	url = {https://link.aps.org/doi/10.1103/PhysRevLett.76.1055},
	doi = {},
	number = {7},
	urldate = {2025-03-25},
	journal = {Physical Review Letters},
	author = {Law, C. K. and Eberly, J. H.},
	month = feb,
	year = {1996},
	note = {Publisher: American Physical Society},
	pages = {1055--1058},
}

@article{hastrup2023analysis,
	title = {Analysis of loss correction with the {Gottesman}-{Kitaev}-{Preskill} code},
	volume = {108},
	url = {https://link.aps.org/doi/10.1103/PhysRevA.108.052413},
	doi = {},
	number = {5},
	urldate = {2025-03-25},
	journal = {Physical Review A},
	author = {Hastrup, Jacob and Andersen, Ulrik Lund},
	month = nov,
	year = {2023},
	note = {Publisher: American Physical Society},
	pages = {052413},
}

@article{Johansson2013,
	title = {{QuTiP} 2: {A} {Python} framework for the dynamics of open quantum systems},
	volume = {184},
	issn = {0010-4655},
	shorttitle = {{QuTiP} 2},
	url = {https://www.sciencedirect.com/science/article/pii/S0010465512003955},
	doi = {},
	number = {4},
	urldate = {2025-03-25},
	journal = {Computer Physics Communications},
	author = {Johansson, J. R. and Nation, P. D. and Nori, Franco},
	month = apr,
	year = {2013},
	pages = {1234--1240},
}

@article{garcia2019single,
	title = {Single-{Shot} {Nondestructive} {Detection} of {Rydberg}-{Atom} {Ensembles} by {Transmission} {Measurement} of a {Microwave} {Cavity}},
	volume = {123},
	url = {https://link.aps.org/doi/10.1103/PhysRevLett.123.193201},
	doi = {},
	number = {19},
	urldate = {2025-03-25},
	journal = {Physical Review Letters},
	author = {Garcia, S. and Stammeier, M. and Deiglmayr, J. and Merkt, F. and Wallraff, A.},
	month = nov,
	year = {2019},
	note = {Publisher: American Physical Society},
	pages = {193201},
}

@article{fluhmann2019encoding,
		title        = {Encoding a qubit in a trapped-ion mechanical oscillator},
	author       = {Flühmann, C. and Nguyen, T. L. and Marinelli, M. and Negnevitsky, V. and Mehta, K. and Home, J. P.},
	year         = 2019,
	journal      = {Nature},
	volume       = 566,
	pages        = {513--517},
	doi          = {10.1038/s41586-019-0960-6},
	issn         = {1476-4687},
	url          = {https://doi.org/10.1038/s41586-019-0960-6},
	type         = {Journal Article}
}

@misc{de2024modular,
	title = {Modular variable laser cooling for efficient entropy extraction},
	url = {http://arxiv.org/abs/2408.16128},
	doi = {},
	urldate = {2025-03-28},
	publisher = {arXiv},
	author = {Neeve, Brennan de and Nguyen, Thanh-Long and Ferk, Alexander and Behrle, Tanja and Lancellotti, Francesco and Simoni, Matteo and Welte, Stephan and Home, Jonathan},
	month = aug,
	year = {2024},
	note = {arXiv:2408.16128 [quant-ph]},
}

@article{de2022error,
	title = {Error correction of a logical grid state qubit by dissipative pumping},
	volume = {18},
	copyright = {2022 The Author(s), under exclusive licence to Springer Nature Limited},
	issn = {1745-2481},
	url = {https://www.nature.com/articles/s41567-021-01487-7},
	doi = {},
	language = {en},
	number = {3},
	urldate = {2025-03-25},
	journal = {Nature Physics},
	author = {de Neeve, Brennan and Nguyen, Thanh-Long and Behrle, Tanja and Home, Jonathan P.},
	month = mar,
	year = {2022},
	note = {Publisher: Nature Publishing Group},
	pages = {296--300},
}

@article{hastrup2021unconditional,
	title = {Unconditional {Preparation} of {Squeezed} {Vacuum} from {Rabi} {Interactions}},
	volume = {126},
	url = {https://link.aps.org/doi/10.1103/PhysRevLett.126.153602},
	doi = {},
	number = {15},
	urldate = {2025-03-25},
	journal = {Physical Review Letters},
	author = {Hastrup, Jacob and Park, Kimin and Filip, Radim and Andersen, Ulrik Lund},
	month = apr,
	year = {2021},
	note = {Publisher: American Physical Society},
	pages = {153602},
}

@article{haroche2013nobel,
	title = {Nobel {Lecture}: {Controlling} photons in a box and exploring the quantum to classical boundary},
	volume = {85},
	shorttitle = {Nobel {Lecture}},
	url = {https://link.aps.org/doi/10.1103/RevModPhys.85.1083},
	doi = {},
	number = {3},
	urldate = {2025-03-25},
	journal = {Reviews of Modern Physics},
	author = {Haroche, Serge},
	month = jul,
	year = {2013},
	note = {Publisher: American Physical Society},
	pages = {1083--1102},
}

@article{cummins2003tackling,
	title = {Tackling systematic errors in quantum logic gates with composite rotations},
	volume = {67},
	url = {https://link.aps.org/doi/10.1103/PhysRevA.67.042308},
	doi = {},
	number = {4},
	urldate = {2025-03-25},
	journal = {Physical Review A},
	author = {Cummins, Holly K. and Llewellyn, Gavin and Jones, Jonathan A.},
	month = apr,
	year = {2003},
	note = {Publisher: American Physical Society},
	pages = {042308},
}

@article{tycko1985composite,
	title = {Composite pulses without phase distortion},
	volume = {61},
	issn = {0022-2364},
	url = {https://www.sciencedirect.com/science/article/pii/0022236485902707},
	doi = {},
	number = {1},
	urldate = {2025-03-25},
	journal = {Journal of Magnetic Resonance (1969)},
	author = {Tycko, R and Cho, H. M and Schneider, E and Pines, A},
	month = jan,
	year = {1985},
	pages = {90--101},
}

@misc{nemeth2023variants,
	title = {On variants of multivariate quantum signal processing and their characterizations},
	url = {http://arxiv.org/abs/2312.09072},
	doi = {},
	urldate = {2025-03-25},
	publisher = {arXiv},
	author = {Németh, Balázs and Kövér, Blanka and Kulcsár, Boglárka and Miklósi, Roland Botond and Gilyén, András},
	month = dec,
	year = {2023},
	note = {arXiv:2312.09072 [quant-ph]},
}

@article{wimperis1994broadband,
	title = {Broadband, {Narrowband}, and {Passband} {Composite} {Pulses} for {Use} in {Advanced} {NMR} {Experiments}},
	volume = {109},
	issn = {1064-1858},
	url = {https://www.sciencedirect.com/science/article/pii/S1064185884711594},
	doi = {},
	number = {2},
	urldate = {2025-03-25},
	journal = {Journal of Magnetic Resonance, Series A},
	author = {Wimperis, S.},
	month = aug,
	year = {1994},
	pages = {221--231},
}

@article{kumar2023quantum,
	title = {Quantum-enabled millimetre wave to optical transduction using neutral atoms},
	volume = {615},
	copyright = {2023 The Author(s), under exclusive licence to Springer Nature Limited},
	issn = {1476-4687},
	url = {https://www.nature.com/articles/s41586-023-05740-2},
	doi = {},
	language = {en},
	number = {7953},
	urldate = {2025-03-25},
	journal = {Nature},
	author = {Kumar, Aishwarya and Suleymanzade, Aziza and Stone, Mark and Taneja, Lavanya and Anferov, Alexander and Schuster, David I. and Simon, Jonathan},
	month = mar,
	year = {2023},
	note = {Publisher: Nature Publishing Group},
	pages = {614--619},
}

@article{sinanan2023single,
	title = {Single-shot {Quantum} {Signal} {Processing} {Interferometry}},
	volume = {8},
	url = {https://quantum-journal.org/papers/q-2024-07-30-1427/},
	doi = {},
	language = {en-GB},
	urldate = {2025-03-25},
	journal = {Quantum},
	author = {Sinanan-Singh, Jasmine and Mintzer, Gabriel L. and Chuang, Isaac L. and Liu, Yuan},
	month = jul,
	year = {2024},
	note = {Publisher: Verein zur Förderung des Open Access Publizierens in den Quantenwissenschaften},
	pages = {1427},
}

@article{acharya2024quantum,
	title = {Quantum error correction below the surface code threshold},
	volume = {638},
	copyright = {2024 The Author(s)},
	issn = {1476-4687},
	url = {https://www.nature.com/articles/s41586-024-08449-y},
	doi = {},
	language = {en},
	number = {8052},
	urldate = {2025-03-25},
	journal = {Nature},
	author = {{Google Quantum AI and Collaborators}},
	month = feb,
	year = {2025},
	note = {Publisher: Nature Publishing Group},
	pages = {920--926},
}

@article{brock2024quantum,
	title = {Quantum error correction of qudits beyond break-even},
	volume = {641},
	copyright = {2025 The Author(s)},
	issn = {1476-4687},
	url = {https://www.nature.com/articles/s41586-025-08899-y},
	doi = {10.1038/s41586-025-08899-y},
	language = {en},
	number = {8063},
	urldate = {2025-05-14},
	journal = {Nature},
	author = {Brock, Benjamin L. and Singh, Shraddha and Eickbusch, Alec and Sivak, Volodymyr V. and Ding, Andy Z. and Frunzio, Luigi and Girvin, Steven M. and Devoret, Michel H.},
	month = may,
	year = {2025},
	note = {Publisher: Nature Publishing Group},
	pages = {612--618},
}

@article{brady2024advances,
	title = {Advances in bosonic quantum error correction with {Gottesman}–{Kitaev}–{Preskill} {Codes}: {Theory}, engineering and applications},
	volume = {93},
	issn = {0079-6727},
	shorttitle = {Advances in bosonic quantum error correction with {Gottesman}–{Kitaev}–{Preskill} {Codes}},
	url = {https://www.sciencedirect.com/science/article/pii/S0079672723000459},
	doi = {},
	urldate = {2025-03-25},
	journal = {Progress in Quantum Electronics},
	author = {Brady, Anthony J. and Eickbusch, Alec and Singh, Shraddha and Wu, Jing and Zhuang, Quntao},
	month = jan,
	year = {2024},
	pages = {100496},
}

@article{royer2020stabilization,
	title = {Stabilization of {Finite}-{Energy} {Gottesman}-{Kitaev}-{Preskill} {States}},
	volume = {125},
	url = {https://link.aps.org/doi/10.1103/PhysRevLett.125.260509},
	doi = {},
	number = {26},
	urldate = {2025-03-25},
	journal = {Physical Review Letters},
	author = {Royer, Baptiste and Singh, Shraddha and Girvin, S. M.},
	month = dec,
	year = {2020},
	note = {Publisher: American Physical Society},
	pages = {260509},
}

@article{sivak2023real,
	title = {Real-time quantum error correction beyond break-even},
	volume = {616},
	copyright = {2023 The Author(s), under exclusive licence to Springer Nature Limited},
	issn = {1476-4687},
	url = {https://www.nature.com/articles/s41586-023-05782-6},
	doi = {},
	language = {en},
	number = {7955},
	urldate = {2025-03-25},
	journal = {Nature},
	author = {Sivak, V. V. and Eickbusch, A. and Royer, B. and Singh, S. and Tsioutsios, I. and Ganjam, S. and Miano, A. and Brock, B. L. and Ding, A. Z. and Frunzio, L. and Girvin, S. M. and Schoelkopf, R. J. and Devoret, M. H.},
	month = apr,
	year = {2023},
	note = {Publisher: Nature Publishing Group},
	pages = {50--55},
}

@article{mirrahimi2014dynamically,
	title = {Dynamically protected cat-qubits: a new paradigm for universal quantum computation},
	volume = {16},
	issn = {1367-2630},
	shorttitle = {Dynamically protected cat-qubits},
	url = {https://dx.doi.org/10.1088/1367-2630/16/4/045014},
	doi = {},
	language = {en},
	number = {4},
	urldate = {2025-03-25},
	journal = {New Journal of Physics},
	author = {Mirrahimi, Mazyar and Leghtas, Zaki and Albert, Victor V and Touzard, Steven and Schoelkopf, Robert J and Jiang, Liang and Devoret, Michel H},
	month = apr,
	year = {2014},
	note = {Publisher: IOP Publishing},
	pages = {045014},
}

@article{puri2019stabilized,
	title = {Stabilized {Cat} in a {Driven} {Nonlinear} {Cavity}: {A} {Fault}-{Tolerant} {Error} {Syndrome} {Detector}},
	volume = {9},
	shorttitle = {Stabilized {Cat} in a {Driven} {Nonlinear} {Cavity}},
	url = {https://link.aps.org/doi/10.1103/PhysRevX.9.041009},
	doi = {},
	number = {4},
	urldate = {2025-03-25},
	journal = {Physical Review X},
	author = {Puri, Shruti and Grimm, Alexander and Campagne-Ibarcq, Philippe and Eickbusch, Alec and Noh, Kyungjoo and Roberts, Gabrielle and Jiang, Liang and Mirrahimi, Mazyar and Devoret, Michel H. and Girvin, S. M.},
	month = oct,
	year = {2019},
	note = {Publisher: American Physical Society},
	pages = {041009},
}

@misc{ding2024quantum,
	title = {Quantum {Control} of an {Oscillator} with a {Kerr}-cat {Qubit}},
	url = {http://arxiv.org/abs/2407.10940},
	doi = {},
	urldate = {2025-03-25},
	publisher = {arXiv},
	author = {Ding, Andy Z. and Brock, Benjamin L. and Eickbusch, Alec and Koottandavida, Akshay and Frattini, Nicholas E. and Cortinas, Rodrigo G. and Joshi, Vidul R. and Graaf, Stijn J. de and Chapman, Benjamin J. and Ganjam, Suhas and Frunzio, Luigi and Schoelkopf, Robert J. and Devoret, Michel H.},
	month = jul,
	year = {2024},
	note = {arXiv:2407.10940 [quant-ph]},
}

@article{grimm2020stabilization,
	title = {Stabilization and operation of a {Kerr}-cat qubit},
	volume = {584},
	copyright = {2020 The Author(s), under exclusive licence to Springer Nature Limited},
	issn = {1476-4687},
	url = {https://www.nature.com/articles/s41586-020-2587-z},
	doi = {},
	language = {en},
	number = {7820},
	urldate = {2025-03-25},
	journal = {Nature},
	author = {Grimm, A. and Frattini, N. E. and Puri, S. and Mundhada, S. O. and Touzard, S. and Mirrahimi, M. and Girvin, S. M. and Shankar, S. and Devoret, M. H.},
	month = aug,
	year = {2020},
	note = {Publisher: Nature Publishing Group},
	pages = {205--209},
}

@article{hastrup2022universal,
	title = {Universal {Unitary} {Transfer} of {Continuous}-{Variable} {Quantum} {States} into a {Few} {Qubits}},
	volume = {128},
	url = {https://link.aps.org/doi/10.1103/PhysRevLett.128.110503},
	doi = {},
	number = {11},
	urldate = {2025-03-25},
	journal = {Physical Review Letters},
	author = {Hastrup, Jacob and Park, Kimin and Brask, Jonatan Bohr and Filip, Radim and Andersen, Ulrik Lund},
	month = mar,
	year = {2022},
	note = {Publisher: American Physical Society},
	pages = {110503},
}

@article{wineland2013nobel,
	title = {Nobel {Lecture}: {Superposition}, entanglement, and raising {Schr}{\textbackslash}"odinger's cat},
	volume = {85},
	shorttitle = {Nobel {Lecture}},
	url = {https://link.aps.org/doi/10.1103/RevModPhys.85.1103},
	doi = {},
	number = {3},
	urldate = {2025-03-25},
	journal = {Reviews of Modern Physics},
	author = {Wineland, David J.},
	month = jul,
	year = {2013},
	note = {Publisher: American Physical Society},
	pages = {1103--1114},
}

@article{sun2014tracking,
	title = {Tracking photon jumps with repeated quantum non-demolition parity measurements},
	volume = {511},
	copyright = {2014 Springer Nature Limited},
	issn = {1476-4687},
	url = {https://www.nature.com/articles/nature13436},
	doi = {},
	language = {en},
	number = {7510},
	urldate = {2025-03-25},
	journal = {Nature},
	author = {Sun, L. and Petrenko, A. and Leghtas, Z. and Vlastakis, B. and Kirchmair, G. and Sliwa, K. M. and Narla, A. and Hatridge, M. and Shankar, S. and Blumoff, J. and Frunzio, L. and Mirrahimi, M. and Devoret, M. H. and Schoelkopf, R. J.},
	month = jul,
	year = {2014},
	note = {Publisher: Nature Publishing Group},
	pages = {444--448},
}

@article{ma2020path,
	title = {Path-{Independent} {Quantum} {Gates} with {Noisy} {Ancilla}},
	volume = {125},
	url = {https://link.aps.org/doi/10.1103/PhysRevLett.125.110503},
	doi = {},
	number = {11},
	urldate = {2025-03-25},
	journal = {Physical Review Letters},
	author = {Ma, Wen-Long and Zhang, Mengzhen and Wong, Yat and Noh, Kyungjoo and Rosenblum, Serge and Reinhold, Philip and Schoelkopf, Robert J. and Jiang, Liang},
	month = sep,
	year = {2020},
	note = {Publisher: American Physical Society},
	pages = {110503},
}

@article{ma2022algebraic,
	title = {Algebraic structure of path-independent quantum control},
	volume = {4},
	url = {https://link.aps.org/doi/10.1103/PhysRevResearch.4.023102},
	doi = {},
	number = {2},
	urldate = {2025-03-25},
	journal = {Physical Review Research},
	author = {Ma, Wen-Long and Li, Shu-Shen and Jiang, Liang},
	month = may,
	year = {2022},
	note = {Publisher: American Physical Society},
	pages = {023102},
}

@book{paris2004quantum,
	address = {Berlin, Heidelberg},
	series = {Lecture {Notes} in {Physics}},
	title = {Quantum {State} {Estimation}},
	volume = {649},
	copyright = {http://www.springer.com/tdm},
	isbn = {978-3-540-22329-0 978-3-540-44481-7},
	url = {http://link.springer.com/10.1007/b98673},
	language = {en},
	urldate = {2025-03-25},
	publisher = {Springer},
	editor = {Paris, Matteo and Řeháček, Jaroslav},
	year = {2004},
	doi = {},
}

@article{matsos_universal_2025,
	title = {Universal quantum gate set for {Gottesman}–{Kitaev}–{Preskill} logical qubits},
	volume = {21},
	issn = {1745-2481},
	url = {https://doi.org/10.1038/s41567-025-03002-8},
	doi = {10.1038/s41567-025-03002-8},
	number = {10},
	journal = {Nature Physics},
	author = {Matsos, V. G. and Valahu, C. H. and Millican, M. J. and Navickas, T. and Kolesnikow, X. C. and Biercuk, M. J. and Tan, T. R.},
	month = oct,
	year = {2025},
	pages = {1664--1669},
}

@article{PhysRevLett.133.050602,
  title = {Robust and Deterministic Preparation of Bosonic Logical States in a Trapped Ion},
  author = {Matsos, V. G. and Valahu, C. H. and Navickas, T. and Rao, A. D. and Millican, M. J. and Kolesnikow, X. C. and Biercuk, M. J. and Tan, T. R.},
  journal = {Phys. Rev. Lett.},
  volume = {133},
  issue = {5},
  pages = {050602},
  numpages = {7},
  year = {2024},
  month = {Jul},
  publisher = {American Physical Society},
  doi = {10.1103/PhysRevLett.133.050602},
  url = {https://link.aps.org/doi/10.1103/PhysRevLett.133.050602}
}

@misc{singh_quantum_2025,
	title = {Quantum {Computing} in {Discrete}- and {Continuous}-{Variable} {Architectures}},
	url = {http://arxiv.org/abs/2507.01146},
	doi = {10.48550/arXiv.2507.01146},
	urldate = {2025-12-15},
	publisher = {arXiv},
	author = {Singh, Shraddha},
	month = jul,
	year = {2025},
	note = {arXiv:2507.01146 [quant-ph]},
}

@misc{singh_shraggynaqsp_simulations_2025,
	title = {Towards Non-Abelian Quantum Signal Processing: Efficient Control of Hybrid Continuous- and Discrete-Variable Architectures [{GitHub}]},
	url = {https://github.com/shraggy/NA-QSP_sims},
	urldate = {2026-01-02},
	author = {Singh, Shraddha},
	month = jan,
	year = {2026},
}

@misc{singh_shraggybb1gcr_simulations_2026,
	title = {Towards Non-Abelian Quantum Signal Processing: BB1(GCR) analysis [{GitHub}]},
	url = {https://github.com/shraggy/BB1_GCR},
	urldate = {2026-09-23},
	author = {Singh, Shraddha},
	month = sep,
	year = {2026},
}

@article{xsgf-txnq,
  title = {Generalized parity measurements and efficient large multicomponent cat state preparation with quantum signal processing},
  author = {Zeytino\ifmmode \breve{g}\else \u{g}\fi{}lu, Sina},
  journal = {Phys. Rev. Res.},
  volume = {8},
  issue = {2},
  pages = {023208},
  numpages = {8},
  year = {2026},
  month = {May},
  publisher = {American Physical Society},
  doi = {10.1103/xsgf-txnq},
  url = {https://link.aps.org/doi/10.1103/xsgf-txnq}
}

@article{HigginsRPE2007,
  title={Entanglement-free Heisenberg-limited phase estimation},
  author={Higgins, B. L. and Berry, D. W. and Bartlett, S. D. and Wiseman, H. M. and Pryde, G. J.},
  journal={Nature},
  volume={450},
  pages={393--396},
  year={2007}
}

@article{KimmelRPE2015,
  title={Robust calibration of a universal single-qubit gate set via robust phase estimation},
  author={Kimmel, Shelby and Low, Guang Hao and Yoder, Theodore J.},
  journal={Phys. Rev. A},
  volume={92},
  pages={062315},
  year={2015}
}

\end{document}